\documentclass[reprint, showpacs,
preprintnumbers,
amsmath,amssymb,
aps, floatfix,
superscriptaddress,
prd, nofootinbib]{revtex4-1}

\usepackage{graphicx}
\usepackage{dcolumn}
\usepackage{bm}
\usepackage{hyperref} 
\usepackage{siunitx} 
\usepackage[utf8]{inputenc}
\usepackage{amsmath}
\usepackage{accents}
\usepackage[version=4]{mhchem} 
\usepackage{enumitem}
\usepackage{xcolor}

\newcommand{\Lik}[3]{\mathcal{L}_{#1}\left(#2 \middle| #3\right)}

\newlength{\dhatheight}
\newcommand{\doublehat}[1]{%
    \settoheight{\dhatheight}{\ensuremath{\hat{#1}}}%
    \addtolength{\dhatheight}{-0.15ex}%
    \hat{\vphantom{\rule{1pt}{\dhatheight}}%
    \smash{\hat{#1}}}}
\newcommand{\PDF}{\mathcal{P}}
\newcommand*\diff{\mathop{}\!\mathrm{d}}

\newcommand{\D}[2][]{\ensuremath{\operatorname{d}^{#1}\!{#2}}}

\begin{document}

\title{Constraints on effective field theory couplings using 311.2 days of LUX data}
\author{D.S.~Akerib} \affiliation{SLAC National Accelerator Laboratory, 2575 Sand Hill Road, Menlo Park, CA 94205, USA} \affiliation{Kavli Institute for Particle Astrophysics and Cosmology, Stanford University, 452 Lomita Mall, Stanford, CA 94309, USA} 
\author{S.~Alsum}  \email{salsum@wisc.edu} \affiliation{University of Wisconsin-Madison, Department of Physics, 1150 University Ave., Madison, WI 53706, USA}  
\author{H.M.~Ara\'{u}jo} \affiliation{Imperial College London, High Energy Physics, Blackett Laboratory, London SW7 2BZ, United Kingdom}  
\author{X.~Bai} \affiliation{South Dakota School of Mines and Technology, 501 East St Joseph St., Rapid City, SD 57701, USA}  
\author{J.~Balajthy} \affiliation{University of California Davis, Department of Physics, One Shields Ave., Davis, CA 95616, USA}  
\author{J.~Bang} \affiliation{Brown University, Department of Physics, 182 Hope St., Providence, RI 02912, USA}  
\author{A.~Baxter} \affiliation{University of Liverpool, Department of Physics, Liverpool L69 7ZE, UK}  
\author{E.P.~Bernard} \affiliation{University of California Berkeley, Department of Physics, Berkeley, CA 94720, USA}  
\author{A.~Bernstein} \affiliation{Lawrence Livermore National Laboratory, 7000 East Ave., Livermore, CA 94551, USA}  
\author{T.P.~Biesiadzinski} \affiliation{SLAC National Accelerator Laboratory, 2575 Sand Hill Road, Menlo Park, CA 94205, USA} \affiliation{Kavli Institute for Particle Astrophysics and Cosmology, Stanford University, 452 Lomita Mall, Stanford, CA 94309, USA} 
\author{E.M.~Boulton} \affiliation{University of California Berkeley, Department of Physics, Berkeley, CA 94720, USA} \affiliation{Lawrence Berkeley National Laboratory, 1 Cyclotron Rd., Berkeley, CA 94720, USA} \affiliation{Yale University, Department of Physics, 217 Prospect St., New Haven, CT 06511, USA}
\author{B.~Boxer} \affiliation{University of Liverpool, Department of Physics, Liverpool L69 7ZE, UK}  
\author{P.~Br\'as} \affiliation{LIP-Coimbra, Department of Physics, University of Coimbra, Rua Larga, 3004-516 Coimbra, Portugal}  
\author{S.~Burdin} \affiliation{University of Liverpool, Department of Physics, Liverpool L69 7ZE, UK}  
\author{D.~Byram} \affiliation{University of South Dakota, Department of Physics, 414E Clark St., Vermillion, SD 57069, USA} \affiliation{South Dakota Science and Technology Authority, Sanford Underground Research Facility, Lead, SD 57754, USA} 
\author{M.C.~Carmona-Benitez} \affiliation{Pennsylvania State University, Department of Physics, 104 Davey Lab, University Park, PA  16802-6300, USA}  
\author{C.~Chan} \affiliation{Brown University, Department of Physics, 182 Hope St., Providence, RI 02912, USA}  
\author{J.E.~Cutter} \affiliation{University of California Davis, Department of Physics, One Shields Ave., Davis, CA 95616, USA}  
\author{L.~de\,Viveiros}  \affiliation{Pennsylvania State University, Department of Physics, 104 Davey Lab, University Park, PA  16802-6300, USA}  
\author{E.~Druszkiewicz} \affiliation{University of Rochester, Department of Physics and Astronomy, Rochester, NY 14627, USA}  
\author{A.~Fan} \affiliation{SLAC National Accelerator Laboratory, 2575 Sand Hill Road, Menlo Park, CA 94205, USA} \affiliation{Kavli Institute for Particle Astrophysics and Cosmology, Stanford University, 452 Lomita Mall, Stanford, CA 94309, USA} 
\author{S.~Fiorucci} \affiliation{Lawrence Berkeley National Laboratory, 1 Cyclotron Rd., Berkeley, CA 94720, USA} \affiliation{Brown University, Department of Physics, 182 Hope St., Providence, RI 02912, USA} 
\author{R.J.~Gaitskell} \affiliation{Brown University, Department of Physics, 182 Hope St., Providence, RI 02912, USA}  
\author{C.~Ghag} \affiliation{Department of Physics and Astronomy, University College London, Gower Street, London WC1E 6BT, United Kingdom}  
\author{M.G.D.~Gilchriese} \affiliation{Lawrence Berkeley National Laboratory, 1 Cyclotron Rd., Berkeley, CA 94720, USA}  
\author{C.~Gwilliam} \affiliation{University of Liverpool, Department of Physics, Liverpool L69 7ZE, UK}  
\author{C.R.~Hall} \affiliation{University of Maryland, Department of Physics, College Park, MD 20742, USA}  
\author{S.J.~Haselschwardt} \affiliation{University of California Santa Barbara, Department of Physics, Santa Barbara, CA 93106, USA}  
\author{S.A.~Hertel} \affiliation{University of Massachusetts, Amherst Center for Fundamental Interactions and Department of Physics, Amherst, MA 01003-9337 USA} \affiliation{Lawrence Berkeley National Laboratory, 1 Cyclotron Rd., Berkeley, CA 94720, USA} 
\author{D.P.~Hogan} \affiliation{University of California Berkeley, Department of Physics, Berkeley, CA 94720, USA}  
\author{M.~Horn} \affiliation{South Dakota Science and Technology Authority, Sanford Underground Research Facility, Lead, SD 57754, USA} \affiliation{University of California Berkeley, Department of Physics, Berkeley, CA 94720, USA} 
\author{D.Q.~Huang} \affiliation{Brown University, Department of Physics, 182 Hope St., Providence, RI 02912, USA}  
\author{C.M.~Ignarra} \affiliation{SLAC National Accelerator Laboratory, 2575 Sand Hill Road, Menlo Park, CA 94205, USA} \affiliation{Kavli Institute for Particle Astrophysics and Cosmology, Stanford University, 452 Lomita Mall, Stanford, CA 94309, USA} 
\author{R.G.~Jacobsen} \affiliation{University of California Berkeley, Department of Physics, Berkeley, CA 94720, USA}  
\author{O.~Jahangir} \affiliation{Department of Physics and Astronomy, University College London, Gower Street, London WC1E 6BT, United Kingdom}  
\author{W.~Ji} \affiliation{SLAC National Accelerator Laboratory, 2575 Sand Hill Road, Menlo Park, CA 94205, USA} \affiliation{Kavli Institute for Particle Astrophysics and Cosmology, Stanford University, 452 Lomita Mall, Stanford, CA 94309, USA} 
\author{K.~Kamdin} \affiliation{University of California Berkeley, Department of Physics, Berkeley, CA 94720, USA} \affiliation{Lawrence Berkeley National Laboratory, 1 Cyclotron Rd., Berkeley, CA 94720, USA} 
\author{K.~Kazkaz} \affiliation{Lawrence Livermore National Laboratory, 7000 East Ave., Livermore, CA 94551, USA}  
\author{D.~Khaitan} \affiliation{University of Rochester, Department of Physics and Astronomy, Rochester, NY 14627, USA}  
\author{E.V.~Korolkova} \affiliation{University of Sheffield, Department of Physics and Astronomy, Sheffield, S3 7RH, United Kingdom}  
\author{S.~Kravitz} \affiliation{Lawrence Berkeley National Laboratory, 1 Cyclotron Rd., Berkeley, CA 94720, USA}  
\author{V.A.~Kudryavtsev} \affiliation{University of Sheffield, Department of Physics and Astronomy, Sheffield, S3 7RH, United Kingdom}  
\author{E.~Leason} \affiliation{SUPA, School of Physics and Astronomy, University of Edinburgh, Edinburgh EH9 3FD, United Kingdom}  
\author{B.G.~Lenardo} \affiliation{University of California Davis, Department of Physics, One Shields Ave., Davis, CA 95616, USA} \affiliation{Lawrence Livermore National Laboratory, 7000 East Ave., Livermore, CA 94551, USA} 
\author{K.T.~Lesko} \affiliation{Lawrence Berkeley National Laboratory, 1 Cyclotron Rd., Berkeley, CA 94720, USA}  
\author{J.~Liao} \affiliation{Brown University, Department of Physics, 182 Hope St., Providence, RI 02912, USA}  
\author{J.~Lin} \affiliation{University of California Berkeley, Department of Physics, Berkeley, CA 94720, USA}  
\author{A.~Lindote} \affiliation{LIP-Coimbra, Department of Physics, University of Coimbra, Rua Larga, 3004-516 Coimbra, Portugal}  
\author{M.I.~Lopes} \affiliation{LIP-Coimbra, Department of Physics, University of Coimbra, Rua Larga, 3004-516 Coimbra, Portugal}  
\author{A.~Manalaysay} \affiliation{Lawrence Berkeley National Laboratory, 1 Cyclotron Rd., Berkeley, CA 94720, USA} \affiliation{University of California Davis, Department of Physics, One Shields Ave., Davis, CA 95616, USA} 
\author{R.L.~Mannino} \affiliation{Texas A \& M University, Department of Physics, College Station, TX 77843, USA} \affiliation{University of Wisconsin-Madison, Department of Physics, 1150 University Ave., Madison, WI 53706, USA} 
\author{N.~Marangou} \affiliation{Imperial College London, High Energy Physics, Blackett Laboratory, London SW7 2BZ, United Kingdom}  
\author{D.N.~McKinsey} \affiliation{University of California Berkeley, Department of Physics, Berkeley, CA 94720, USA} \affiliation{Lawrence Berkeley National Laboratory, 1 Cyclotron Rd., Berkeley, CA 94720, USA} 
\author{D.-M.~Mei} \affiliation{University of South Dakota, Department of Physics, 414E Clark St., Vermillion, SD 57069, USA}  
\author{J.A.~Morad} \affiliation{University of California Davis, Department of Physics, One Shields Ave., Davis, CA 95616, USA}  
\author{A.St.J.~Murphy} \affiliation{SUPA, School of Physics and Astronomy, University of Edinburgh, Edinburgh EH9 3FD, United Kingdom}  
\author{A.~Naylor} \affiliation{University of Sheffield, Department of Physics and Astronomy, Sheffield, S3 7RH, United Kingdom}  
\author{C.~Nehrkorn} \affiliation{University of California Santa Barbara, Department of Physics, Santa Barbara, CA 93106, USA}  
\author{H.N.~Nelson} \affiliation{University of California Santa Barbara, Department of Physics, Santa Barbara, CA 93106, USA}  
\author{F.~Neves} \affiliation{LIP-Coimbra, Department of Physics, University of Coimbra, Rua Larga, 3004-516 Coimbra, Portugal}  
\author{A.~Nilima} \affiliation{SUPA, School of Physics and Astronomy, University of Edinburgh, Edinburgh EH9 3FD, United Kingdom}  
\author{K.C.~Oliver-Mallory} \affiliation{Imperial College London, High Energy Physics, Blackett Laboratory, London SW7 2BZ, United Kingdom} \affiliation{University of California Berkeley, Department of Physics, Berkeley, CA 94720, USA} \affiliation{Lawrence Berkeley National Laboratory, 1 Cyclotron Rd., Berkeley, CA 94720, USA}
\author{K.J.~Palladino} \affiliation{University of Oxford, Department of Physics, Oxford OX1 3RH, United Kingdon} \affiliation{University of Wisconsin-Madison, Department of Physics, 1150 University Ave., Madison, WI 53706, USA} 
\author{C.~Rhyne} \affiliation{Brown University, Department of Physics, 182 Hope St., Providence, RI 02912, USA}  
\author{Q.~Riffard} \affiliation{University of California Berkeley, Department of Physics, Berkeley, CA 94720, USA} \affiliation{Lawrence Berkeley National Laboratory, 1 Cyclotron Rd., Berkeley, CA 94720, USA} 
\author{G.R.C.~Rischbieter} \email{grischbieter@albany.edu} \affiliation{University at Albany, State University of New York, Department of Physics, 1400 Washington Ave., Albany, NY 12222, USA}  
\author{P.~Rossiter} \affiliation{University of Sheffield, Department of Physics and Astronomy, Sheffield, S3 7RH, United Kingdom}  
\author{S.~Shaw} \affiliation{University of California Santa Barbara, Department of Physics, Santa Barbara, CA 93106, USA} \affiliation{Department of Physics and Astronomy, University College London, Gower Street, London WC1E 6BT, United Kingdom} 
\author{T.A.~Shutt} \affiliation{SLAC National Accelerator Laboratory, 2575 Sand Hill Road, Menlo Park, CA 94205, USA} \affiliation{Kavli Institute for Particle Astrophysics and Cosmology, Stanford University, 452 Lomita Mall, Stanford, CA 94309, USA} 
\author{C.~Silva} \affiliation{LIP-Coimbra, Department of Physics, University of Coimbra, Rua Larga, 3004-516 Coimbra, Portugal}  
\author{M.~Solmaz} \affiliation{University of California Santa Barbara, Department of Physics, Santa Barbara, CA 93106, USA}  
\author{V.N.~Solovov} \affiliation{LIP-Coimbra, Department of Physics, University of Coimbra, Rua Larga, 3004-516 Coimbra, Portugal}  
\author{P.~Sorensen} \affiliation{Lawrence Berkeley National Laboratory, 1 Cyclotron Rd., Berkeley, CA 94720, USA}  
\author{T.J.~Sumner} \affiliation{Imperial College London, High Energy Physics, Blackett Laboratory, London SW7 2BZ, United Kingdom}  
\author{N.~Swanson} \affiliation{Brown University, Department of Physics, 182 Hope St., Providence, RI 02912, USA}  
\author{M.~Szydagis} \affiliation{University at Albany, State University of New York, Department of Physics, 1400 Washington Ave., Albany, NY 12222, USA}  
\author{D.J.~Taylor} \affiliation{South Dakota Science and Technology Authority, Sanford Underground Research Facility, Lead, SD 57754, USA}  
\author{R.~Taylor} \affiliation{Imperial College London, High Energy Physics, Blackett Laboratory, London SW7 2BZ, United Kingdom}  
\author{W.C.~Taylor} \affiliation{Brown University, Department of Physics, 182 Hope St., Providence, RI 02912, USA}  
\author{B.P.~Tennyson} \affiliation{Yale University, Department of Physics, 217 Prospect St., New Haven, CT 06511, USA}  
\author{P.A.~Terman} \affiliation{Texas A \& M University, Department of Physics, College Station, TX 77843, USA}  
\author{D.R.~Tiedt} \affiliation{University of Maryland, Department of Physics, College Park, MD 20742, USA}  
\author{W.H.~To} \affiliation{California State University Stanislaus, Department of Physics, 1 University Circle, Turlock, CA 95382, USA}  
\author{L.~Tvrznikova} \affiliation{University of California Berkeley, Department of Physics, Berkeley, CA 94720, USA} \affiliation{Lawrence Berkeley National Laboratory, 1 Cyclotron Rd., Berkeley, CA 94720, USA} \affiliation{Yale University, Department of Physics, 217 Prospect St., New Haven, CT 06511, USA}
\author{U.~Utku} \affiliation{Department of Physics and Astronomy, University College London, Gower Street, London WC1E 6BT, United Kingdom}  
\author{A.~Vacheret} \affiliation{Imperial College London, High Energy Physics, Blackett Laboratory, London SW7 2BZ, United Kingdom}  
\author{A.~Vaitkus} \affiliation{Brown University, Department of Physics, 182 Hope St., Providence, RI 02912, USA}  
\author{V.~Velan} \affiliation{University of California Berkeley, Department of Physics, Berkeley, CA 94720, USA}  
\author{R.C.~Webb} \affiliation{Texas A \& M University, Department of Physics, College Station, TX 77843, USA}  
\author{J.T.~White} \affiliation{Texas A \& M University, Department of Physics, College Station, TX 77843, USA}  
\author{T.J.~Whitis} \affiliation{SLAC National Accelerator Laboratory, 2575 Sand Hill Road, Menlo Park, CA 94205, USA} \affiliation{Kavli Institute for Particle Astrophysics and Cosmology, Stanford University, 452 Lomita Mall, Stanford, CA 94309, USA} 
\author{M.S.~Witherell} \affiliation{Lawrence Berkeley National Laboratory, 1 Cyclotron Rd., Berkeley, CA 94720, USA}  
\author{F.L.H.~Wolfs} \affiliation{University of Rochester, Department of Physics and Astronomy, Rochester, NY 14627, USA}  
\author{D.~Woodward} \affiliation{Pennsylvania State University, Department of Physics, 104 Davey Lab, University Park, PA  16802-6300, USA}  
\author{X.~Xiang} \affiliation{Brown University, Department of Physics, 182 Hope St., Providence, RI 02912, USA}  
\author{J.~Xu} \affiliation{Lawrence Livermore National Laboratory, 7000 East Ave., Livermore, CA 94551, USA}  
\author{C.~Zhang} \affiliation{University of South Dakota, Department of Physics, 414E Clark St., Vermillion, SD 57069, USA}

\collaboration{LUX Collaboration}

\date{\today}

\begin{abstract}
\small
\noindent We report here the results of a nonrelativistic effective field theory (EFT) WIMP search analysis using LUX data. We build upon previous LUX analyses by extending the search window to include nuclear recoil energies up to $\sim$180~keV$_{nr}$, requiring a reassessment of data quality criteria and background models. In order to use an unbinned profile likelihood statistical framework, the development of new analysis techniques to account for higher-energy backgrounds was required. With a 3.14$\times10^4$~kg$\cdot$day exposure using data collected between 2014 and 2016, we find our data is compatible with the background expectation and set 90\%~C.L. exclusion limits on nonrelativistic EFT WIMP-nucleon couplings, improving upon previous LUX results and providing constraints on a EFT WIMP interactions using the $\{$neutron,proton$\}$ interaction basis. Additionally, we report exclusion limits on inelastic EFT WIMP-isoscalar recoils that are competitive and world-leading for several interaction operators. 
\normalsize
\end{abstract}

\maketitle

\section{Introduction}\label{sec:introduction}

Over the last century, an abundance of evidence suggests that nonbaryonic, nonluminous ``dark matter'' comprises approximately 25\% of the universe's energy density~\cite{RubinFord1970,clowe2006,SDSS2014, PLANCK2015overview, hu2002}. A popular dark matter candidate has been the weakly interacting massive particle (WIMP) with masses between 10 GeV and several TeV~\cite{feng2010}. However, nongravitational interactions with dark matter have never been definitively observed, despite many dedicated experiments over the last several decades~\cite{xenon10_results, cresst, Agnese_2019,  ZEPLIN_WIMPsearch2,Armengaud_2011, Agnes_2016,Akerib:2016FullLux, Aprile_2018, Cui_2017}. 

In an attempt to detect dark matter, the Large Underground Xenon Experiment (LUX) collected data between 2013 and 2016, while being hosted 4850 feet underground in the Davis Cavern at the Sanford Underground Research Facility (SURF) in Lead, South Dakota. The LUX detector was a dual-phase time projection chamber (TPC) equipped with an active xenon mass of 250~kg to detect the possible interactions between WIMP dark matter and Standard Model nucleons. Liquid xenon is a promising target medium for dark matter searches, as it constitutes a dense, stable target with well-developed purification techniques to minimize background contamination that may overwhelm a potential WIMP signal or hinder detection of xenon scintillation and ionization~\cite{purification, JonThesis}. LUX set world-leading limits in the mass range of $\mathcal{O}$(GeV)-$\mathcal{O}$(TeV) for Spin-Independent (SI) WIMP interactions and Spin-Dependent (SD) interactions with neutrons~\cite{luxrun3_2013,luxrun42016,Akerib:2016FullLux,LUX_SD_Run3_Run4}. Those results were confirmed and improved upon by other Xe TPC-based experiments: XENON1T and PandaX~\cite{Aprile_2018,Cui_2017}.

In this paper, following theoretical work by Fan \textit{et al.}~\cite{Fan_2010} and Fitzpatrick \textit{et al.}~\cite{Fitzpatrick:EFT}, with conventions set by Anand \textit{et al.}~\cite{Anand:MathematicaEFT}, we extend prior analyses by utilizing a generalized nonrelativistic EFT approach going beyond simple SI and SD couplings, with the inclusion of momentum-dependent and velocity-dependent operators. All operators in the elastic WIMP-nucleon interaction, under momentum conservation and Galilean invariance, can be reduced to a basis of four Hermitian quantities:
\begin{equation}
i \frac{\vec{q}}{m_N}, \quad \vec{v}^\perp \equiv \vec{v} + \frac{\vec{q}}{2\mu}, \quad \vec{S}_{\chi}, \quad \vec{S}_{N}
\label{basis}
\end{equation}
where $\vec{q}$ is the momentum transferred from the WIMP to the nucleus, $m_N$ is the nucleon mass, $\vec{v}^\perp$ is the component of the relative velocity between the WIMP and the nucleon that is perpendicular to the momentum transfer, $\vec{S_\chi}$ is the spin of the WIMP, and $\vec{S_N}$ the spin of the relevant nucleon.

Linear combinations of these quantities up to second-order in $\vec{q}$ are combined and result in fifteen independent and dimensionless EFT operators:
\begin{equation}
\begin{array}{l}{\mathcal{O}_{1}=1_\chi 1_N} \\ 
{\mathcal{O}_{2}=\left(v^{\perp}\right)^{2}} \\ 
{\mathcal{O}_{3}=i \vec{S}_{N} \cdot\left(\frac{\vec{q}}{m_N} \times \vec{v}^{\perp}\right)} \\ 
{\mathcal{O}_{4}=\vec{S}_{\chi} \cdot \vec{S}_{N}} \\ 
{\mathcal{O}_{5}=i \vec{S}_{\chi} \cdot(\frac{\vec{q}}{m_N} \times \vec{v}^{\perp})} \\ 
{\mathcal{O}_{6}=\left(\vec{S}_{\chi} \cdot \frac{\vec{q}}{m_N}\right)\left(\vec{S}_{N} \cdot \frac{\vec{q}}{m_N}\right)} \\
{\mathcal{O}_{7}=\vec{S}_{N} \cdot \vec{v}^{\perp}} \\
{\mathcal{O}_{8}=\vec{S}_{\chi} \cdot \vec{v}^{\perp}} \\ 
\mathcal{O}_{9}=i \vec{S}_{\chi} \cdot\left(\vec{S}_{N} \times \frac{\vec{q}}{m_N}\right) \\
{\mathcal{O}_{10}=i \vec{S}_{N} \cdot \frac{\vec{q}}{m_N}} \\ 
{\mathcal{O}_{11}=i \vec{S}_{\chi} \cdot \frac{\vec{q}}{m_N}} \\
\mathcal{O}_{12} =\vec{S}_{\chi} \cdot\left(\vec{S}_{N} \times \vec{v}^{\perp}\right) \\
\mathcal{O}_{13} =i\left(\vec{S}_{\chi} \cdot \vec{v}^{\perp}\right)\left(\vec{S}_{N} \cdot \frac{\vec{q}}{m_N}\right) \\ 
\mathcal{O}_{14} =i\left(\vec{S}_{\chi} \cdot \frac{\vec{q}}{m_N}\right)\left(\vec{S}_{N} \cdot \vec{v}^{\perp}\right) \\ 
\mathcal{O}_{15} =-\left(\vec{S}_{\chi} \cdot \frac{\vec{q}}{m_N}\right)\left(\left(\vec{S}_{N} \times \vec{v}^{\perp}\right) \cdot \frac{\vec{q}}{m_N}\right) .
\end{array}
\label{operators}
\end{equation}

\begin{figure*}
    \includegraphics[width=0.315\textwidth]{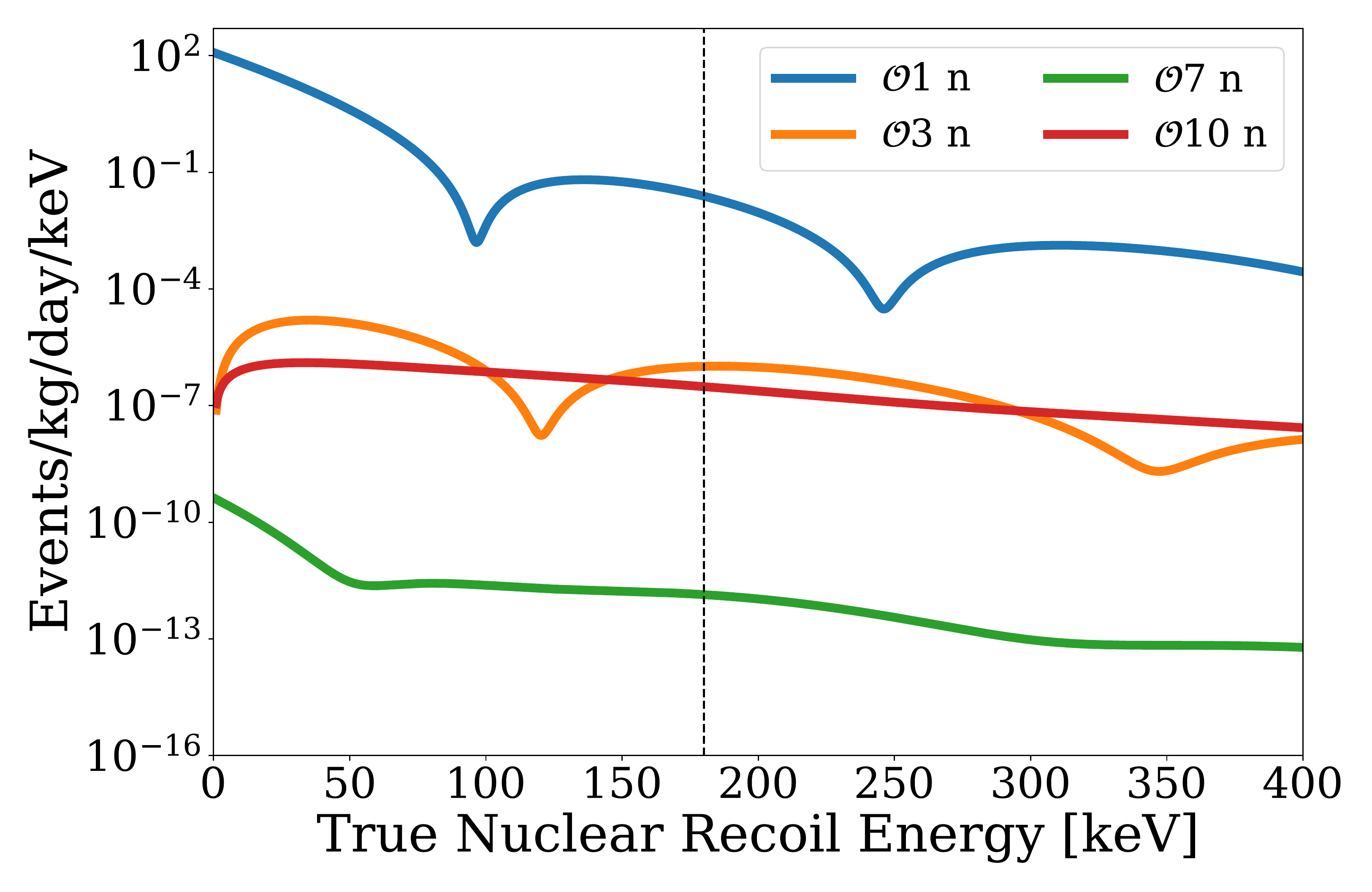}
  	\includegraphics[width=0.315\textwidth]{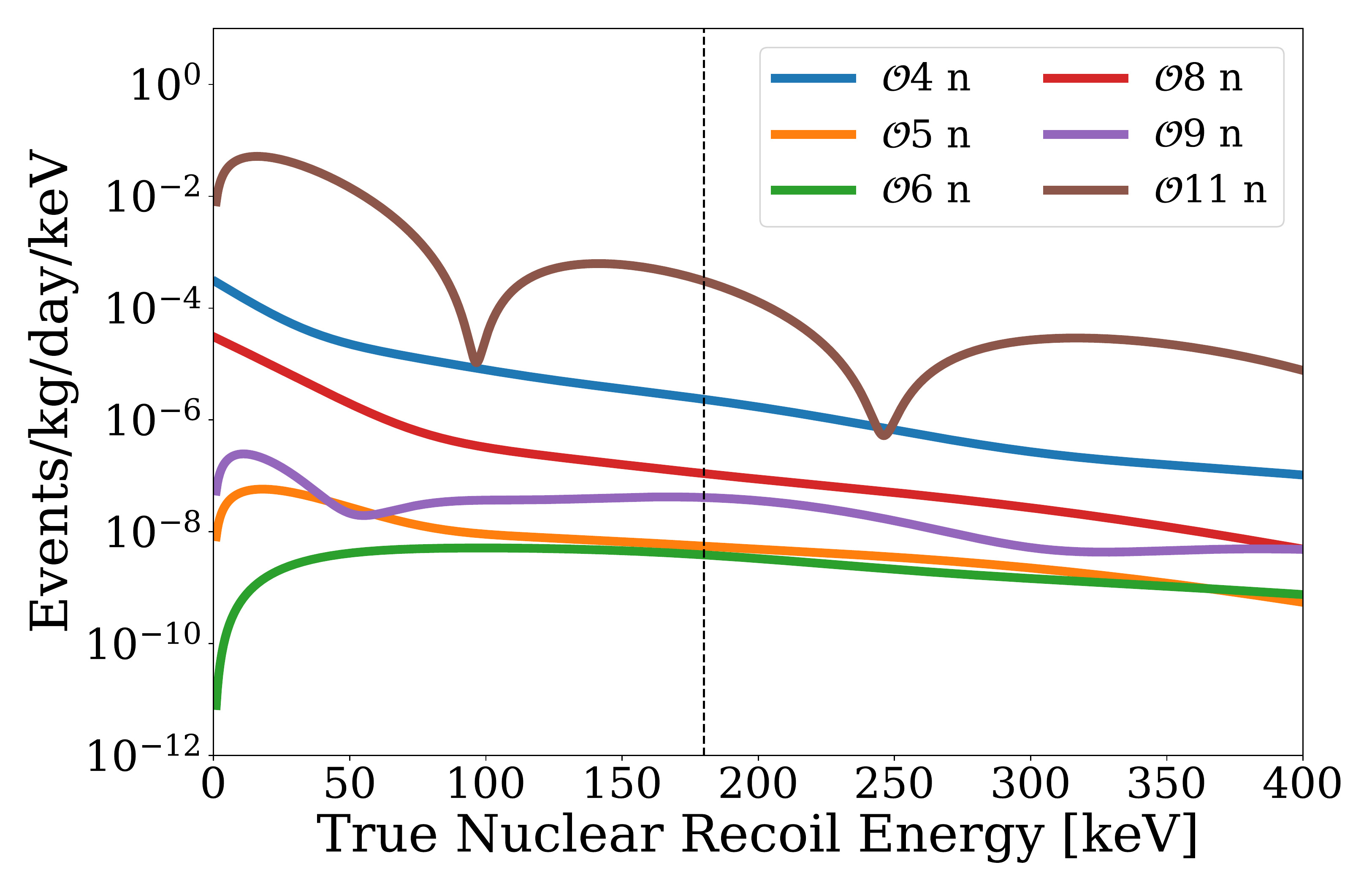}
  	\includegraphics[width=0.315\textwidth]{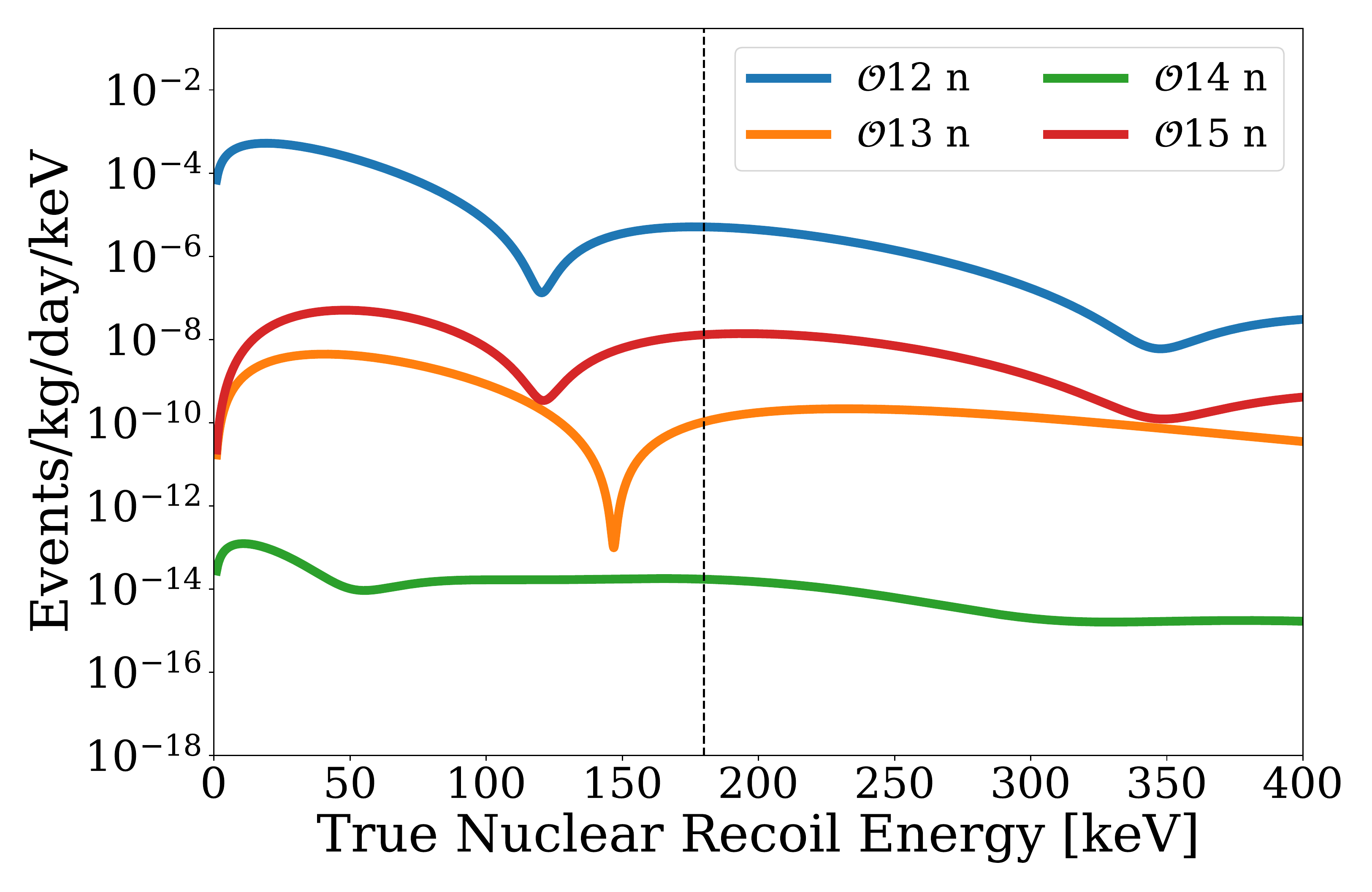}
  	\includegraphics[width=0.315\textwidth]{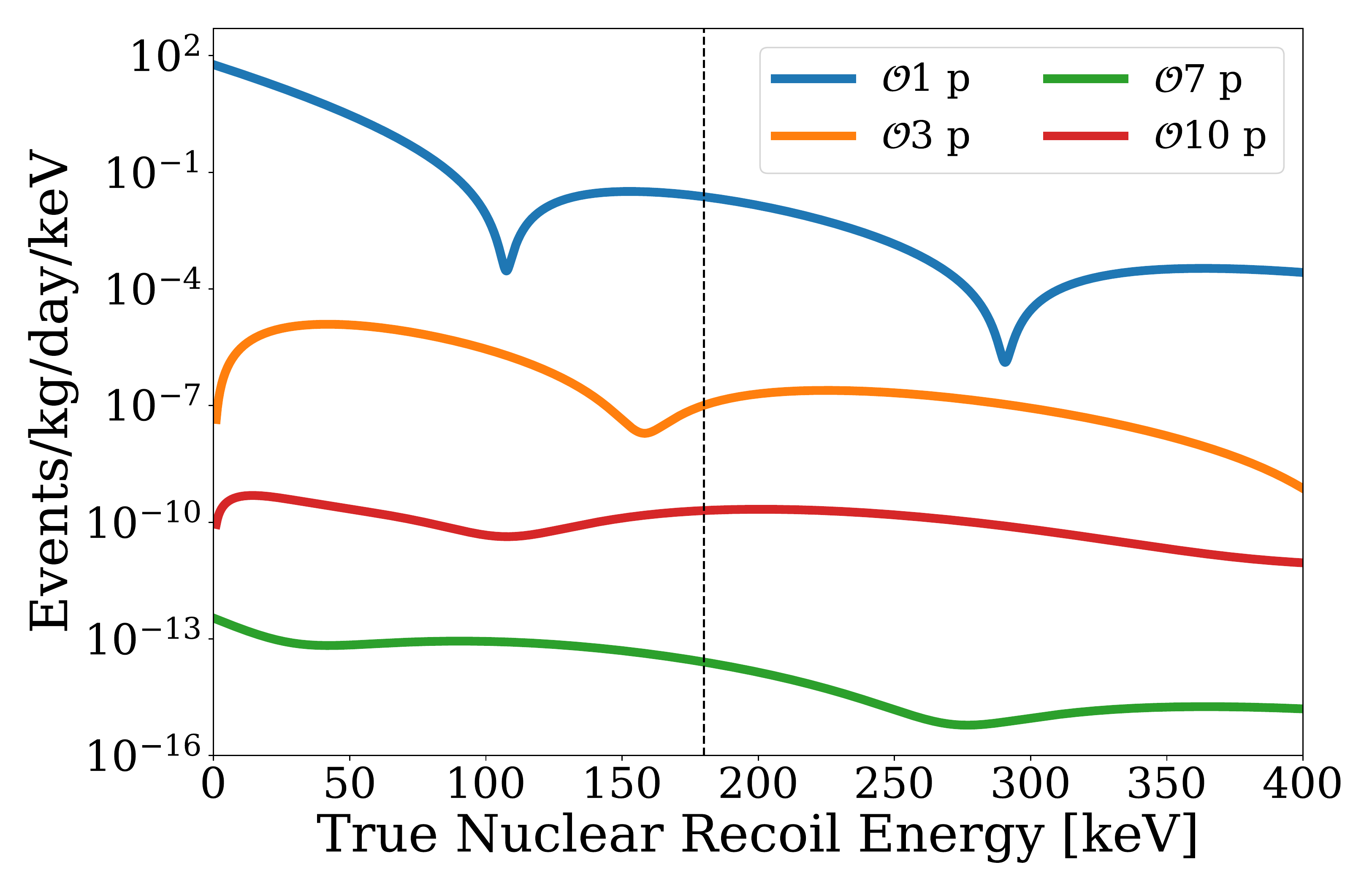}
  	\includegraphics[width=0.315\textwidth]{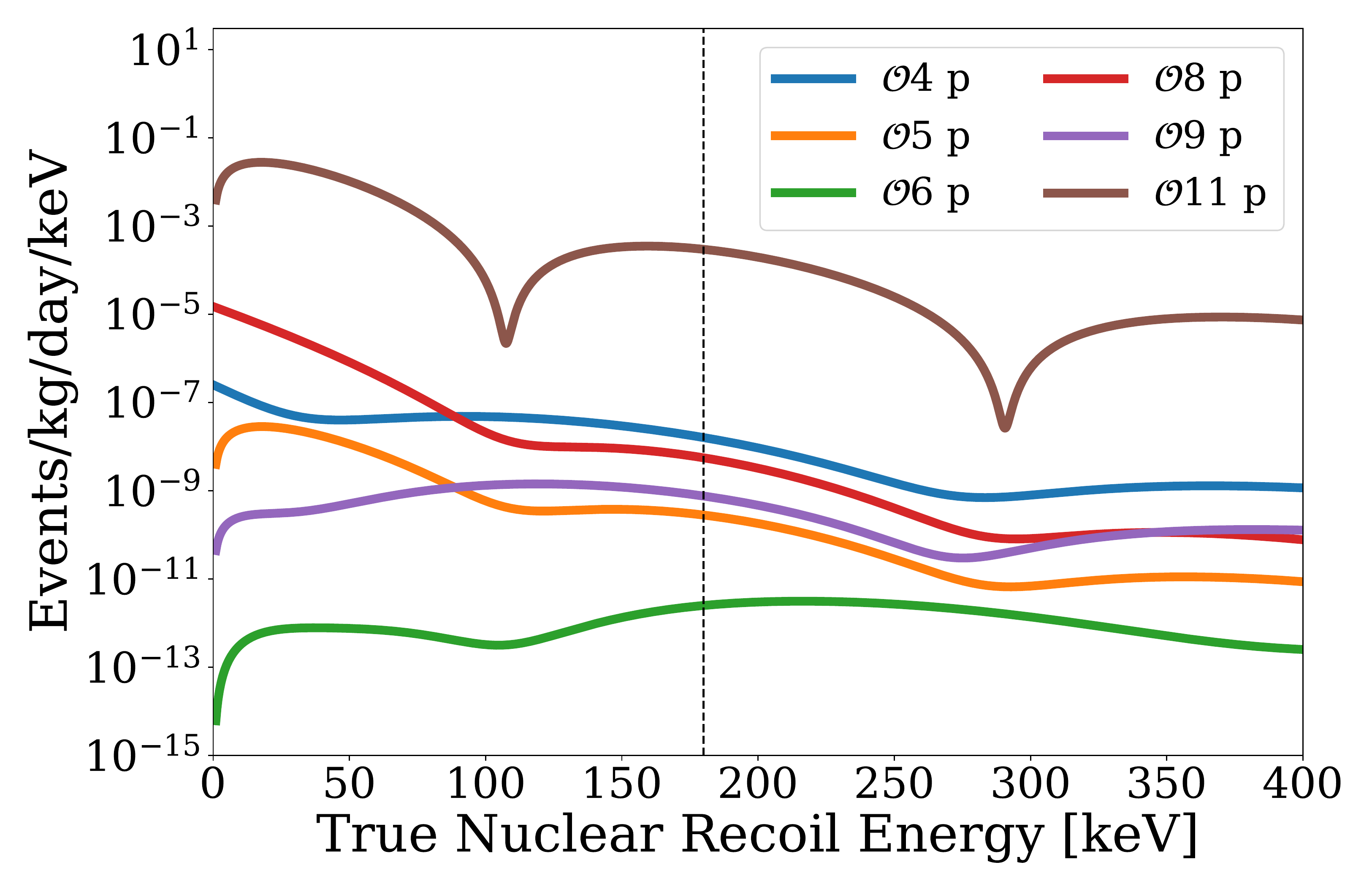}
  	\includegraphics[width=0.315\textwidth]{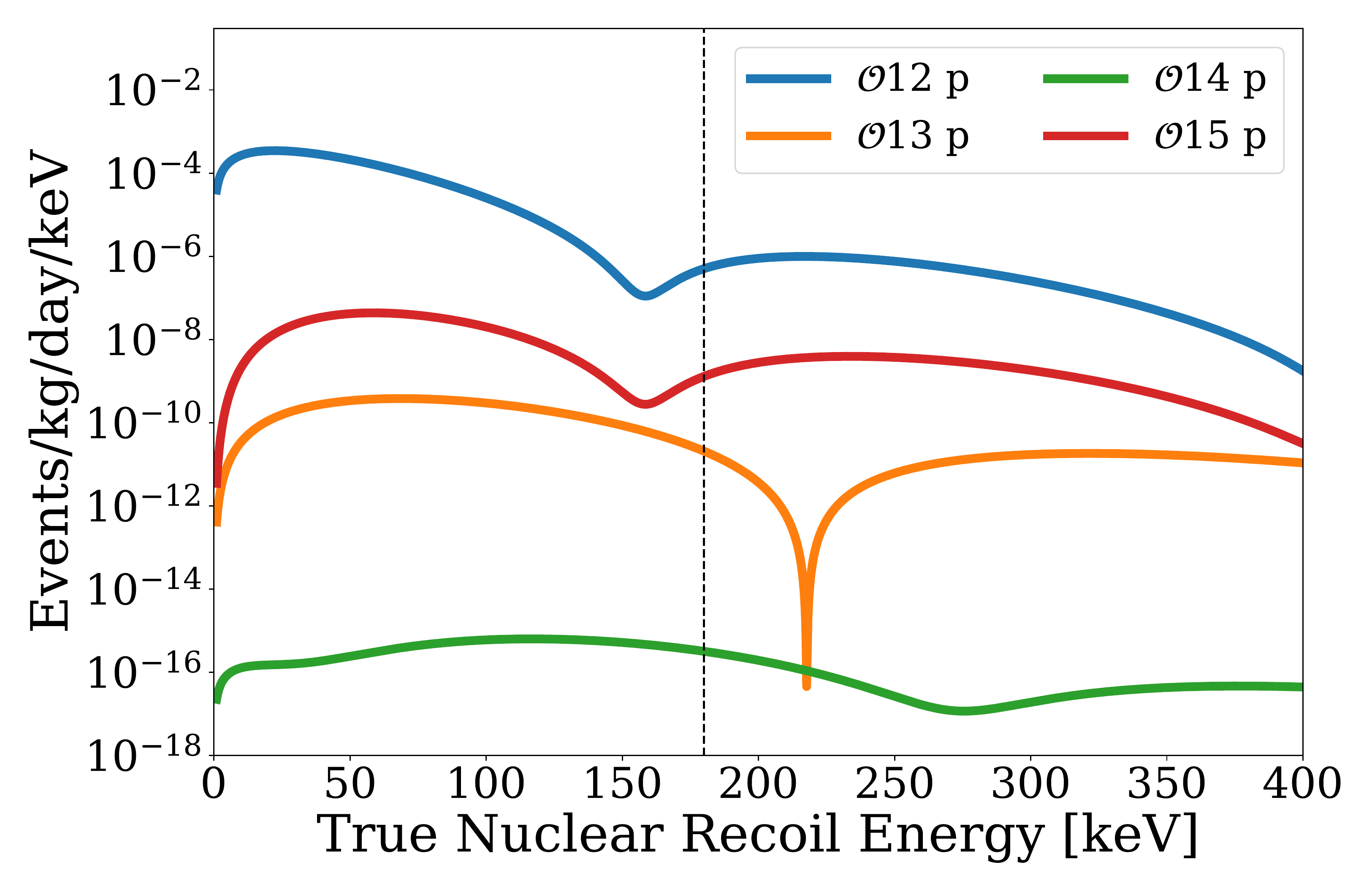}
  \caption{Differential event rates versus true nuclear recoil energy for the fourteen nonrelativistic EFT operators. This example is a 400~GeV WIMP. From left to right: $\vec{S}_{\chi}$-independent operators, $\vec{S}_{\chi}$-dependent operators, and $\vec{S}_{\chi}$-dependent operators that arise only in interactions which do not involve exchange of a spin-0 or spin-1 mediator. Plots on the top row are WIMP-n rates, while the bottom consists of WIMP-p spectra. Vertical dashed black lines correspond to the energy above which the detection efficiency for the analyses presented here falls below 50\% (see Fig.~\ref{efficiency}). For each spectrum, it is assumed that WIMPs only interact with the relevant nucleon through a single operator with the coupling strength set to unity, ignoring the possibility of interference between different operators.}
  \label{signalSpectra}
\end{figure*}

\noindent Dividing each instance of $\vec{q}$ by $m_N$ leaves each operator conveniently dimensionless without compromising the operator's Hermiticity. We neglect operator $\mathcal{O}_2$ in this analysis, as it cannot arise in the nonrelativistic limit from a relativistic operator to leading order~\cite{Anand:MathematicaEFT}. Each of these operators can in principle be coupled differently to protons versus neutrons (or equivalently, to isoscalars versus isovectors); therefore, we consider 28 different couplings in this analysis. In an actual experiment the dark matter would not couple to an individual nucleon, but to a composite nucleus. This leads to a series of nuclear responses that can vary by target isotope causing certain targets to be better at probing certain operator couplings. Additionally, while the recoil energy spectrum for momentum-independent interactions peaks at zero energy due to kinematics, momentum-dependent operators can have significant contributions at energies well above nuclear recoil energies of 100~keV, motivating analysis of a larger energy window than that used in other LUX analyses \cite{luxrun3_2013,luxrun42016,Akerib:2016FullLux,LUX_SD_Run3_Run4}. Figure~\ref{signalSpectra} shows the differential rate spectra for each of the nonrelativistic operators.

Additional WIMP models exist that allow for the masses of the incoming and outgoing dark matter particles to differ~\cite{Smith_2001}, typically where the WIMP transitions into a more massive state during the scattering interaction. If the value for the mass difference between the outgoing and incoming states, $\delta_m \equiv m_{\chi,out} - m_{\chi,in}$, is nonzero, the recoil rate at lower energies becomes suppressed, thus causing any observed signal to be more contained at higher energies. In certain models where the elastic scattering process is suppressed~\cite{Han_1997, Hall_1998}, inelastic transitions between WIMP states becomes the primary method of interaction.

For the case of inelastic WIMP-nucleon interactions in this report, only a slight modification of the Hermitian basis vectors is required. From conservation of energy, its required that $\vec{v}^{\perp} \cdot \vec{q} = 0$ for elastic recoils. To account for nonzero mass splitting, $\delta_m$, its required that 

\begin{equation}\label{massSplitting}
           \delta_m + \vec{v} \cdot \vec{q} +\frac{|\vec{q}|^2}{2\mu_N }= 0.
           \end{equation}
 
\noindent This requirement is included into our basis of Hermitian quantities by replacing the perpendicular velocity in Eq.~\ref{basis} with
      \begin{equation}\label{idmVelocity}
           \vec{v}^\perp_{inel} \equiv \vec{v} + \frac{\vec{q}}{2\mu} + \frac{\delta_m}{|\vec{q}|^2}\vec{q} =v^\perp + \frac{\delta_m}{|\vec{q}|^2}\vec{q},
           \end{equation}
\noindent and a similar replacement is made for each operator, $\mathcal{O}_i$. Mass splitting values of order $\mathcal{O}$(100~keV) are well-motivated~\cite{Smith_2001, Barello_2014}, therefore values between 50-200 keV are considered in this report. 

A previous EFT analysis was conducted on LUX's first WIMP search (WS) i.e. WS2013~\cite{run3eft}, consisting of 95 live-days of data collected in 2013. In our current analysis, however, we utilize the longer-duration WS2014--16: 332 live-days collected between 2014 and 2016. Additionally, we extend our focus to both elastic and inelastic EFT interactions. We focus solely on WS2014--16 data, as the detector experienced significantly different data-collection conditions between the two science runs, as described in the following section. This creates different systematics and independent analysis frameworks between the two runs, making it difficult to combine both science runs in a single analysis.  While a typical WIMP search region is restricted to lower energies, such as $\sim$40~keV$_{nr}$\footnote{ We distinguish reconstructed energies from true recoil energies in this report using the subscripts ``nr" and ``ee" in the units for reconstructed energies, referring to ``nuclear recoil" and ``electron equivalent" energies, respectively.} in LUX's SI and SD analyses~\cite{Akerib:2016FullLux,LUX_SD_Run3_Run4}, this analysis extends the region of interest (ROI) to approximately 180~keV$_{nr}$, corresponding to detected scintillation signals (S1) of up to 300 detected photons (phd). As reported in \cite{run3eft}, the extension of the WIMP ROI leads to the inclusion of backgrounds considered negligible in the traditional WIMP paradigm. In this work, we describe in detail the necessary steps to take these backgrounds into account.

\section{The LUX Experiment}\label{sec:exp}

As a two-phase TPC utilizing both liquid and gaseous Xe, LUX measures signals by extracting electrons and collecting light released by the Xe target after a recoil event. The initial interaction excites and ionizes electrons from multiple Xe atoms; some ionized electrons recombine with Xe ions, producing additional scintillation light, while others are extracted to the gas layer by an applied electric field of order $\mathcal{O}$(100~V/cm) where they produce an electroluminesence signal~\cite{LUXinstrument2013}. In LUX, initial scintillation light collection takes place on timescales of $\mathcal{O}$(10-100 ns), while the electron drift takes 0--325~$\mu$s, creating two distinct signals: S1 and S2, respectively. LUX detected S1 and S2 light via 122 photomultiplier tubes (PMTs) separated into two arrays at the top and bottom of the detector, with a photon detection efficiency of $\sim$10\%. The hit-pattern of S2 light in the top PMT array provides \{x,y\} coordinate reconstruction of the original event, while the drift time between the S1 and S2 signals provides information regarding event depth.

It is important to note that the amount of primary and secondary scintillation light collected for a given event depends on the location in the detector in which the energy deposition occurred. Because of this, ${}^{83\text{m}}$Kr dissolved in the liquid Xe (providing a spatially uniform, effectively monoenergetic 41.5~keV electron recoil calibration) was used to construct S1 and S2 detection maps in order to correct for the position-dependence in the observed S1 and S2 signals~\cite{LUX_Kr83m}. This allows us to take advantage of the following linear conversions:
\begin{equation}
    S1_{c} = g_{1} \cdot n_{\gamma};  \hspace{0.5cm} S2_c = g_{2} \cdot n_{e},
\end{equation}
where $S1_c$ and $S2_c$ are the position-corrected S1 and S2 signals, $n_{\gamma}$ and $n_{e}$ are the initial numbers of photons and electrons leaving the interaction site, and $g_1$ and $g_2$ are the scintillation and electroluminescence gains, respectively. We note that while $g_1$ is simply a geometric light collection efficiency multiplied by PMT quantum efficiency for the prompt scintillation light S1, $g_2$ is a product of the efficiency to extract electrons from the liquid to gaseous xenon, photons produced per extracted electron in the gas layer, and the S2 photon detection efficiency in the gas~\cite{GR_ERmodel}.

Discrimination between electronic recoil (ER) and nuclear recoil (NR) interactions is possible in a dual-phase xenon TPC, as the total produced quanta, the ratio between excited and ionized electrons for an energy deposition, as well as the recombination probability for ionized electrons, all differ between the two interaction types. However, discrimination is not 100\% efficient, as ER events with a stochastically lower charge-to-light ratio can ``leak'' into the expected NR signal region in \{S1, S2\} space. As we expect WIMPs to primarily produce NR, it is paramount that we minimize ER leakage, while fully characterizing all backgrounds, in order to distinguish a possible WIMP signal from them.

\begin{figure}[h]
\begin{center}
\includegraphics[width=0.5\textwidth,clip]{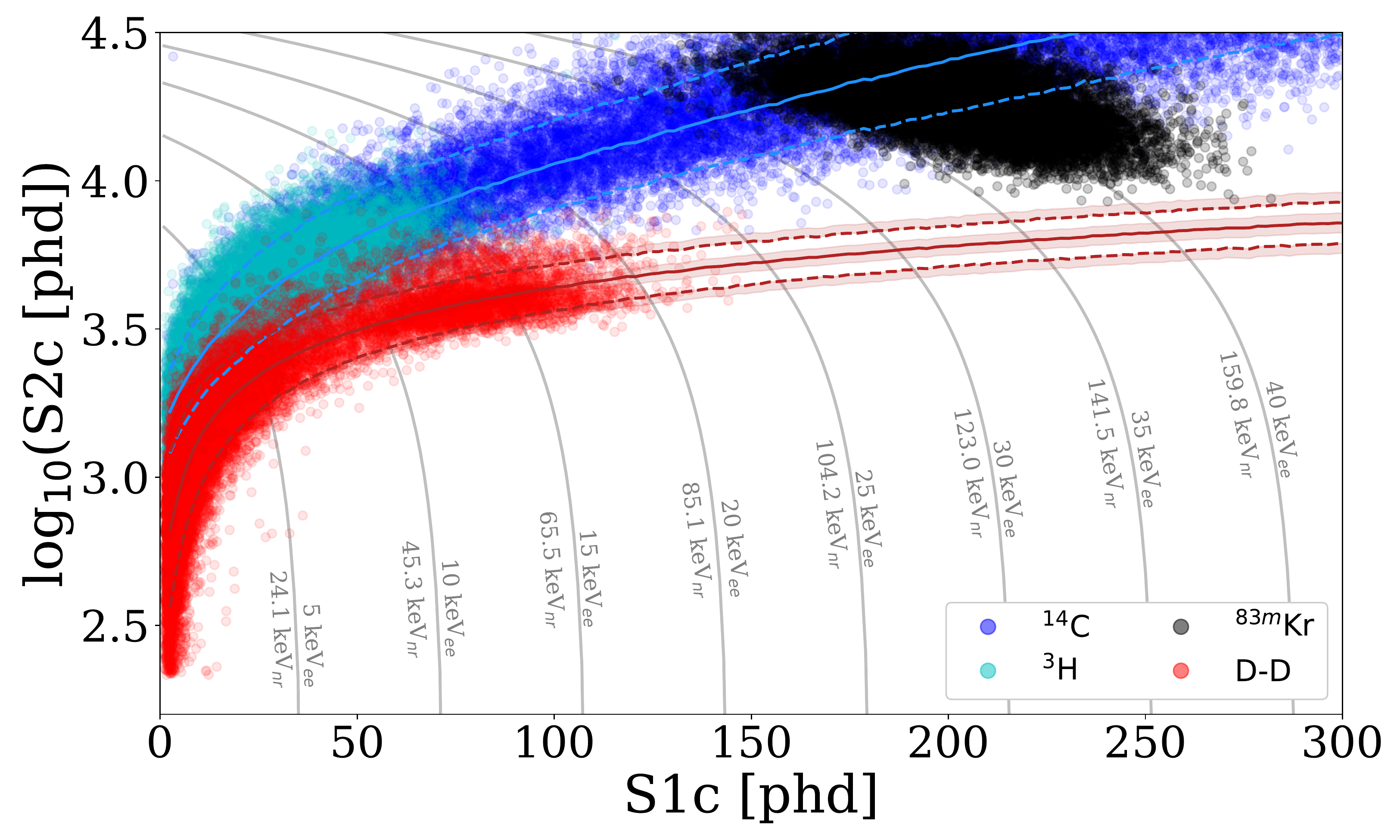}
\vspace{-10pt}
\caption{A sample of single-scatter calibration events taken near the end of WS2014--16 with drift times between 40-105~$\mu$s. Cyan points correspond to the ${}^3$H $\beta$ ER calibration; blue points correspond to the ${}^{14}$C $\beta$ ER calibration; red markers are events associated with the D-D NR calibration; and black markers are ${}^{83\text{m}}$Kr events. Each population consists of a random selection of 20,000 events. The light blue solid and dashed lines show the expected mean and 90\%~C.L. ER response region, while red solid and dashed lines show the expected mean and 90\%~C.L. NR response from NEST~v2.1.0. The shaded red region shows the uncertainty in the NR expectation based on \textit{ex situ} NR calibrations reported in the literature (see Sec.~\ref{sec:modeling}). Grey contours show lines of constant recoil energy, each labeled for both electronic recoils (keV$_{ee}$) and nuclear recoils (keV$_{nr}$).}
\vspace{-22pt}
\label{calibrations}
\end{center}
\end{figure}

To characterize the \{S1, S2\} response of liquid Xe (LXe) in LUX for both ER and NR interactions, LUX underwent periodic calibrations. For ER, tritiated methane (0--18.6~keV $\beta$ decay) was injected into the detector several times over LUX's lifetime, providing the LXe response for energies relevant to most typical lower-energy WIMP searches~\cite{tritium}. Additionally, at the end of LUX's tenure in the Davis Cavern, a $^{14}$C calibration took place (0--156~keV $\beta$ decay), allowing for characterization of the ER response out to much higher energies~\cite{C14}. For NR, an external deuterium-deuterium (D-D) fusion neutron generator was used to provide \textit{in situ} characterization of nuclear recoils between 0.7--74~keV$_{nr}$~\cite{LUXDD2015}. We note here that a nuclear recoil with a given energy produces smaller S1 and S2 signals than an ER event of the same energy; this is due to the fraction of energy being transferred to the electrons to produce ionized and excited atoms being smaller for NRs than ERs. Figure~\ref{calibrations} shows a sample of the \{S1, S2\} response for LUX's calibrations compared to expected ER and NR responses from simulation.

Before WS2014--16, LUX underwent a grid conditioning campaign to significantly increase the allowed applied drift field and extraction efficiency. However, this had the unintended consequence of creating a significant amount of trapped charge on the inner walls of the TPC, creating a spatially distorted and temporally varying drift field, varying between 50-550~V/cm as function of time and position.  3-D electrostatic models of the built-up charge density were created using the COMSOL Multiphysics software~\cite{comsol}, providing a spatial map of the electric field configuration. Field and charge maps were updated monthly, which allows for a robust understanding of the temporal features of the applied drift field. More details are reported in Ref~\cite{luxFieldModeling}. Additionally, WS2014--16 data were collected with temporally changing gain factors, where $g_1$ gradually decreased from 0.100$\pm$0.002 to 0.097$\pm$0.001~phd/photon and $g_2$ varied between 18.9$\pm$0.8 and 19.7$\pm$0.2~phd/e${}^-$~\cite{Akerib:2016FullLux}.

\section{ Data Selection }\label{dataSelection}

For this analysis, data from WS2014--16 are used. Despite the challenges from the temporally varying gain factors ($g_1$ and $g_2$) and electric field distortions, WS2014--16 has been well-characterized by multiple analyses since LUX's decommissioning~\cite{luxFieldModeling,vvelanDisc,GR_ERmodel}. As the EFT ROI is significantly larger in \{S1, S2\} space than in the SI and SD WIMP analyses, implementation of data selection criteria are crucial for removing backgrounds, including: events with poor position reconstruction; multiple scatters with merged S2 signals; events with gaseous xenon interactions classified as the event's S2; and events with an overabundance of non-S1 and non-S2 pulses such as single photons and electrons not associated with the observed S1 or S2. To minimize potential bias when creating these selection criteria (described in more detail below), the WS2014--16 data were ``salted" with artificial WIMP-like events at early stages of the data-processing pipeline. Salt was added to the data in an early stage of the data processing pipeline and was manufactured from ${}^3$H calibration data, resulting in salted signal region out to ~80 keV$_{nr}$; the remainder of the ROI did not contain salt events. Additional details of the salting procedure are described in Ref.~\cite{Akerib:2016FullLux}. These events are only removed from the dataset after all data quality criteria and models (described in Sec.~\ref{sec:modeling}) had been finalized. Additionally, energy depositions from LUX's ${}^{83\text{m}}$Kr calibrations fall into this extended-energy ROI. To combat the additional leakage from the regular high-statistics calibration injections, data acquisitions corresponding to significant ${}^{83\text{m}}$Kr contamination are omitted from this analysis. A similar exclusion was reported in Ref.~\cite{annualMod}, however, this resulted in a significant loss of live-time. To increase the exposure of this analysis while also maintaining low ${}^{83\text{m}}$Kr activity, each exclusion period was reduced by 17 ${}^{83\text{m}}$Kr half-lives (31.1 hours). The final amount of exposure excluded was 20.8 live-days, resulting in a 311.2-day science run. 

To account for the temporal and spatial variation of the detector response, the WS2014--16 data are divided into four temporal bins, each further subdivided into four spatial bins corresponding to 65~$\mu$s windows of drift time.
Selecting periods when the field configuration was approximately static, we approximate each of the resulting 16 bins as temporally static with near-uniform electric field distribution. This results in negligible loss of accuracy for reproduction of light and charge yields. This same division of the dataset into 16 date and drift bins was is further described in Refs.~\cite{Akerib:2016FullLux, GR_ERmodel, luxFieldModeling}. Bins near the bottom of the detector experienced weaker electric fields (50--100 V/cm), while the strongest fields were in the topmost portion of LXe (400--550 V/cm). The four temporal bins result in unequal live times: 43.9, 43.8, 85.8, and 137.7~days. An illustration of the data divided into these 16 time and drift time bins is provided in Appendix~\ref{binSeparation}.

The fiducial volume is defined as the region for which the electron drift time (vertical coordinate) lies between 40 and 300~$\mu$s and (in the radial dimension) the region that is greater than 3~cm inward from the TPC wall. The distorted electric field also caused the electron drift paths to bend significantly inward as the electrons drift from the interaction vertex to the liquid surface. This effect is strongest for events originating near the bottom of the TPC. As a result, near-wall events at the bottom of the TPC have more centralized S2 hit-patterns in the top PMT array than near-wall events at the top of the TPC. Effectively, this moves the observed wall position inward at the bottom of the TPC, requiring that the fiducial LXe target volume is reduced as a function of drift time. In temporal order, the resultant fiducial masses for each WS2014--16 date bin are: 105.4$\pm$5.3, 107.2$\pm$5.4, 99.2$\pm$5.0, and 98.4$\pm$4.9~kg. These volumes are determined by counting remaining ${}^{83\text{m}}$Kr events in the fiducial volume, while using the knowledge that the full TPC volume contains 250~kg of LXe. The total exposure used in this analysis therefore is 3.14$\times$10$^{4}$~kg$\cdot$days. 

\begin{figure}[h]
\begin{center}
\includegraphics[width=0.445\textwidth,clip]{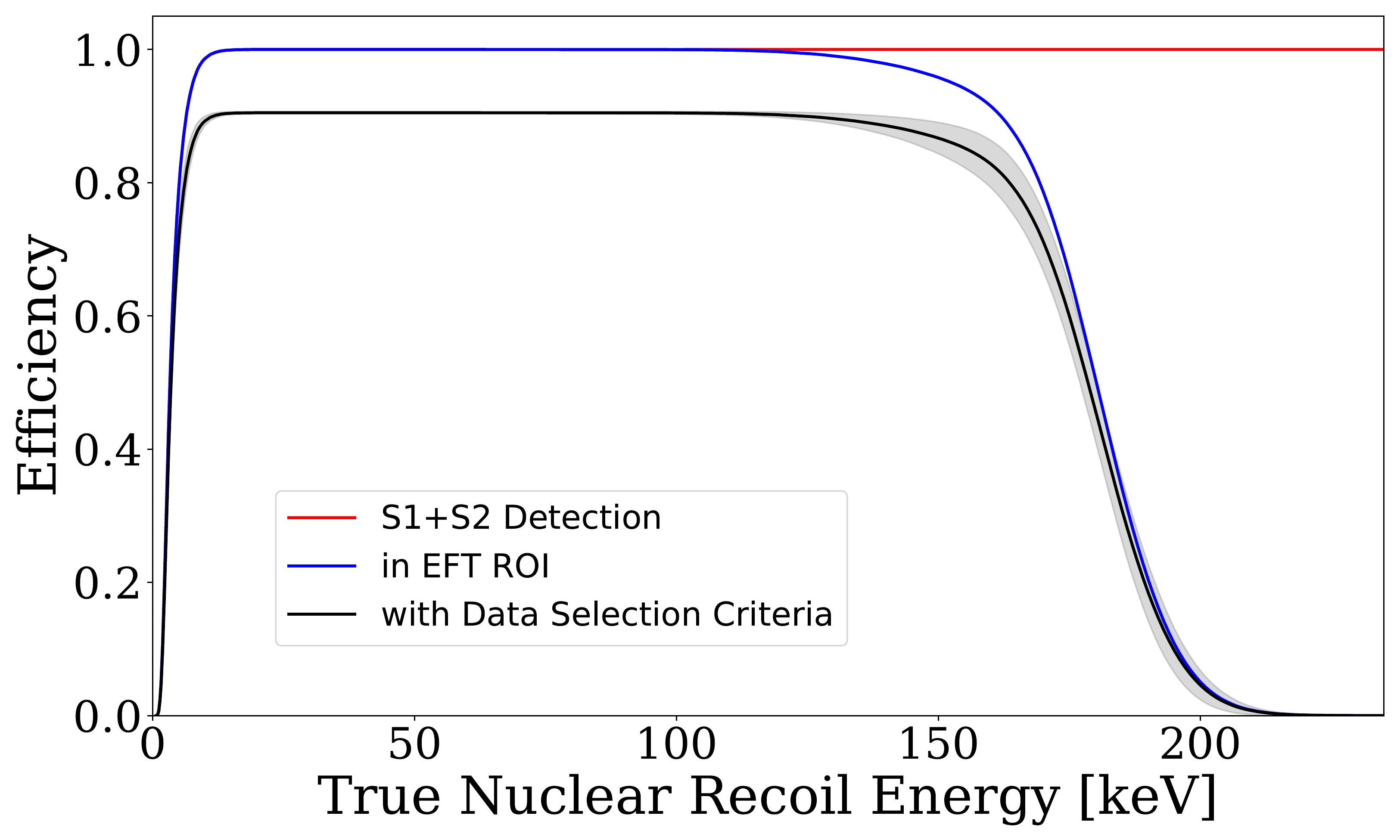}
\vspace{-10pt}
\caption{Nuclear recoil detection and selection efficiencies based on calibration data and simulations. Red corresponds to S1+S2 detection efficiency in the fiducial target without any data selection criteria implemented, while blue corresponds to the mean detection efficiency in the EFT \{S1, S2\} ROI. The black curve corresponds to the mean overall efficiency after the application of standard selection criteria, and the grey band signifies the standard deviation of the efficiency due to differing temporal and spatial detector conditions.}
\vspace{-22pt}
\label{efficiency}
\end{center}
\end{figure}

To remove adverse events that could potentially be incorrectly classified as single scatters from the dataset, a series of data selection criteria was implemented. While similar criteria were used in previous analyses~\cite{lux_cuts}, a complete reassessment of WS2014-16 data selection was done to properly characterize the high energy region. Data from the ${}^{14}$C calibration was used to validate each criterion. Events with an overabundance of pulses preceding or following either the S1 or S2 --- such as single photons or single electrons emitted from the detector's grids or delayed releases from impurities~\cite{jxu_SE} --- were removed, as these events are more likely to have misidentified S1 or S2 signals. Data selection criteria are applied based on the S1 PMT hit-patterns as well as the shape of the S1 pulse; these remove events where S1s may originate from light leaking in from outside the TPC walls and misidentified S1s, respectively. For S2s, selection criteria are based on the pulse width and shape as a function of area and drift time. As bulk S2s are expected to be approximately Gaussian in shape~\cite{luxS2only}, events were removed if a Gaussian fit to the pulse shape returned a poor goodness of fit value. The mean single-scatter selection efficiency of these selection criteria based on ER and NR calibration data and simulations is 96\% between 0-300 phd, while the full NR detection efficiency for these selection criteria is shown in Fig.~\ref{efficiency}. 

\subsection{Removing $\gamma-X$ events using a boosted decision tree classifier}

Unmodeled backgrounds in and below the signal region were reported in Ref~\cite{run3eft}, motivating the inclusion of novel sources to the background model used in this analysis. There is a 5.6~cm gap between the cathode and the bottom PMT array where the mean electric field has opposite direction than that of the bulk LXe; scintillation produced in this ``reverse field region" (RFR) is visible to the PMTs, but emitted electrons are carried downward (instead of upward to produce an S2 signal). If a $\gamma$-ray scatters in the RFR in addition to the fiducial volume, both scatters contribute to the S1, while only the fiducial scatter produces S2 light. The result is a ``$\gamma-X$'' event with an S1--S2 ratio anomalously low for an ER event~\cite{zep_gammaX,xenon10_gX,Akerib:2018dfk}. Combined with the reduced recombination due to having the weakest electric fields at the bottom of the fiducial region, these events could significantly increase the leakage of ER events into and below the NR signal region.

$\gamma-X$ events pose a unique challenge because they can appear as typical single scatters. Any hints of their anomalous behavior could in principle be captured in the S1 signal.  However, due to the timescales at which light collection takes place ($\mathcal{O}$(10~ns)) being longer than transit time between scatters (typically less than 1~ns), these S1s are not readily separable from single scatters using simple one-dimensional or two-dimensional criteria, such as those described in the preceding subsection. Instead, a six-dimensional parameter space is utilized, with the intent of using a boosted decision tree (BDT) machine learning event classifier to identify and remove $\gamma-X$-like events. BDTs are becoming more commonly used in particle physics analyses, and they provide an efficient way to draw distinctions between two populations in higher dimensional spaces~\cite{luxS2only,BDT}. The six features used are \begin{itemize}
\small
    \item Position-corrected S1 area;
    \item Position-corrected S2 area;
    \item \textit{Bottom array cluster size} -- the mean spatial extent of the hit pattern in the bottom PMT array, normalized by total S1 area;
    \item \textit{Max peak area fraction} -- the fraction of the total S1 light detected by the PMT registering the largest contribution to the S1;
    \item \textit{Top-bottom light collection ratio} -- the ratio of collected scintillation of top and bottom PMT arrays;
    \item Reconstructed S2 event depth.
    \normalsize
\end{itemize} S1 hit-patterns in the bottom PMT array will be more localized for interactions below the cathode, therefore the cluster sizes and max peak area fractions should differ between RFR energy deposits and bulk single scatters. Thus by using these light collection features in addition to position reconstruction information from the S2, we hope to separate $\gamma-X$ events from true single scatters. To train the $\gamma-X$ classifier, a model was made to reproduce ``near-miss" double-scattering events near the cathode but using simulation to extrapolate these double scatters into $\gamma-X$ events with a subcathode energy deposit. The details of the $\gamma-X$ model are discussed in Sec~\ref{sec:modeling}.

A BDT classifier using the Extreme Gradient Boosting (XGBoost) algorithm was chosen~\cite{xgboost}. The BDT was trained on simulated $\gamma-X$ events and simulated ER and NR single scatters, outputting a classification score between 0 and 1, with lower scores indicating a more $\gamma-X$-like event. Separate sets of simulated data were used for training and testing the BDT, and each were comprised of equal amounts of $\gamma-X$, ER single scatters, and NR single scatters. Before training, however, a BDT requires selections of user-defined ``hyperparameters", and judicious hyperparameter choices can improve a BDT's classification power and prevent overtraining.  Classification power is quantified using combinations of the true and false positive and negative classification rates; overtraining was quantified by calculating p-values via Kolmogorov-Smirnov tests for the training and testing datasets, and a lower p-value is indicative of a more overtrained classifier. The number of decision trees used, $N$, and the maximum tree depth (the number of binary decision nodes per tree), $D$, are specified before training begins. To find suitable choices, $N$ and $D$ were tuned to maximize the classification power and minimize overtraining. To prevent unnecessary overtraining to simulated events, final $N$ and $D$ values of 212 and 5, respectively, were chosen by maximizing the acceptance of ${}^3$H single scatters and by maximizing the rejection of near-miss double scatters (while only taking the upper S2 into consideration).  Additionally, the BDT's learning rate, $\varepsilon$, was tuned, as higher learning rates improve classification power but overtrain the algorithm. A final value ($\varepsilon=0.3)$ was chosen by maximizing classification power before there was any indication of overtraining. All other XGBoost hyperparameters were left at their default values.  

\begin{figure}[h!]
\begin{center}
\includegraphics[width=0.445\textwidth,clip]{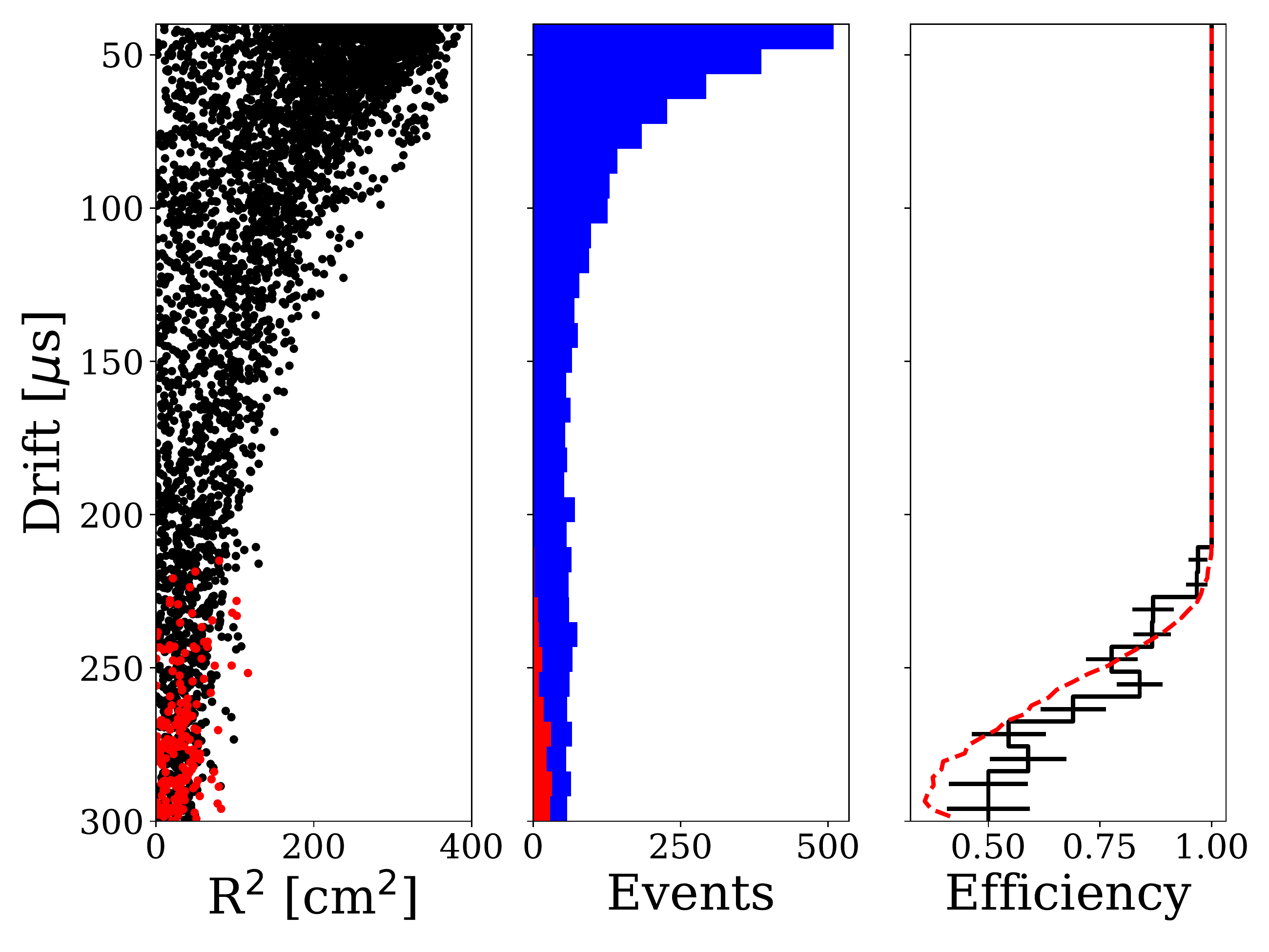}
\vspace{-10pt}
\caption{Left: spatial distribution of the background data. Red markers correspond to events removed after assessment with the BDT $\gamma-X$ classifier. Center: histograms of events as a function of detector depth. Blue corresponds to all events, while red corresponds to the events removed. Right: efficiency as a function of event depth for the background data (black) compared to the ${}^{14}$C and ${}^{3}$H ER calibration data (red). While the efficiency is the poorest near the bottom of the detector, the background data are mostly concentrated near the top, resulting in a 95.5\% overall acceptance for the background data. }
\vspace{-22pt}
\label{gammaXcut}
\end{center}
\end{figure}

A classification score threshold was set by: minimizing the rejection of ${}^{14}$C and ${}^3$H ER single scatters, maintaining perfect acceptance of known WIMP-like salt events (known salt events were available after concluding the WS2014-16 SI analysis~\cite{Akerib:2016FullLux}), and maximizing the rejection of simulated $\gamma-X$ events and multiply scattering events near the cathode. The chosen threshold was a score of 0.36, resulting in 89.1\% rejection of the simulated $\gamma-X$ events, while accepting 95.9\% of simulated single scatters after applying all other data selection criteria. Our final signal detection efficiency including all selection criteria is 90.5\%. The efficiency is highly position-dependent, with 100\% efficiency throughout most of the volume, but the largest loss of efficiency in the bottom-most 20\% of the fiducial target ($\sim$50\% at the poorest). Figure~\ref{gammaXcut} details the effects and efficiency of $\gamma-X$ removal from the background data.

\section{Modeling}\label{sec:modeling}

Using the profile likelihood ratio (PLR) construction (described in Sec.~\ref{sec:stats}), we use statistical inference to quantify the level of sensitivity of our detector to identify or constrain the possibility of WIMPs interacting under a given EFT operator. A likelihood ratio test provides a strong statistical framework when dealing with higher-dimensionality parameter space, and it requires a good model of both the null and alternative hypotheses to be valid. In this section, we describe the construction of each of the models used in the PLR framework. We identified and constructed five-dimensional models ($S1_c$, $\log_{10}(S2_c)$, radius, drift time ($d$), and azimuthal angle ($\phi$) about the TPC's central axis) for the sources that could lead to events in our ROI: EFT WIMPs; ER single scatters; remaining ${}^{83\text{m}}$Kr after the calibration injections; degraded events and ion recoils from the TPC walls; $\gamma-X$; and accidental coincidences of unrelated S1-only or S2-only events. After separation of the data into the 16 date and drift time bins, we make the assumption that the field variation in each drift time bin has minimal impact on the S1 and S2 distributions. Accordingly, we make the simplification of separating the spatial and energetic components of most models, resulting in probability density functions (PDFs) that are the direct product of two ($S1_c$ and $\log_{10}(S2_c)$) and three ($r$, $\phi$, and $d$) dimensions. However, the model for degraded wall events and ions has no such separation as the energy and spatial observables are highly correlated even after separation into 16 drift time bins (see Sec.~\ref{wall}).  We explicitly note that this analysis uses $\log_{10}(S2_c)$ as opposed to $S2_c$ directly, as it allows for finer binning at lower energies, leading to higher resolution PDFs in the region where ER and NR discrimination is poorest. 

\subsection{Signal modeling}
Signal spectra are obtained using the Mathematica package developed by Anand \textit{et al.} \cite{Anand:MathematicaEFT}.  This gives the differential rate of nuclear recoils per recoil energy, $E_R$:
\begin{equation}\label{RateFromMatrixElement}
    \frac{\D{R}}{\D{E_R}} = N_T \frac{\rho_0 m_N^2}{2\pi m_\chi m_A} \int_{v > v_\text{min}} \frac{f(\vec{v})}{v}|\mathcal{M}|^{2} \D[3]{v},
\end{equation}
where $N_T$ is the number of target nuclei, $\rho_{0}$ is the local dark matter density, $m_\chi$ is the mass of a WIMP, $m_A$ is the target nucleus mass, and $f(\vec{v})$ is the galactic WIMP velocity distribution for which we assume a Maxwell-Boltzmann distribution following the standard halo model: characteristic velocity $v_0=220$~km/s and escape velocity $v_{esc}=544$~km/s.
The spin-averaged matrix element $|\mathcal{M}|^{2}$ is calculated via a combination of WIMP velocity and momentum-transfer dependent form factors $F_{i j}^{\left(N, N^{\prime}\right)}\left(v^{2}, q^{2}\right)$ presented in Appendix A.2 of \cite{Fitzpatrick:EFT}, scaled based on the value of the EFT coupling constants $c_i^{(N)}$:
\begin{equation}\label{eq:FormFactor}
\begin{split}
\hspace{-0.25in}\frac{1}{2 j_{\chi}+1} \frac{1}{2 j+1} \sum_{\text {spins}}|\mathcal{M}|^{2} \equiv \\ & \hspace{-1.525in} \frac{m_{A}^{2}}{m_N^2} \sum_{i, j=1}^{15} \sum_{N, N^{\prime}=p, n} c_{i}^{(N)} c_{j}^{\left(N^{\prime}\right)} F_{i j}^{\left(N, N^{\prime}\right)}\left(v^2, q^{2}\right), 
\end{split}
\end{equation}
where $j$ and $j_\chi$ are the spins of the nucleus and WIMP, respectively. Note that this representation of the amplitude differs from Ref~\cite{Fitzpatrick:EFT} by a factor of $(4m_\chi m_N)^2$, accounting for the different normalization conventions and dimensionality of the $c_i$ used in the \textit{Mathematica} package~\cite{Anand:MathematicaEFT}. The form factors are also affected by differing conventions and are scaled to account for this\footnote{Specifically, factors of $\vec{q}$ have been normalized by factors of $m_N$, similar to the normalization used in Eqs. 1 and 2.}.  Putting Eqs.~\ref{RateFromMatrixElement} and \ref{eq:FormFactor} together, the differential rate spectrum becomes
\begin{equation}
\begin{split}
\frac{d R}{d E_{R}}=N_T\frac{\rho_{0}m_A}{2 \pi m_{\chi}}  \int_{v>v_{min}}\biggl[ \frac{f(v)}{v} \\
& \hspace{-1.525in} \cdot \sum_{i, j} \sum_{N, N^{\prime}=n, p} c_{i}^{N} c_{j}^{N^{\prime}} F_{i, j}^{\left(N, N^{\prime}\right)}\left(v^2, q^2\right)  \diff v \biggr] .
\end{split}
\end{equation}


Note that one can just as easily use isoscalars and isovectors in place of the $p$ and $n$ for the proton and neutron. This is also a valid approach, and has been done in analyses by several other experiments as it allows for direct comparisons between experiments with different target compositions~\cite{deap3600,xenon100, PandaX_EFT}. However for consistency with previous LUX results (Ref.~\cite{run3eft}), the \{$n$,$p$\} basis is used for elastic WIMP-nucleon recoils in this analysis. Similarly to a traditional spin-dependent WIMP search, the \{$n$,$p$\} basis provides a more natural representation of the physical interactions that this analysis attempts to identify or constrain. Additionally, due to the presence of two couplings in each term, the possibility for destructive interference exists. For this analysis, we ignore the possibility of interference and make the assumption that one coupling is dominant over all others. As such, the signal spectra that we obtain are the result of setting all but one of the couplings $c_i^{(N)}$ to 0. The resulting differential event rate scales linearly with the remaining nonzero coupling $c_i^{(N)^2}$:
\small
\begin{equation}\label{eq:spectrumReduced}
   \hspace{-0.25in} \frac{d R}{d E_{R}}=N_T\frac{\rho_{0}m_A}{2 \pi m_{\chi}} \int_{v>v_{m i n}} \frac{f(v)}{v} c_{i}^{(N)^2} F_{i, i}^{\left(N, N\right)}\left(v^2, q^{2}\right) \diff v ,
\end{equation}
\normalsize
Due to the linear relation between differential rate and $c_i^{(N)^2}$, the spectrum for any value of the coupling constant can be easily determined by  calculating the $c_i^{(N)^2} = {m_v^{-2}}$ case and then scaling appropriately.  We use the benchmark value $m_v^{-2}$, where $m_v$ = 246.2~GeV and is the Higg's vacuum expectation value, as this is the chosen scaling factor used internally by~\cite{Anand:MathematicaEFT}.

To generate the detector response to the resultant nuclear recoil energy spectra, a recent release of the Noble Element Simulation Technique (NEST v2.1.0) was utilized~\cite{nestv201}, chosen prior to unsalting. An empirical fit to all existing nuclear recoil data in LXe, NEST provides precise light and charge yields resulting from an energy deposition. While the D-D NR calibrations characterize the detector response out to 74~keV$_{nr}$ ($\sim$150~phd), NEST allows for extrapolation to higher energies using reported yields in the literature extending to 330~keV$_{nr}$ from other sources such as AmBe~\cite{Sorensen_2011}. This provides an understanding of the signal region beyond where the detector NR response was directly calibrated. Uncertainty in the signal region for energies beyond the \textit{in situ} D-D calibration was calculated by allowing the NEST v2.1.0 NR model (largely unchanged between versions 2.0.1 through 2.2) to fluctuate within the uncertainties for the total reported quanta of the highest energy data used to fit the model; for 300~phd S1s, the resultant uncertainty of the location of the NR band mean in S2-space is approximately 7.5\%, corresponding to a change in S2 size of roughly 540~phd. Ultimately, the NR band is sufficiently far from the ER band in any scenario to make this difference negligible.

\begin{figure}[h]
    \centering
    \includegraphics[width=0.445\textwidth,clip]{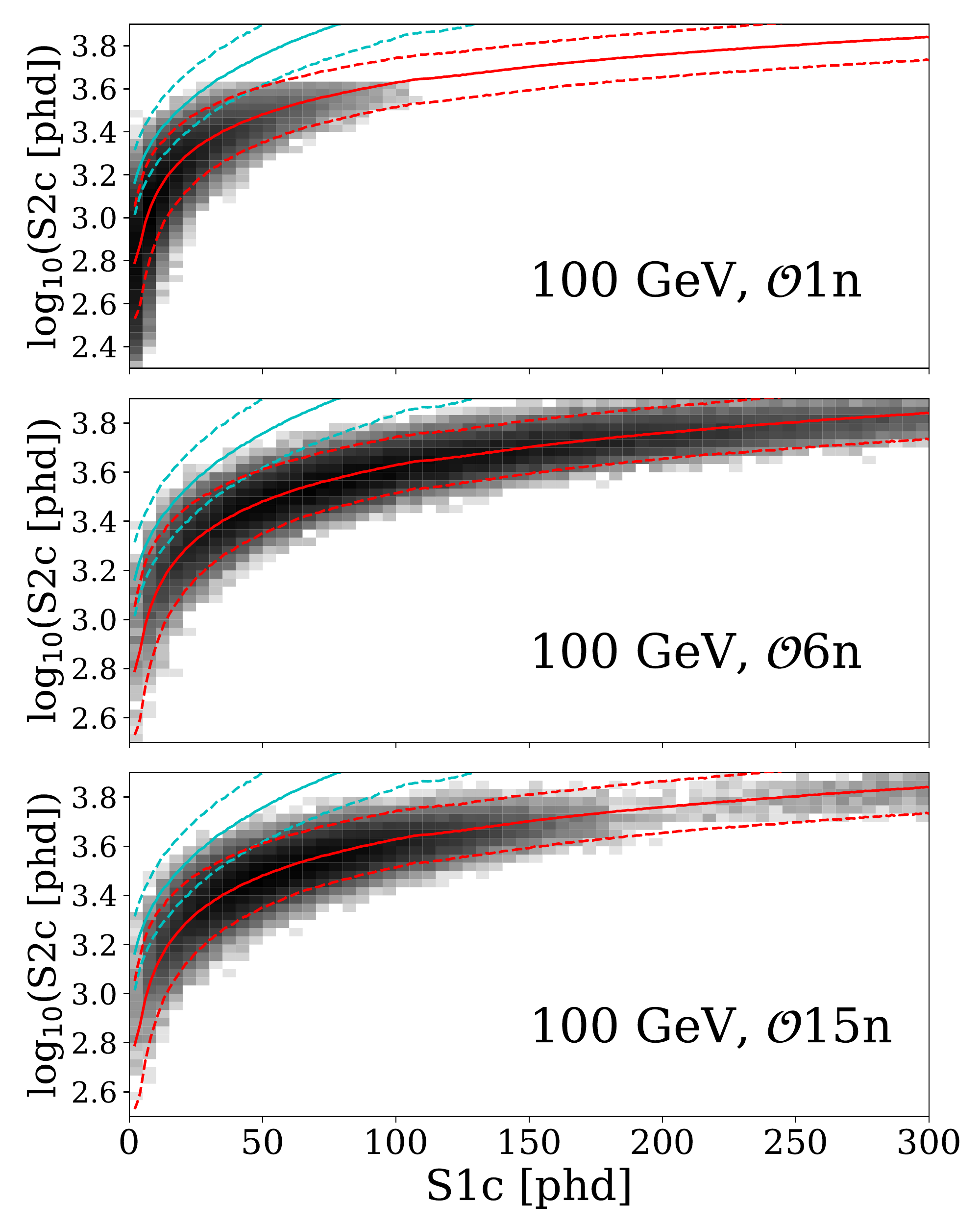}
    \caption{Example $\{$S1,S2$\}$ distributions for 100 GeV WIMP-n interactions for EFT operators $\mathcal{O}_1$ (top), $\mathcal{O}_6$ (middle), and $\mathcal{O}_{15}$ (bottom), highlighting the qualitative differences in the LXe response of various EFT operators. $\mathcal{O}_1$ peaks at low energies and is contained to energies of order 10 keV$_{nr}$; $\mathcal{O}_6$ peaks at medium energies but remains relatively flat throughout our ROI; and $\mathcal{O}_{15}$ exhibits a secondary peak at higher energies. The expected median and $\pm$ 90\% C.L. bands for ER and NR are shown in blue and red, respectively, averaging over the 16 temporal and spatial bins of LUX data. Each pane shows the distribution for 50,000 WIMP nuclear recoils. }
    \label{signalCompare}
\end{figure}

Recoil spectra for different operator-mass combinations are simulated using the LUX Legacy Analysis Monte Carlo Application (LLAMA)~\cite{GR_ERmodel}. LLAMA uses spatial and temporal interpolation between the 16 approximately static WS2014--16 drift time bins, utilizing the NR response from NESTv2.1.0 and the three-dimensional field maps described in Ref.~\cite{luxFieldModeling}. Example distributions for $\mathcal{O}_1$, $\mathcal{O}_6$ and $\mathcal{O}_{15}$ are shown in Fig.~\ref{signalCompare}. Signal spectra are generated homogeneously throughout the detector. 

Although the \{$n$,$p$\} basis is used for the elastic case, there are no previously reported limits on inelastic EFT WIMP-nucleon interactions using LUX data. Therefore, signal models for the inelastic case were generated in the isoscalar basis for the sole purpose of comparing to previously reported results from XENON100 for 1 TeV WIMPs (Ref.~\cite{xenon100}). Recoil spectra were obtained from a modified version of the Anand \textit{et al.} \textit{Mathematica} package developed by Barello et al. \cite{Barello_2014}. As discussed in the Introduction, this requires using operators constructed from a Hermitian basis that takes into account the energy conservation requirement for inelastic recoils [Eq.~\ref{massSplitting}]. Signal models with a range of $\delta_m$ from 0--200~keV were generated for all operators. Other parameters, including astrophysical and nuclear, remain unaltered from those used for the elastic signal models, and the same procedure as described above was applied.  

\subsection{Standard ER backgrounds}

We expect the overwhelming majority of backgrounds to originate from ER-producing contaminants within the LXe, namely ${}^{222}$Rn and ${}^{220}$Rn and their charged daughter isotopes plating-out on the detector surfaces, as well as decays from radioisotopes in the detector components. Decays from the detector components are mostly isotopes originating from ${}^{238}$U, ${}^{232}$Th, ${}^{60}$Co, and ${}^{40}$K, producing $\beta$, $\gamma$, and $\alpha$ radiation at a wide range of energies. A dedicated modeling campaign for reproducing the LXe ER response in LUX was reported in Ref.~\cite{GR_ERmodel}. To summarize, utilization and tuning of NEST ER response models allowed for accurate characterization of the temporal and spatial features of the WS2014--16 detector and precise reproduction of all available LUX ${}^{14}$C and ${}^3$H ER calibration data. While NEST is a global fit to xenon light and charge yields, this LUX-specific version allows for efficient creation of high-statistics LUX ER simulated data for all 16 WS2014--16 drift and date bins for all relevant energies.

Assays of LUX components provide initial expectations for the expected radioactivity from the detector leading to ER backgrounds. However, due to uncertainties in the assay measurements and the modeled response of each detector component and their geometries, the simulated energy depositions from each contributing detector component and radiogenic source was fit to high-energy data, including multiply scattering events, allowing for effective activities from each source. Data below 80~keV$_{ee}$ were excluded when fitting the effective activities. LXe light and charge responses for each source were then simulated using the LUX-specific version of NESTv2.1.0, providing $S1_c$ and $S2_c$ distributions for each expected ER source. 

\subsection{The $^{83\text{m}} \text{Kr}$ model}
$^{83\text{m}}$Kr was injected into the TPC on a weekly basis to ensure proper position corrections.  This source decays in two transitions: $\SI{32.1}{\kilo \electronvolt}$ followed by $\SI{9.4}{\kilo \electronvolt}$. Most often, these deexcitations occur via internal conversion electrons or Auger electrons. The time between the two emissions ranges from $\mathcal{O}$(10 ns) to $\mathcal{O}$(1 $\mu$s), and those on shorter timescales appear as $\SI{41.5}{\kilo \electronvolt}$ single scatters, having only a single detectable S1 and S2.  These quasi-monoenergetic depositions are of high enough energy to be removed in a typical momentum-independent analysis, leading to no loss of exposure time.  However, in an analysis reaching to higher energies, ${}^{83\text{m}}$Kr events can interfere with the signal region.

As a high-statistics monoenergetic peak, ${}^{83\text{m}}$Kr yields are observed with wide recombination fluctuations in the $\frac{S2}{S1}$ ratio, resulting in events near the NR signal region at energies where most other ER backgrounds are well-discriminated (see Fig.~\ref{calibrations}). Additionally, this proximity of ${}^{83\text{m}}$Kr events to the signal region worsens for weaker fields, as the ER and NR bands are less separated than at stronger electric fields. As stated in the Sec.~\ref{dataSelection}, 20.8 live-days were excluded from WS2014--16 that correspond to periods of significant $^{83\text{m}}\text{Kr}$ contamination in order to omit most of these events from this analysis.

Despite this, some ${}^{83\text{m}}$Kr events are expected in the dataset; ${}^{83\text{m}}$Kr has a 1.83 hour half-life, resulting in lingering decays after the injections end. Therefore, a robust characterization of these events was required. For this, the remaining ${}^{83\text{m}}$Kr data excluded from the final search data were used to construct a model for these events. The expected number of events was calculated by measuring the rate of ${}^{83\text{m}}$Kr events at the end of each data exclusion period and extrapolating using the known half-life.

\subsection{The wall model}\label{wall}

Similarly to previous LUX analyses~\cite{lux_improvedLimits,Akerib:2016FullLux}, we construct a model characterizing energy depositions in close proximity to the inner TPC walls. The electron extraction efficiency near the walls is poorer than in the bulk LXe, resulting in degraded S2 signals. Additionally, nuclear recoils from ${}^{206}$Pb (a daughter of ${}^{210}$Po $\alpha$-decay) on the inner TPC walls leads to events with naturally low $\frac{S2}{S1}$ ratios compared to ER backgrounds, resulting in a population of events well-below the signal region in \{$S1_c$,$S2_c$\} space. 

As mentioned briefly at the beginning of the preceding section (Sec~\ref{dataSelection}), the reconstructed position of the detector wall depends on the drift time $d$, azimuthal angle $\phi$, and acquisition time due to the radial field~\cite{luxFieldModeling}. We observed that the reconstructed position of the events fluctuates around the position of the wall according to a Gaussian distribution with width proportional to $1/\sqrt{S2}$~\cite{luxPositionRec}. Therefore, the wall events have a larger uncertainty  for the same deposited energy due to the smaller S2 size, allowing for a fraction of these events to appear within the fiducial volume.

To characterize this background, we selected WS2014--16 events with reconstructed positions beyond the measured position of the TPC wall, counting the number of events for a specific bin in drift time, azimuth, and date bin, as the fluctuations in reconstructed position about the wall should be equal both inside and outside the wall position. Integrating the tail of this empirical fit provides an understanding of the expected number of wall events that leak into the fiducial volume. Since this leakage depends heavily on the observed S2 size, the energy and spatial PDFs are significantly correlated, making the wall model a true five-dimensional PDF. 

\subsection{The $\gamma-X$ model}

As described in Sec.~\ref{dataSelection}, we consider the possibility of multiply scattering $\gamma$-rays with only a single detectable ionization signal due to one or more subcathode energy depositions: $\gamma-X$. We note here that these events would of course be observed near the bottom of the fiducial volume, where the electric field values are the weakest and the ER/NR discrimination is the poorest, combining to create the possibility of excessive ER leakage. The RFR field magnitude is of $\mathcal{O}$(1~kV/cm), which results in significantly lower light yields for a given energy deposition compared to the bottom of fiducial volume: a reduction to approximately 65\% for 50~keV $\gamma$-rays~\cite{nestv201}. This results in higher-energy $\gamma$-rays (which are more likely to traverse a significant portion of the RFR xenon) producing S1s below our 300~phd threshold that would normally be excluded if that interaction occurred in the bulk LXe.

Radiogenic impurities in the bottom PMT array and RFR TPC walls - namely ${}^{238}$U, ${}^{232}$Th, ${}^{60}$Co, ${}^{40}$K and their daughters - may produce $\gamma-X$, in addition to back-scattering events originating from the cathode grid wires.  Because these events appear superficially as normal single scatters, we are unable to obtain a set of known $\gamma-X$ events. However, the presence of double-scatter events near the cathode provides information on multiply scattering $\gamma$-rays near the RFR. We selected a set of double-scatter events that had: at least 3~cm of vertical separation between the two reconstructed interaction locations; $S1_c$ less than 300~phd; the lower-most S2 within 4~cm of the cathode; and the top-most energy deposit within the fiducial radius. The distance between the cathode and the fiducial volume is approximately 3 cm, thus the first condition reproduces the minimum vertex separation for $\gamma-X$ events that may pass other data quality criteria. The remaining criteria allow for selection of events with uppermost S2s similar to single scatters in the background data (as those would be the observed S2s for $\gamma-X$ events). Seventeen of these ``near-miss" double-scatters were found in WS2014--16, with reconstructed energies well-distributed throughout our energy ROI. 

A model was created using the LUX-specific NEST framework, sampling energies and positions within our ROI that were similar to the observed near-miss events and the expected ER background. A surface-based ray-tracing algorithm for the LUX detector was created to efficiently calculate the PMT hit patterns, providing the necessary features used to train the BDT (see Sec.~\ref{dataSelection}). This was possible by not relying on full propagation of photon trajectories, but instead updating the trajectory only when the photon is reflected or refracted. This simplified near-miss model was able to accurately reproduce the features of the observed near-miss events. By translating this model 4 cm downwards, guaranteeing the first simulated scatter to be subcathode, we were able to generate simulated $\gamma-X$ events based on LUX data. This model was used to train the BDT described in Sec.~\ref{dataSelection} in an attempt to remove $\gamma-X$ events from the data. While characterizing the rate of expected $\gamma-X$ events proves challenging, we make the assumption that it should be similar to the rate of near-miss double-scatter events. Taking the efficiency of the BDT classifier into account with respect to simulated $\gamma-X$, we therefore expect $\mathcal{O}$(1) $\gamma-X$ events in our final dataset. 

\subsection{Accidental Coincidences}

Lastly, we take into consideration the coincidental pairing of unrelated S1-only and S2-only events, forming an ``accidental" single scatter (such as those reported in Ref.~\cite{xenon1t_models}). To understand the rate at which to expect these events and their appearance in phase space, LUX data were filtered to obtain two sets of data: events with only one observed S1 and no S2, and events with only one S2 and no S1. The S1-only and S2-only rates and spectra were input into a Monte Carlo generator, and random pairing of S1s and S2s provided a model to characterize these events.

It is possible to have energetic S1-only and S2-only events due to energy depositions in regions of poor light collection and charge extraction efficiencies; however, the most common S1-only and S2-only events consist of only a handful of photons or electrons, respectively. The accidental pairing of these pulses can produce a false event mimicking a lower-energy single scatter, falling in the region of phase space where the expected WIMP recoil rate is the most probable. Using the S1-only and S2-only event rates, we are able to calculate an expectation for accidental coincidence events. However, the data selection criteria described in Sec.~\ref{dataSelection} reduce the expected rate of these events in the ROI considerably, and we expect less than a single accidental event for the exposure in this analysis.

\section{Statistical Methodology}\label{sec:stats}

In setting constraints on the coupling constant for a given mass-operator combination, we use hypothesis test inversion to determine a 2-sided frequentist confidence interval via the Neyman construction \cite{neyman1937x}.  This involves performing a series of hypothesis tests where the null hypothesis $(\mathrm{H_0})$ is our model with the parameter Of interest (POI), $\mu$, fixed at a given value, and the alternative hypothesis $(\mathrm{H_1})$ is allowed to float to all real values:
\begin{equation}\label{eq:Hypotheses}
    \begin{aligned}
    \mathrm{H_0}:& \mu = \mu_0 \\
    \mathrm{H_1}:& \mu \neq \mu_0
    \end{aligned}
\end{equation}
Here, $\mu$ is simply the number of WIMP-nucleon scatters we expect to observe for a given model.  The values of the POI corresponding to hypothesis tests whose p-value is greater than the significance $\alpha = 0.1$ form the $90\%$ confidence interval on the POI for each signal model.

Our test statistic for these hypothesis tests is the profile likelihood ratio (PLR).  More specifically, we use the negative log likelihood, $q = -2 \ln(\lambda)$, where $\lambda$ is the actual PLR:
\begin{equation}
            \lambda(\vec{X}) = \frac{\Lik{P}{(\mu_0,\doublehat{\theta})}{\vec{X}}}{\Lik{P}{(\hat{\mu},\hat{\theta})}{\vec{X}}}.
\end{equation}
Here, $P$ denotes that this likelihood has been modified by the presence of a profile.  $\mu_0$ is just the fixed POI, 
and the terms with hats are allowed to float to maximize the value of the profiled likelihood $\mathcal{L}_P$.  The double hat $\doublehat{\theta}$ indicates that the values of the nuisance parameters, $\theta$, that maximize the likelihood in the case of $\mu = \mu_0$ are not in general the same values that maximize it when $\mu$ is left to float.  $\vec{X}$ represents the dataset used to compare against the model.

For this analysis we use the extended unbinned likelihood as follows:
\begin{equation}\label{eq:theLikelihood}
    \begin{aligned}
    \mathcal{L}\left( \left(\mu, \vec{\theta} \right) \middle| \vec{X} \right) = \text{Pois}\left( n_\text{obs}; n_\text{exp} \right) \\
    & \hspace{-1.5in}  \cdot \prod_{\vec{x}_i \in \vec{X}} \left[n_{\text{sig}}R_{\text{sig},t_i,z_i} \PDF_{\text{sig},t_i,z_i}\left(\vec{\mathcal{O}}_i \right) \right. \\
    & \hspace{-1.75in}  \phantom{ \prod_{\vec{x}_i \in \vec{X}} [} + \sum_{b_j} n_{b_j} R_{b_j,t_i,z_i} \PDF_{b_j,t_i,z_i} \left(\vec{\mathcal{O}}_i\right) \\
    & \hspace{-1.75in} \phantom{ \prod_{\vec{x}_i \in \vec{X}} [} \left. + n_{\text{wall}} R_{\text{wall},t_i,z_i} \PDF_{\text{wall},t_i,z_i} \left(\vec{\mathcal{O}}_i \right) \right] \\
    & \hspace{-1.5in} \cdot \prod_{\theta_i \in \vec{\theta}} \PDF_{i} \left(\theta_i \right)
    \end{aligned}
\end{equation}
Here $n_\text{obs}$ is the number of events contained in the dataset, $\vec{X}$; $n_\text{exp} = n_\text{sig} + \sum_{b_i} n_{b_i} + n_\text{wall}$ is the number of events expected by the model with $b_i$ indicating one of our background models; and $\vec{x}_i$ is a given data point in the set $\vec{X}$.  Each data point $\vec{x}_i$ contains the set of 5 observables: $\{r, d, \phi, S1_c, \text{ and } \log_{10}(S2_c)\} \equiv \vec{\mathcal{O}}$ along with the analysis bin in which it was measured: $\{\text{date bin} (t), \text{drift time bin} (z)\}$.  $n_\text{sig}$ is the number of signal events expected, and is used as a stand-in for our POI as we have not included any nuisance parameters that affect detector thresholds in this analysis, thus $n_\text{sig}$ is a function purely of $c_i^{(N)^2}$.  $n_{b_i}$ is similarly the number of expected events from background source $b_i$, and the same is true of $n_\text{wall}$.  $R_\text{source},t_i,z_i$ is the fraction of the total number of expected events for that source that are expected to occur in the bin ($\text{date bin} = t_i, \text{drift time bin} = z_i$).  Likewise, $\PDF_\text{source},t_i,z_i \left(\vec{\mathcal{O}}_i \right)$ is the probability density function (PDF) modeled for the given source in the given date bin and drift time bin.  The final line in equation \ref{eq:theLikelihood} is the profile term.  $\theta_i$ is a given nuisance parameter, and $\PDF_i (\theta_i)$ is the PDF describing the profile for that nuisance parameter.  In principle, the profiles of multiple nuisance parameters could be correlated, but this was determined to have minimal effect and was not implemented. The set of nuisance parameters $\vec{\theta}$ used in this analysis is simply the number of expected events for each different background source $n_{b_i}$.

We explicitly separate the wall model from the other backgrounds in Eq.~\ref{eq:theLikelihood} because its implementation in our software differs significantly from the others.  As mentioned in Sec.~\ref{sec:modeling}, the spatial observables $\left\{r, d, \phi \right\}$ were determined to be sufficiently independent of the corrected energy observables, $\left\{S1_c, \log_{10}(S2_c) \right\}$, once the detector was split up into its date bins and drift time bins.  This allowed for the implementation of the 5-dimensional PDF to be split into the direct product
\begin{equation}
\begin{split}
    \PDF_{\text{source},t_i,z_i} \left( \vec{\mathcal{O}}_i \right) \equiv 
    \PDF_{\text{source},t_i,z_i} \left( r_i, d_i, \phi_i \right) \\ \cdot
    \PDF_{\text{source},t_i,z_i} \left( S1_{c,i}, \log_{10}(S2_{c,i})\right)
\end{split}
\end{equation}
However, in the case of the wall model, this split is not feasible: the location of the wall as seen by the top PMT array depends significantly on $d$ and $\phi$, while the reconstructed distance from the wall depends strongly on $S2_c$. Therefore, the PDFs for the wall model remain fully 5-dimensional.

We found that our datasets do not lie in the asymptotic regime, and therefore unfortunately cannot make use of the asymptotic formulae that would greatly reduce the computation necessary for performing each hypothesis test~\cite{Cowan_2011}.  Instead, we rely on comparing our test statistic to that of a collection of Monte Carlo psuedo-experiments simulated based on our models. Test statistic distributions are evaluated using a custom-built PLR framework utilizing RooFit \cite{roofit} that has been optimized for the rapid computation of pseudoexperiments in our 5-dimensional regime.

\section{Results}
\label{results} 

\begin{figure}
    \includegraphics[width=1.0\columnwidth]{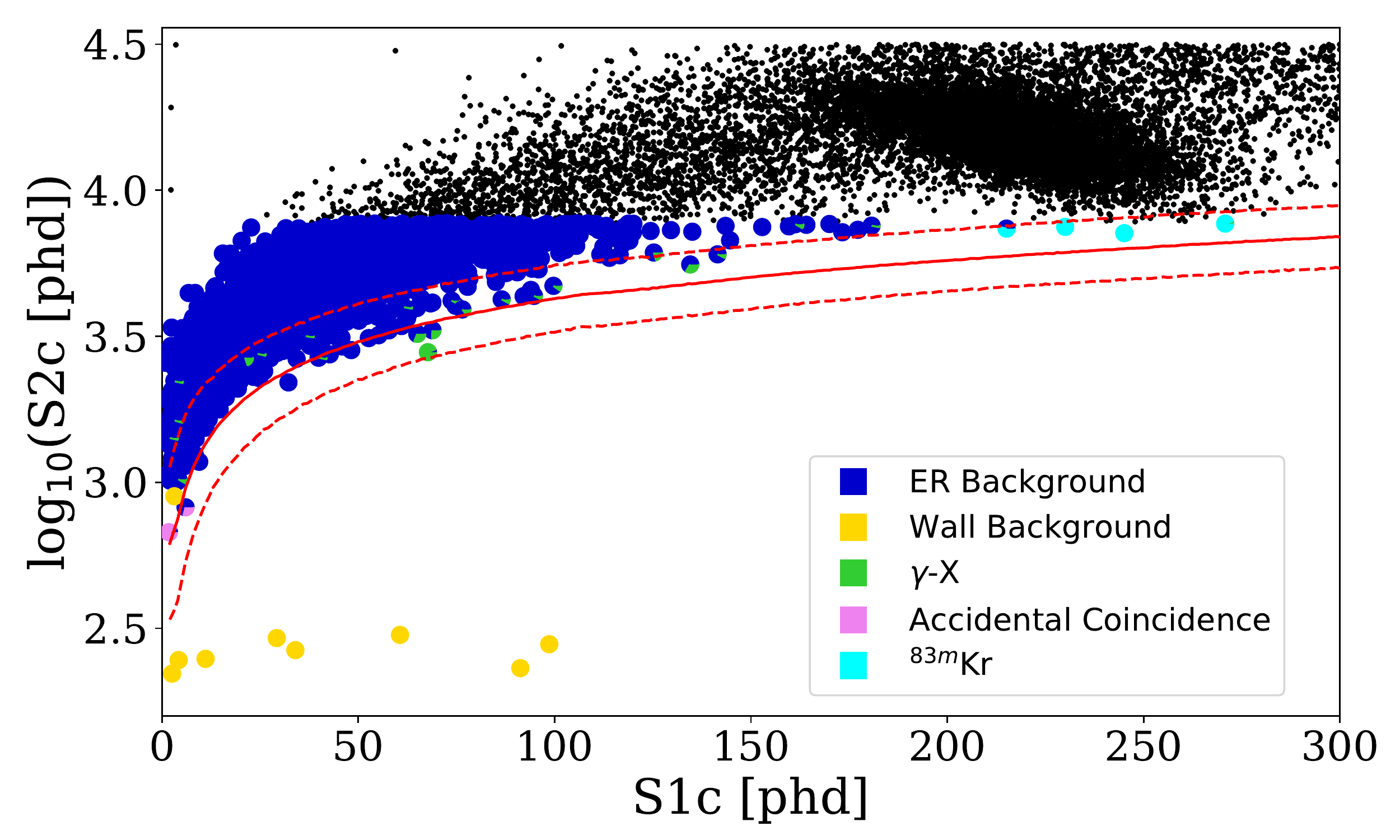}
    \includegraphics[width=1.0\columnwidth]{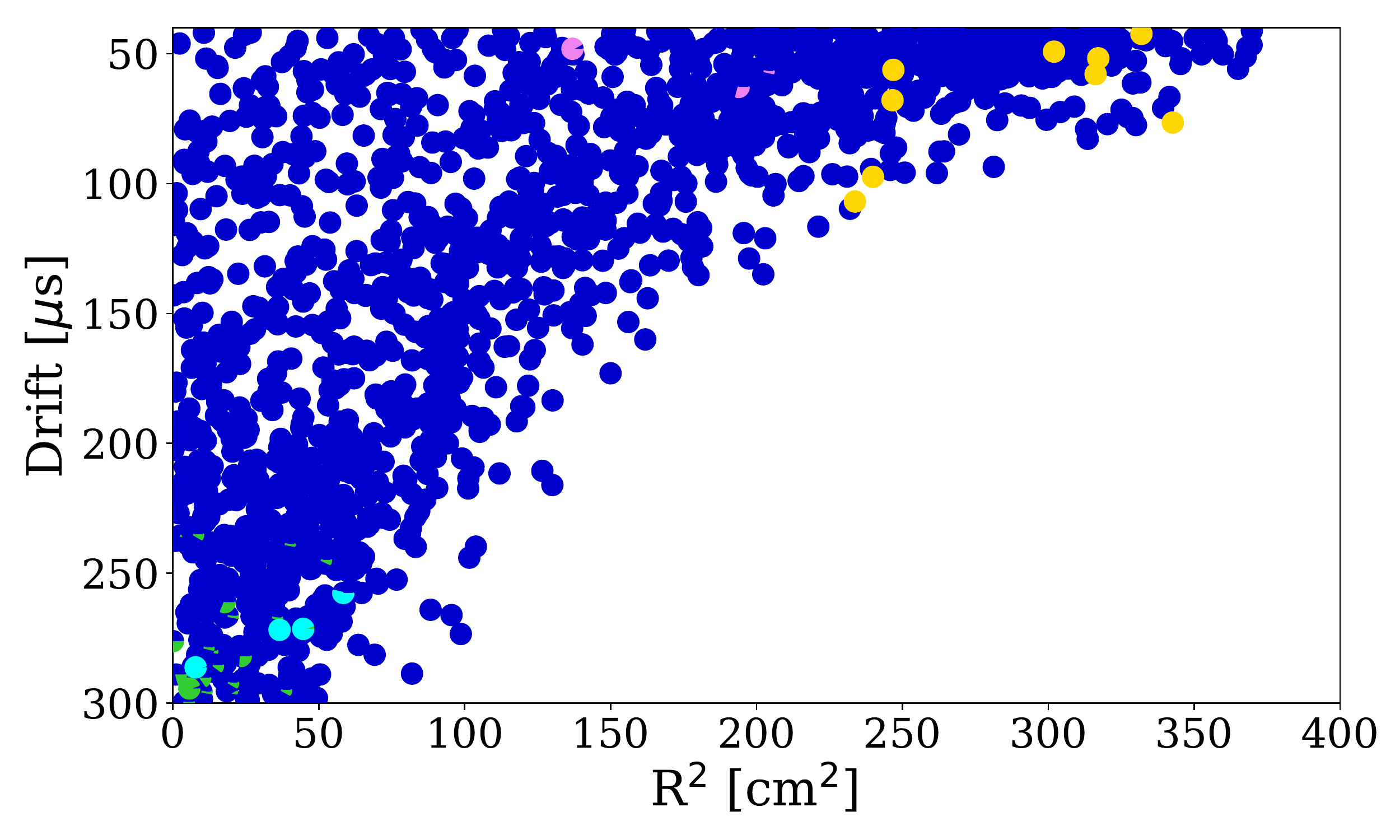}
  \caption{The final unsalted WS2014--16 data used in this analysis. Black markers indicate that the event was outside the final ROI used by the PLR. The remaining data are colored to indicate the values of a given background PDF at that point in phase space (probabilities are calculated using the background-only scenario). Data can have multi-colored markers, indicating that our expected background models overlap in certain regions of phase space. Note that all 16 drift time bins are merged in this plot, and the red solid and dashed lines represent the mean and 90\%~C.L. expected NR signal response for a flat energy spectrum averaged over the 16 drift time bins. Top: distribution of events in \{$S1_c$, $\log_{10}(S2_c)$\} space. Bottom: spatial distribution of final events using radii as seen by the top PMT array and electron drift time. Note that the spatial distribution is not constant as a function of $\phi$.}
  \label{dataPlots}
\end{figure}

\begin{figure*}
    \includegraphics[width=0.5\columnwidth]{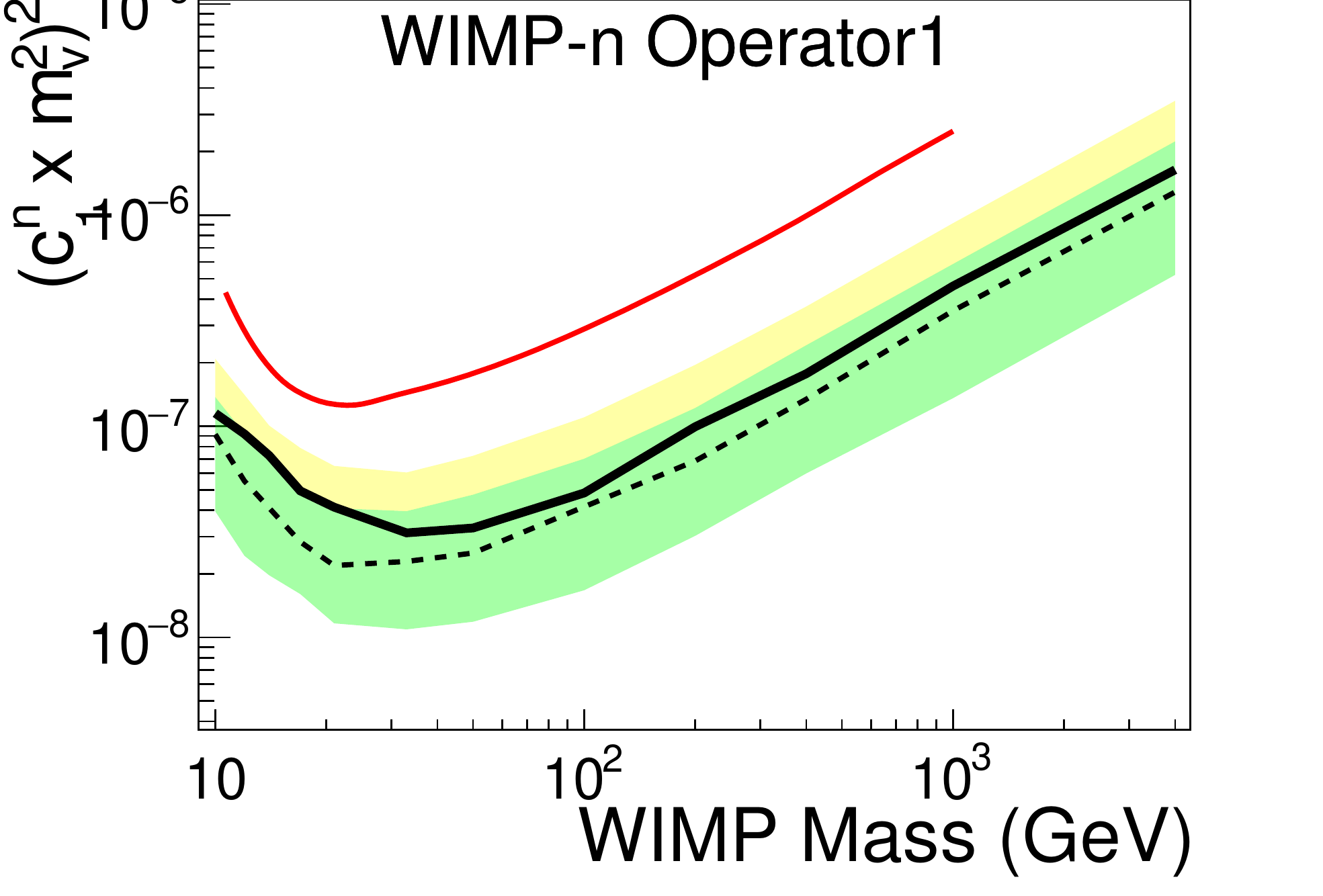}
    \includegraphics[width=0.5\columnwidth]{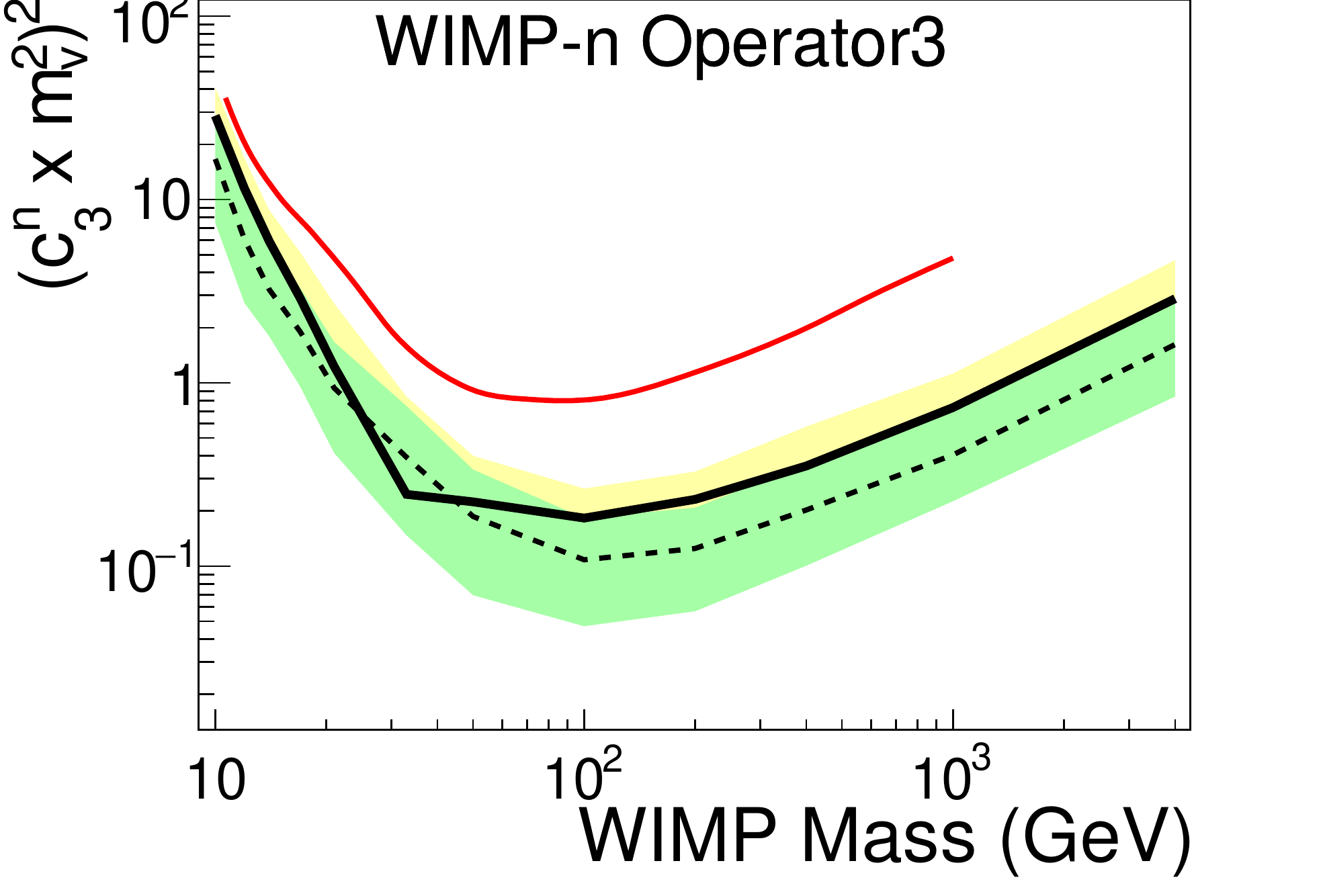}
    \includegraphics[width=0.5\columnwidth]{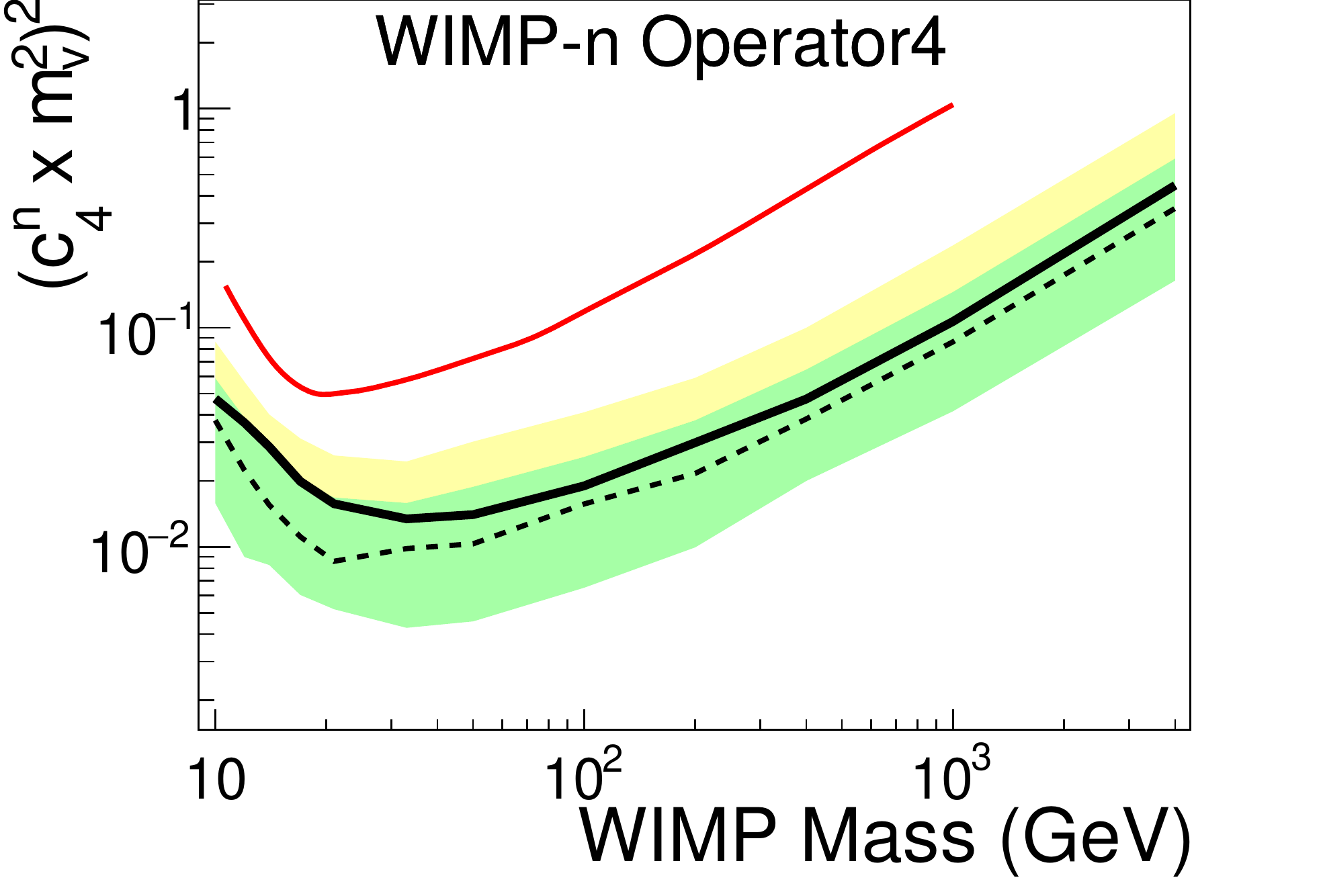}
    \includegraphics[width=0.5\columnwidth]{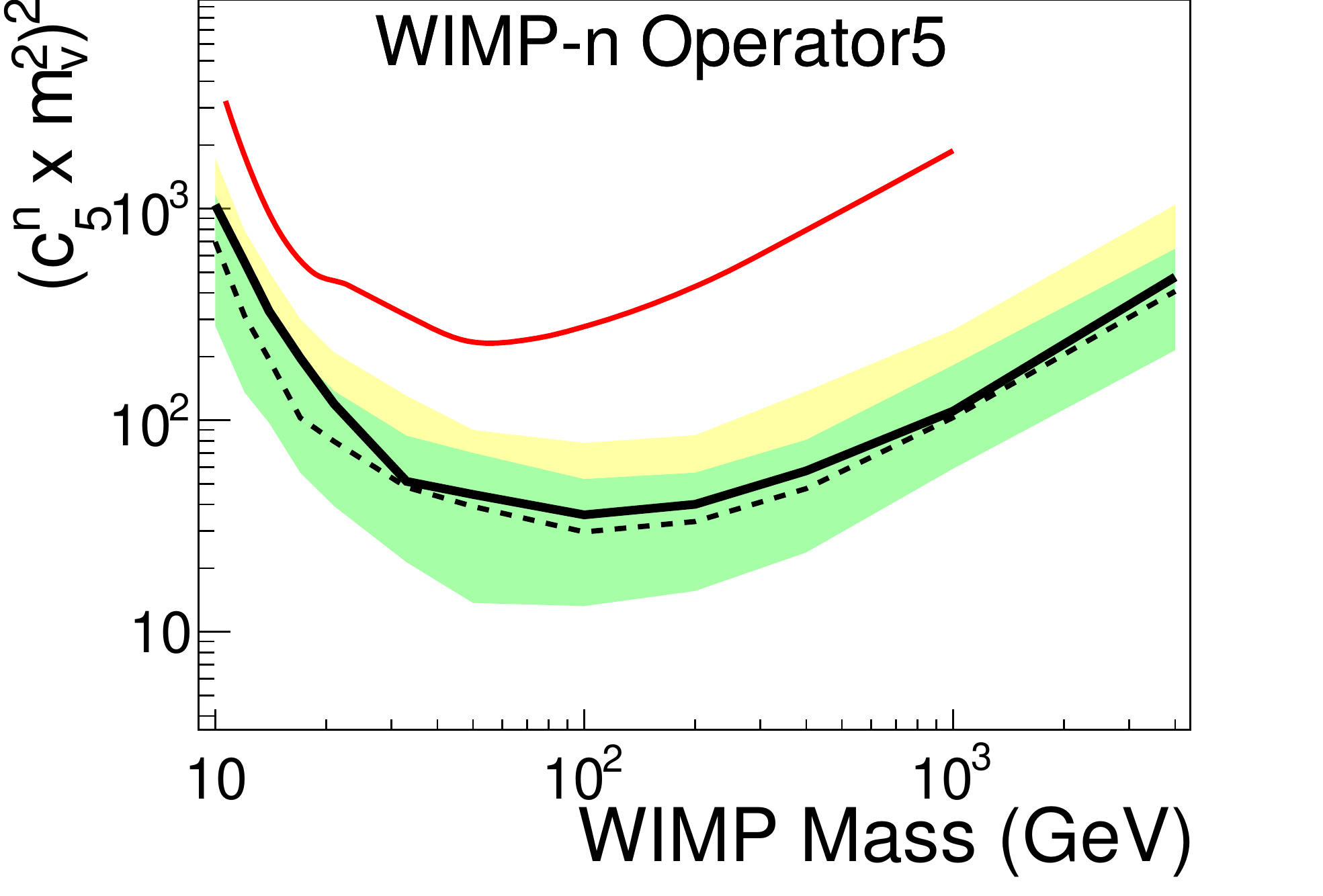}
    \includegraphics[width=0.5\columnwidth]{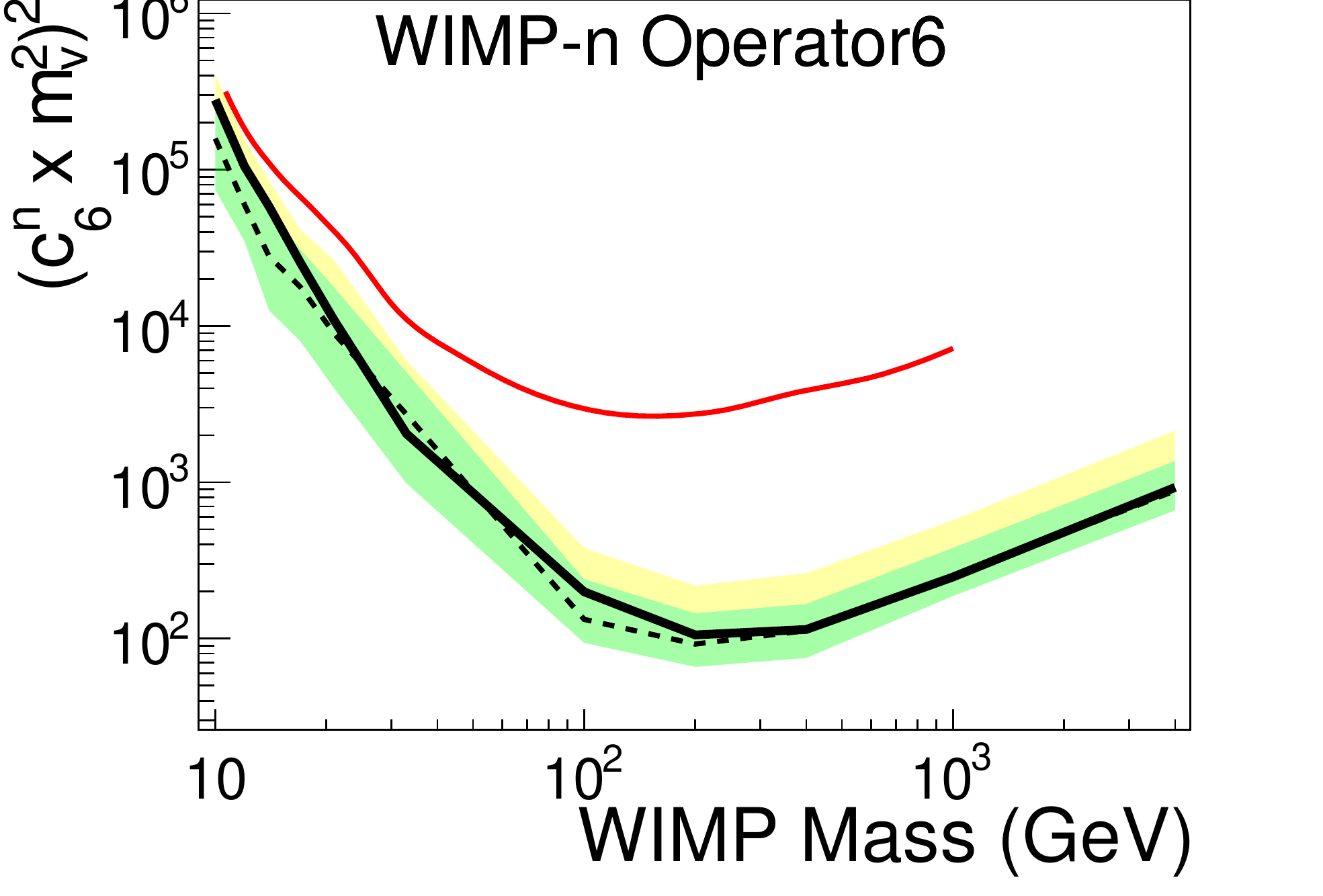}
    \includegraphics[width=0.5\columnwidth]{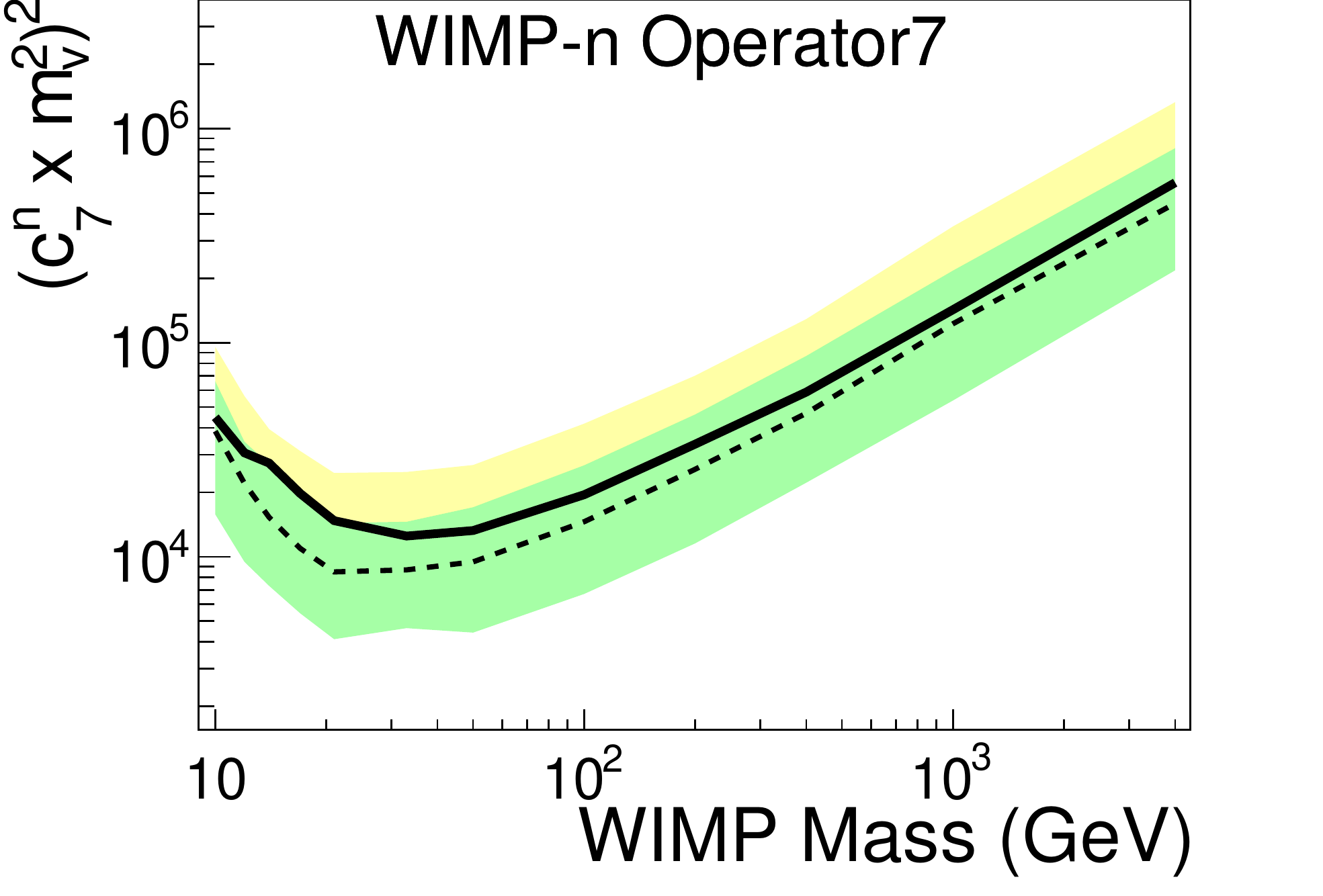}
    \includegraphics[width=0.5\columnwidth]{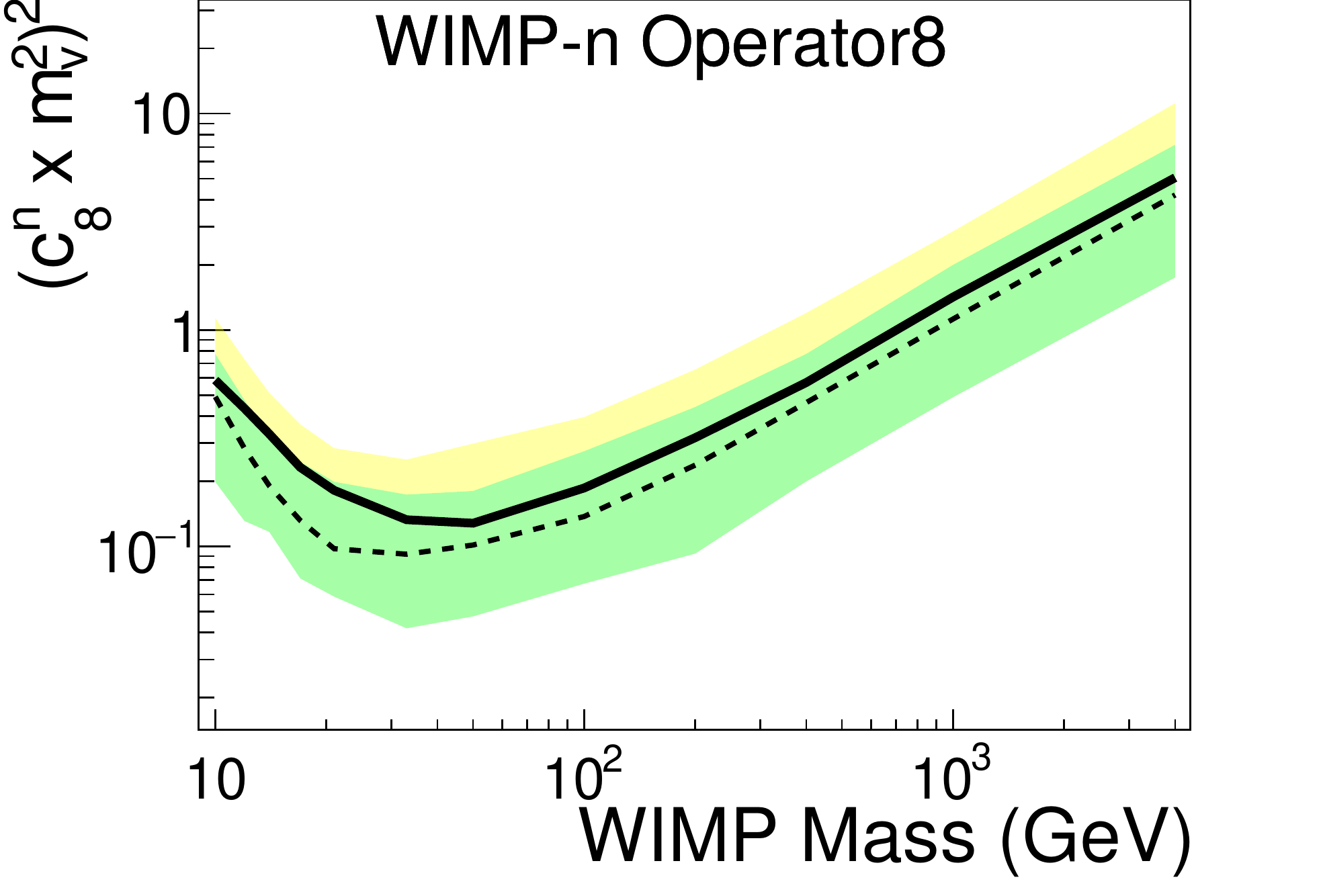}
    \includegraphics[width=0.5\columnwidth]{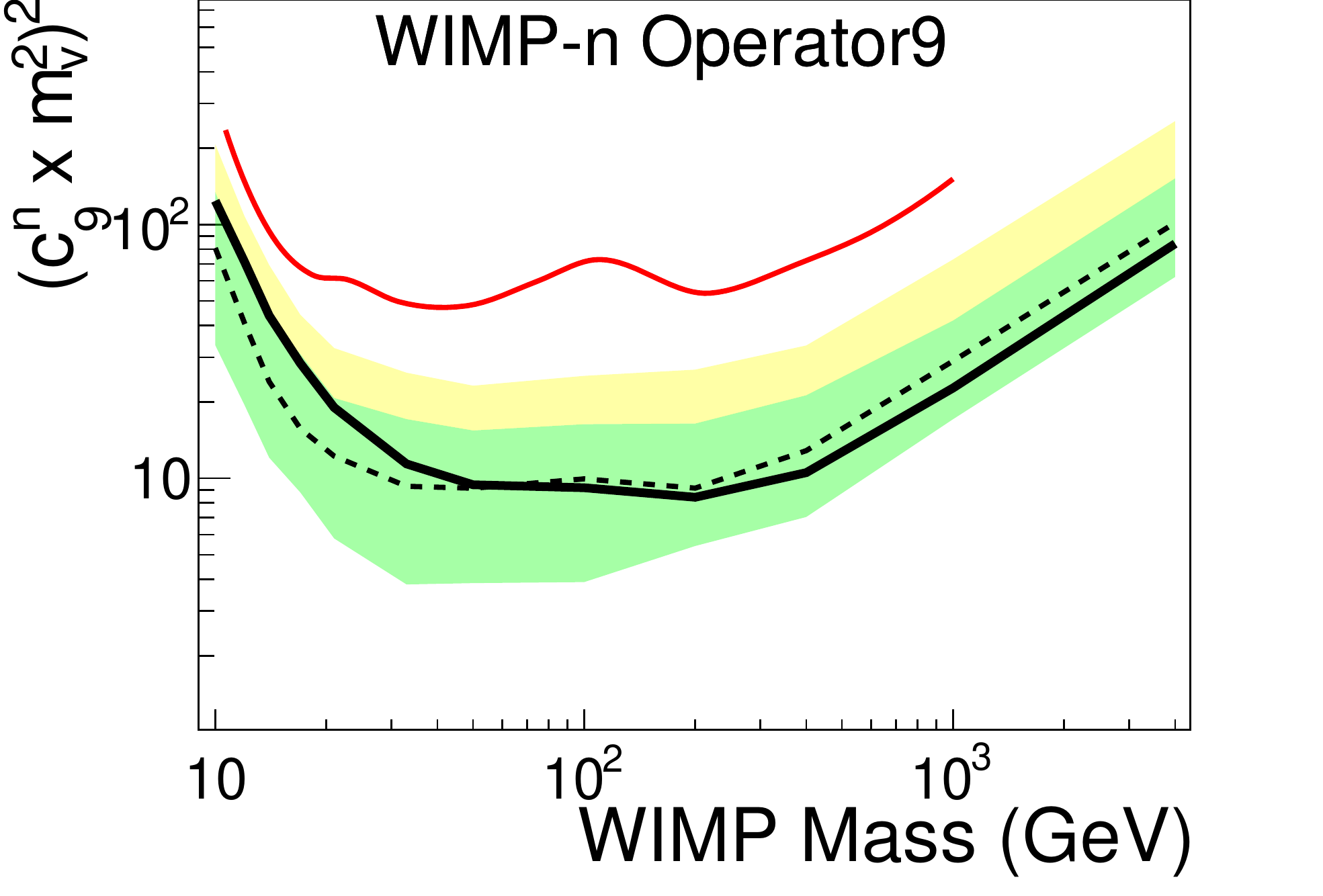}
    \includegraphics[width=0.5\columnwidth]{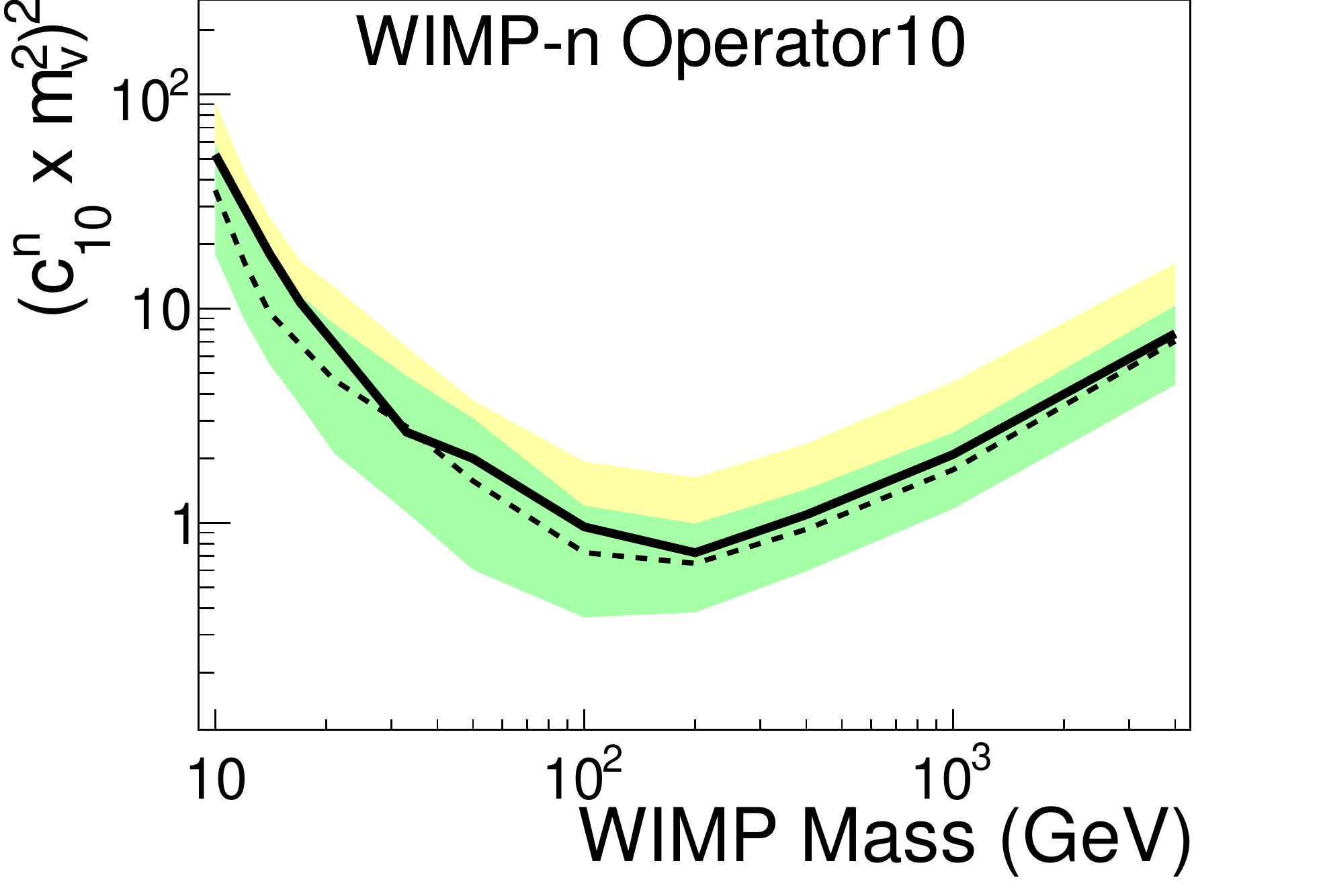}
    \includegraphics[width=0.5\columnwidth]{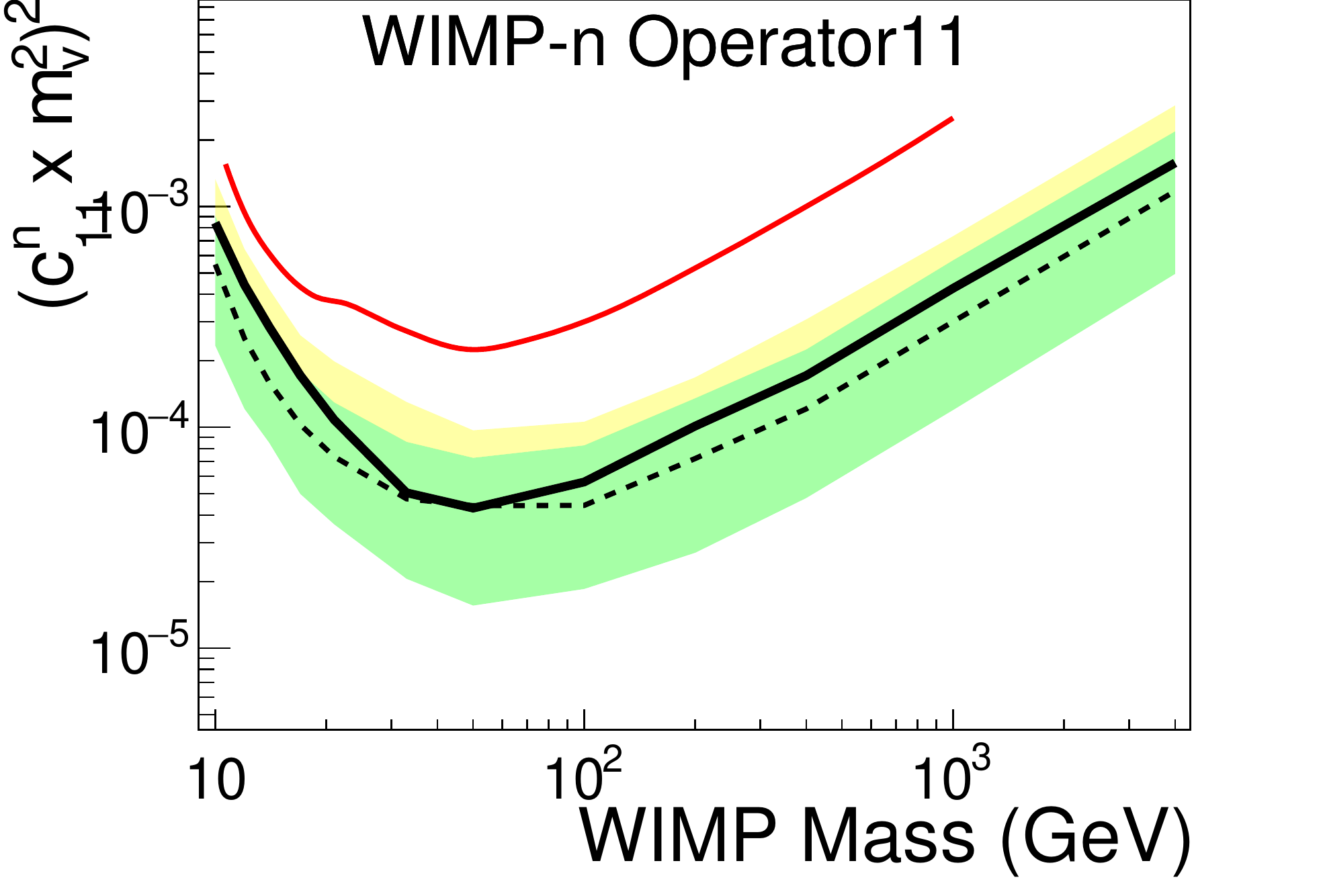}
    \includegraphics[width=0.5\columnwidth]{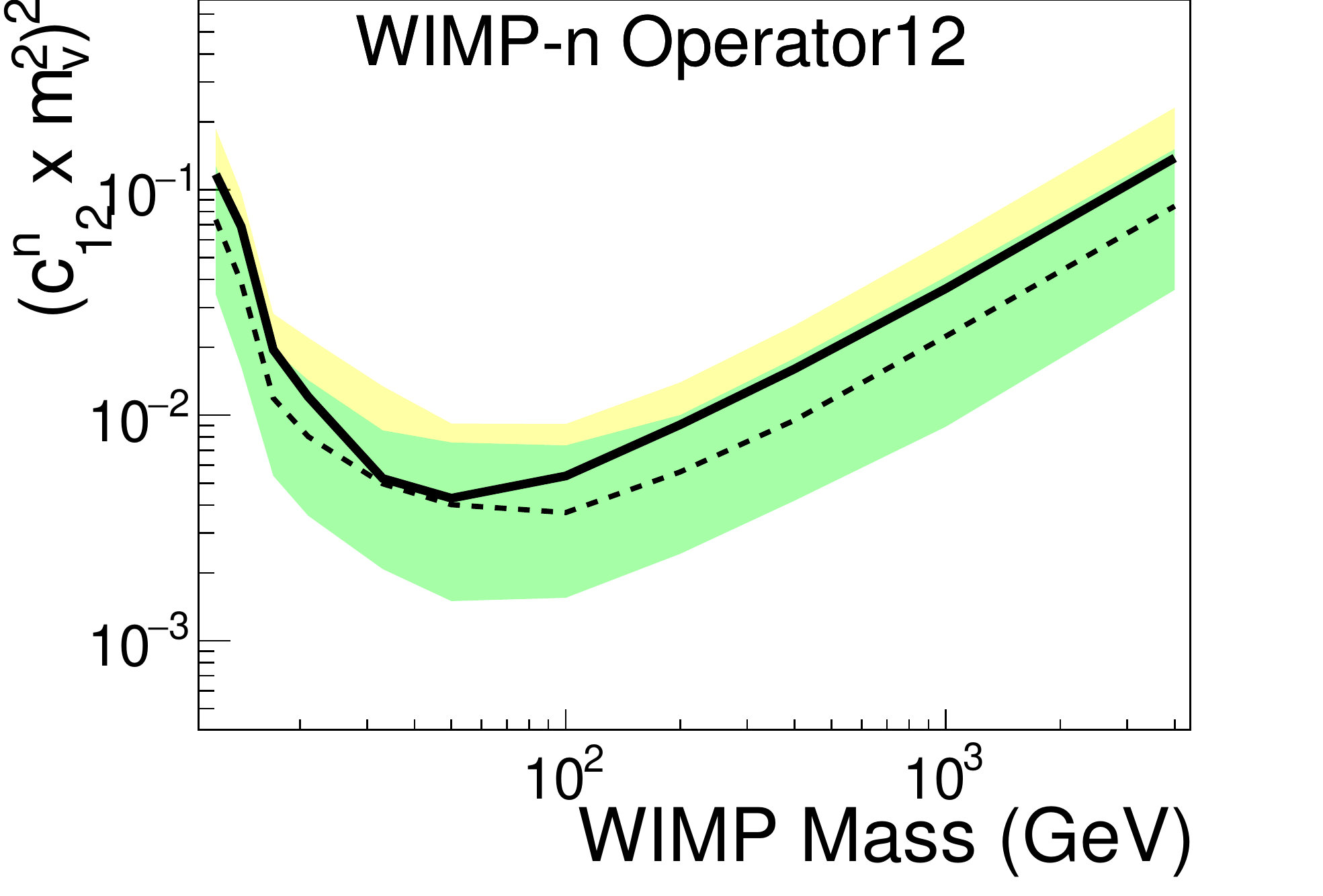}
    \includegraphics[width=0.5\columnwidth]{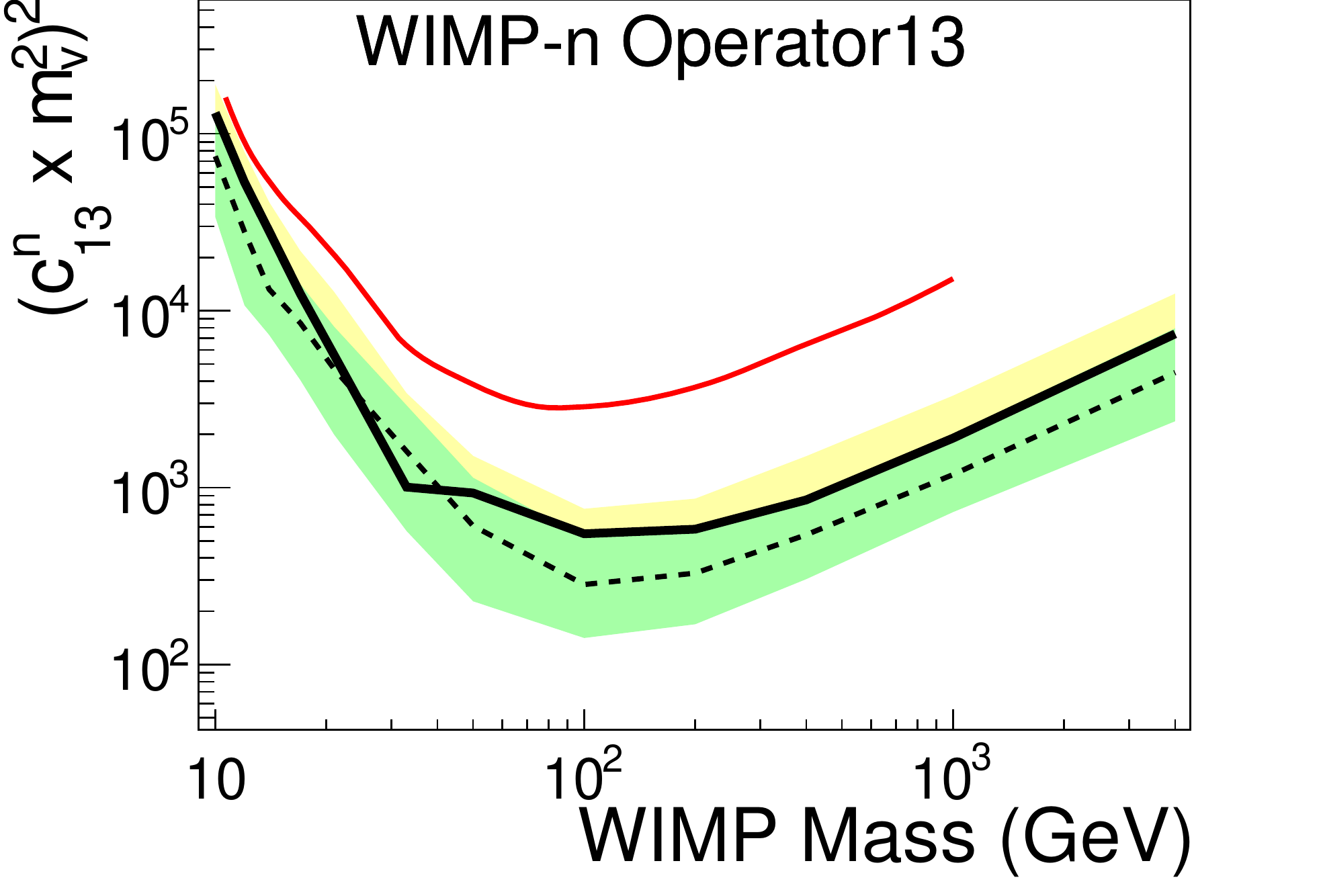}
    \includegraphics[width=0.5\columnwidth]{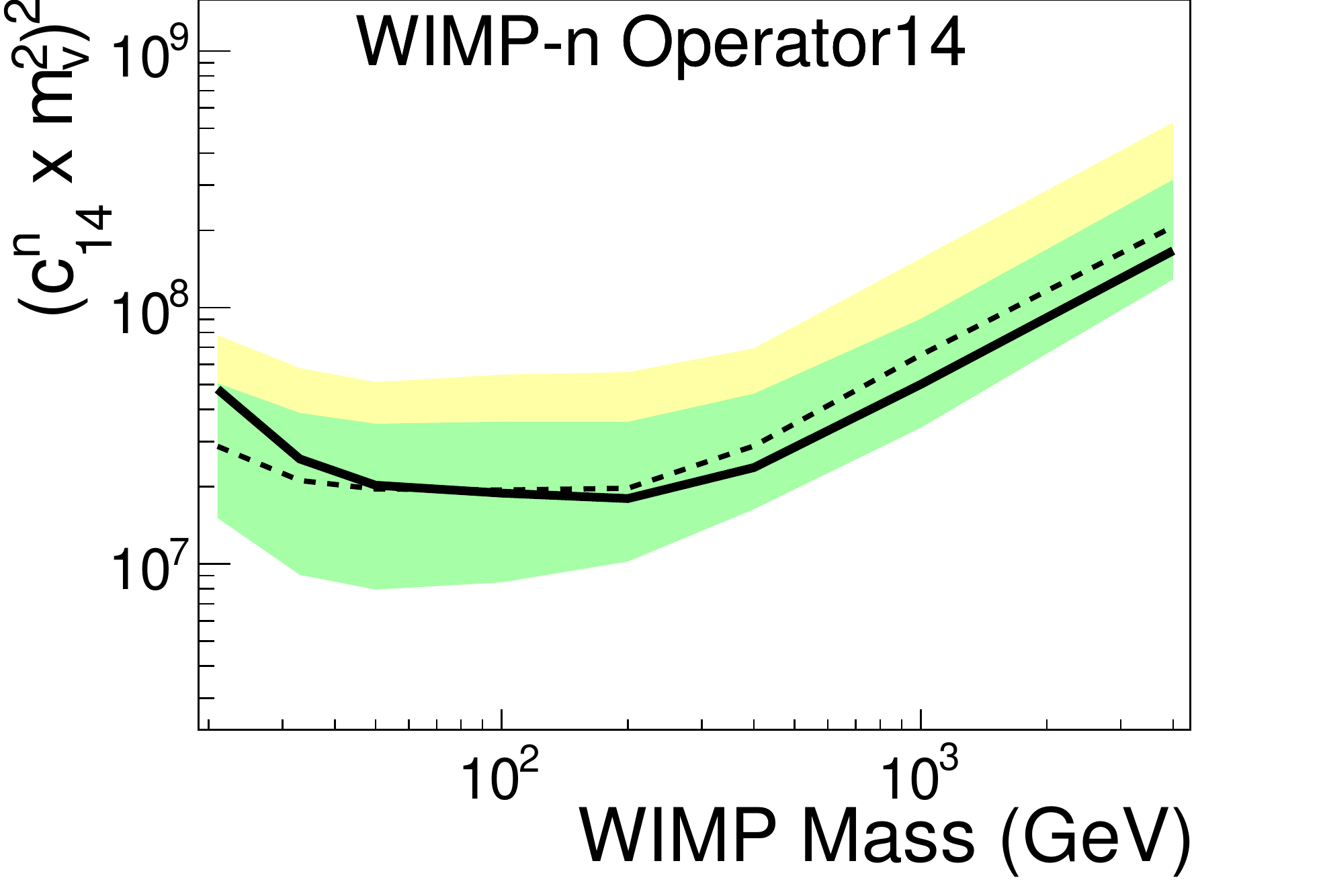}
    \includegraphics[width=0.5\columnwidth]{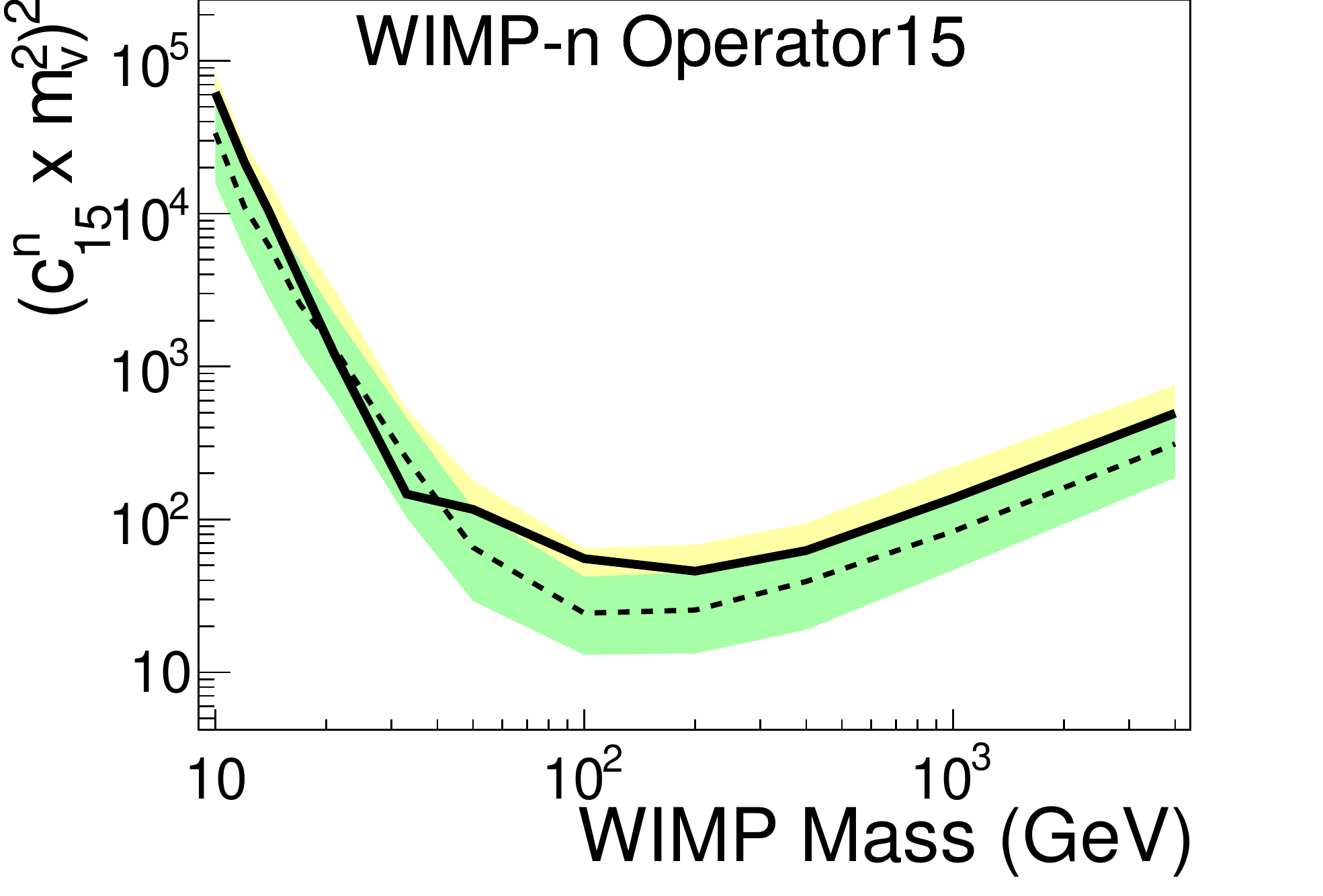}
  \caption{The LUX WS2014--16 90\%~C.L. limits for WIMP-neutron dimensionless couplings for each of the fourteen nonrelativistic EFT operators. Solid black lines show the limit, while dashed black indicate the expectation, with green and yellow bands indicating the $\pm1\sigma$ and $+2\sigma$ sensitivity expectations, respectively. Each plot uses mass values of 10, 12, 14, 17, 21, 33, 50, 100, 200, 400, 1000, and 4000 GeV, except for Operators 12 and 14, which begin at 12 and 21 GeV, respectively. Red lines show the upper limits from the WS2013 analysis~\cite{run3eft}. }
  \label{neutron_limits}
\end{figure*}

\begin{figure*}
    \includegraphics[width=0.5\columnwidth]{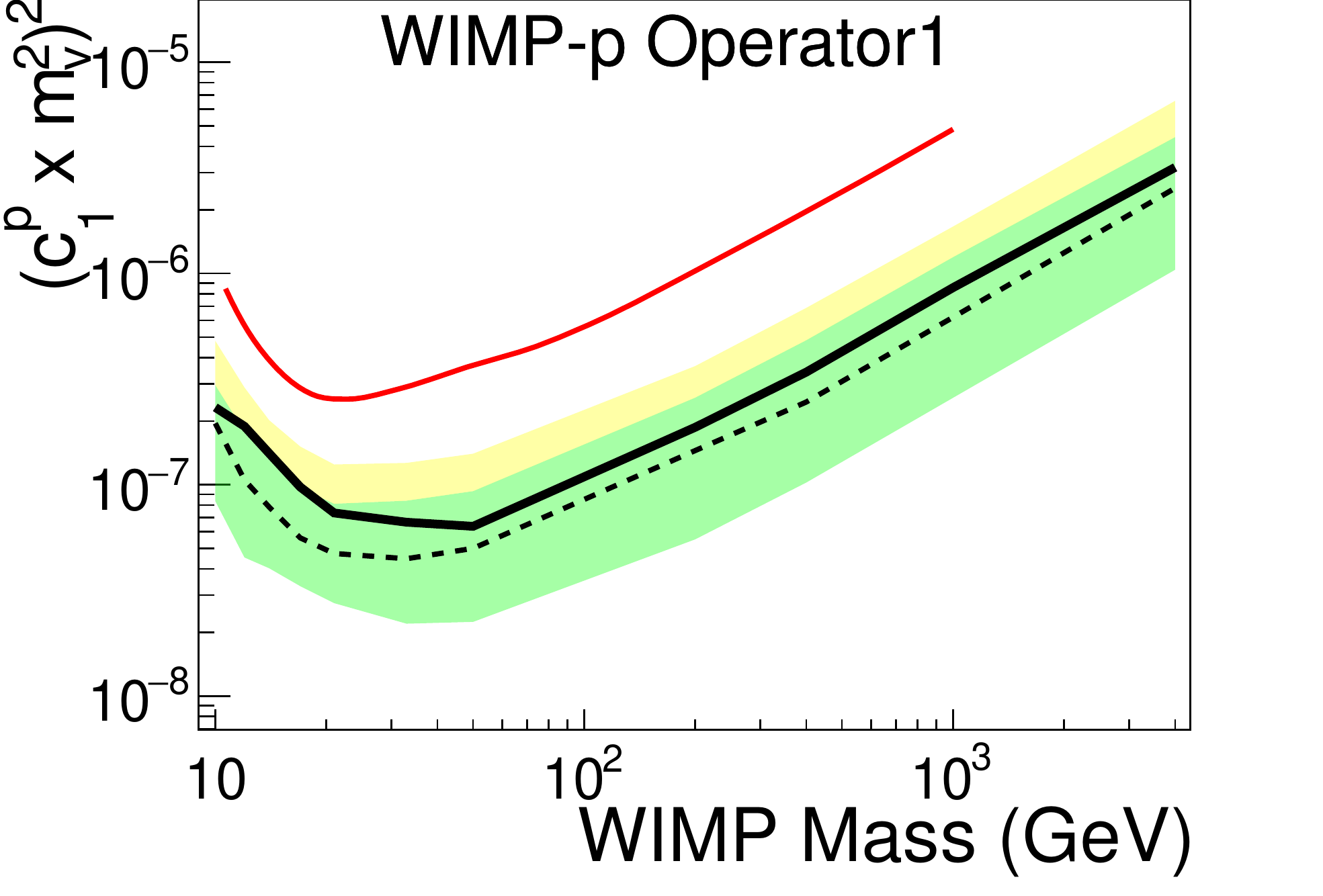}
    \includegraphics[width=0.5\columnwidth]{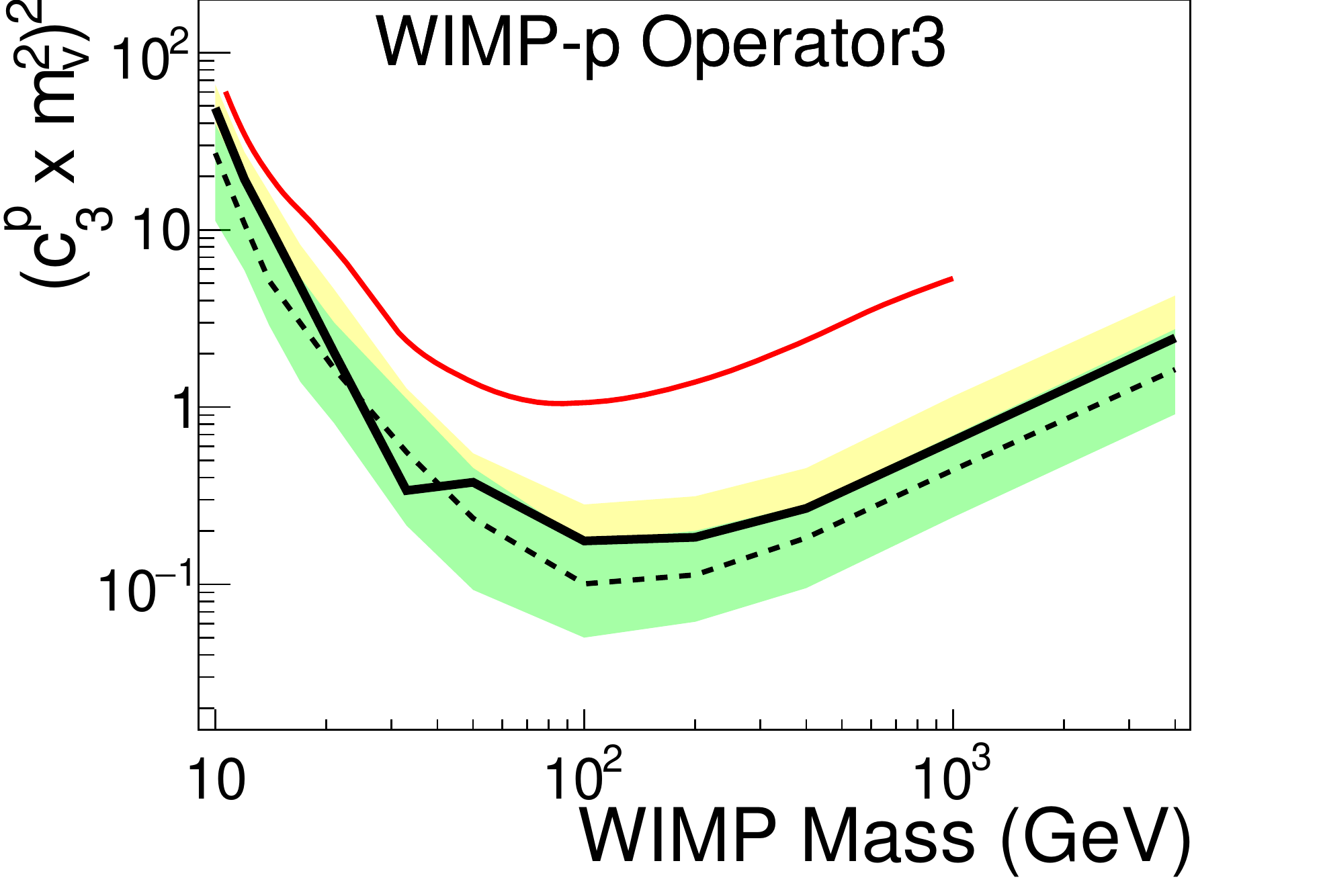}
    \includegraphics[width=0.5\columnwidth]{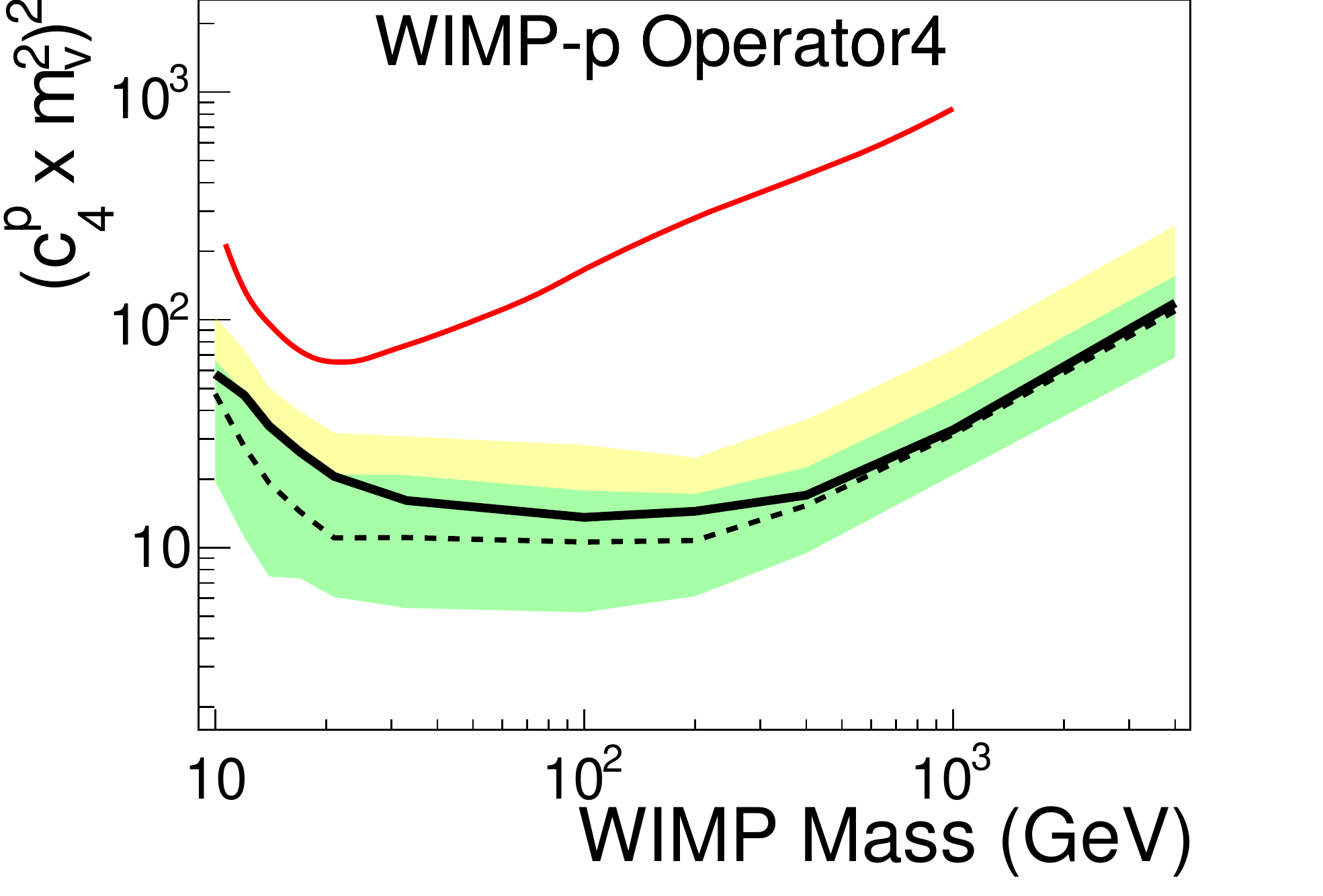}
    \includegraphics[width=0.5\columnwidth]{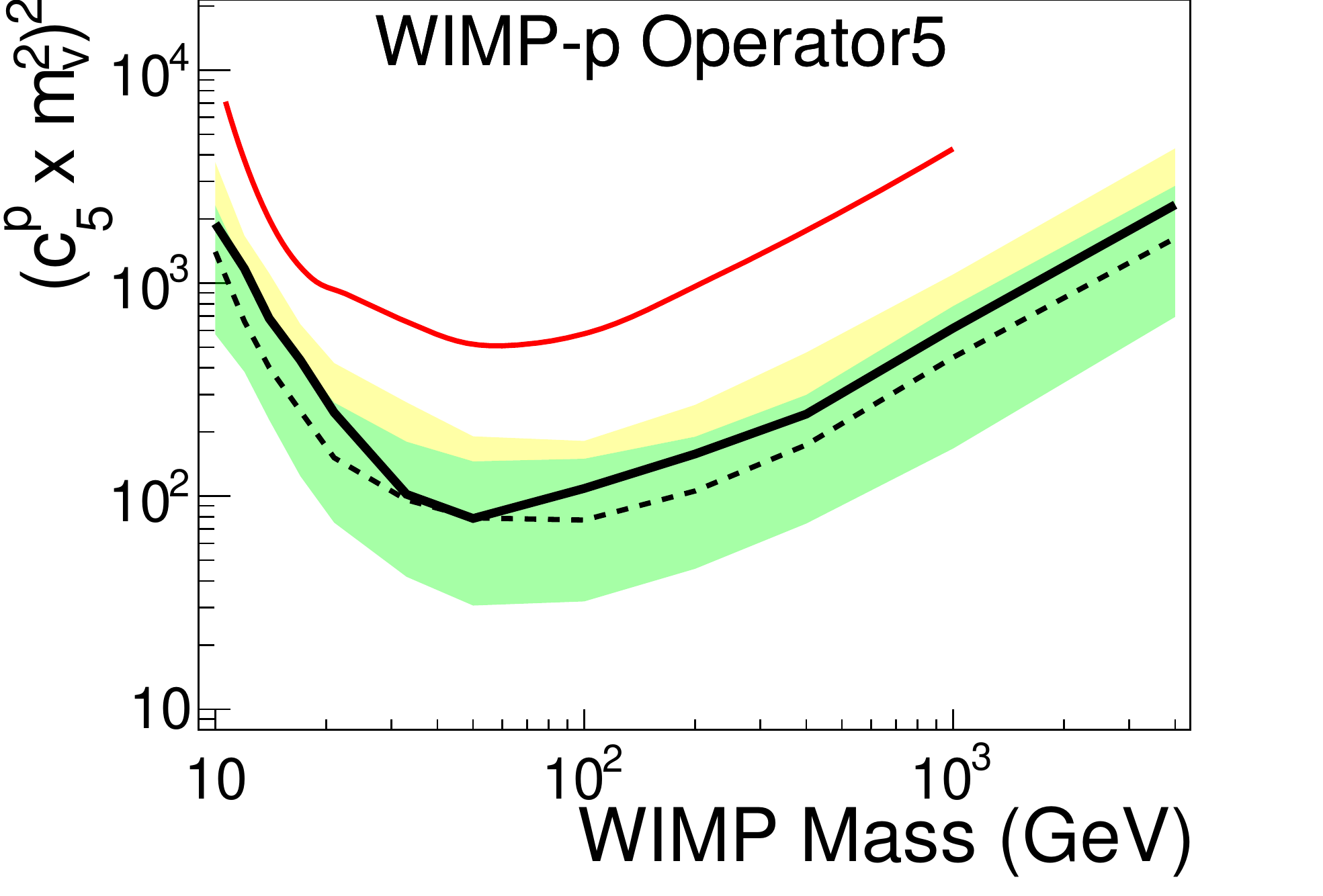}
    \includegraphics[width=0.5\columnwidth]{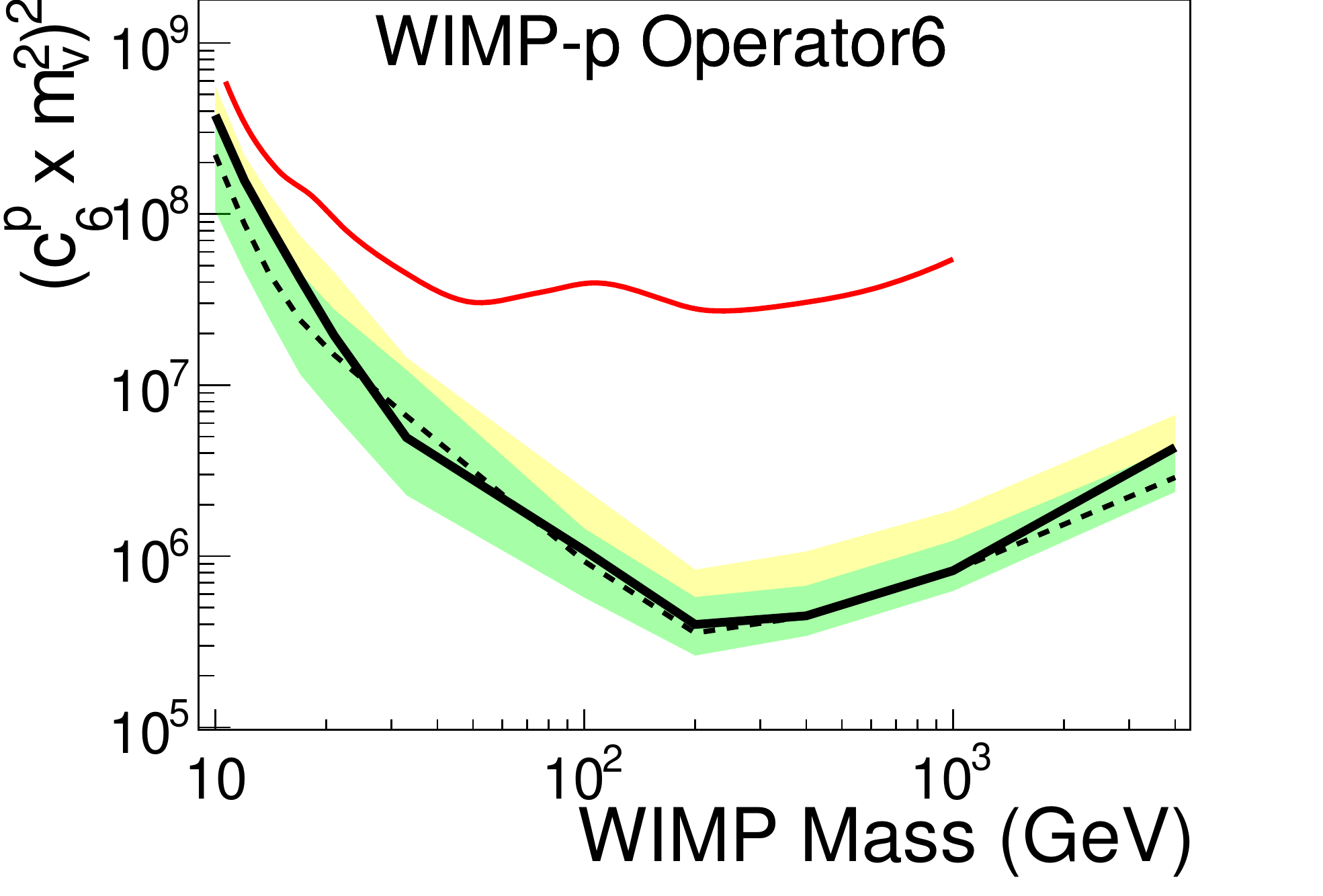}
    \includegraphics[width=0.5\columnwidth]{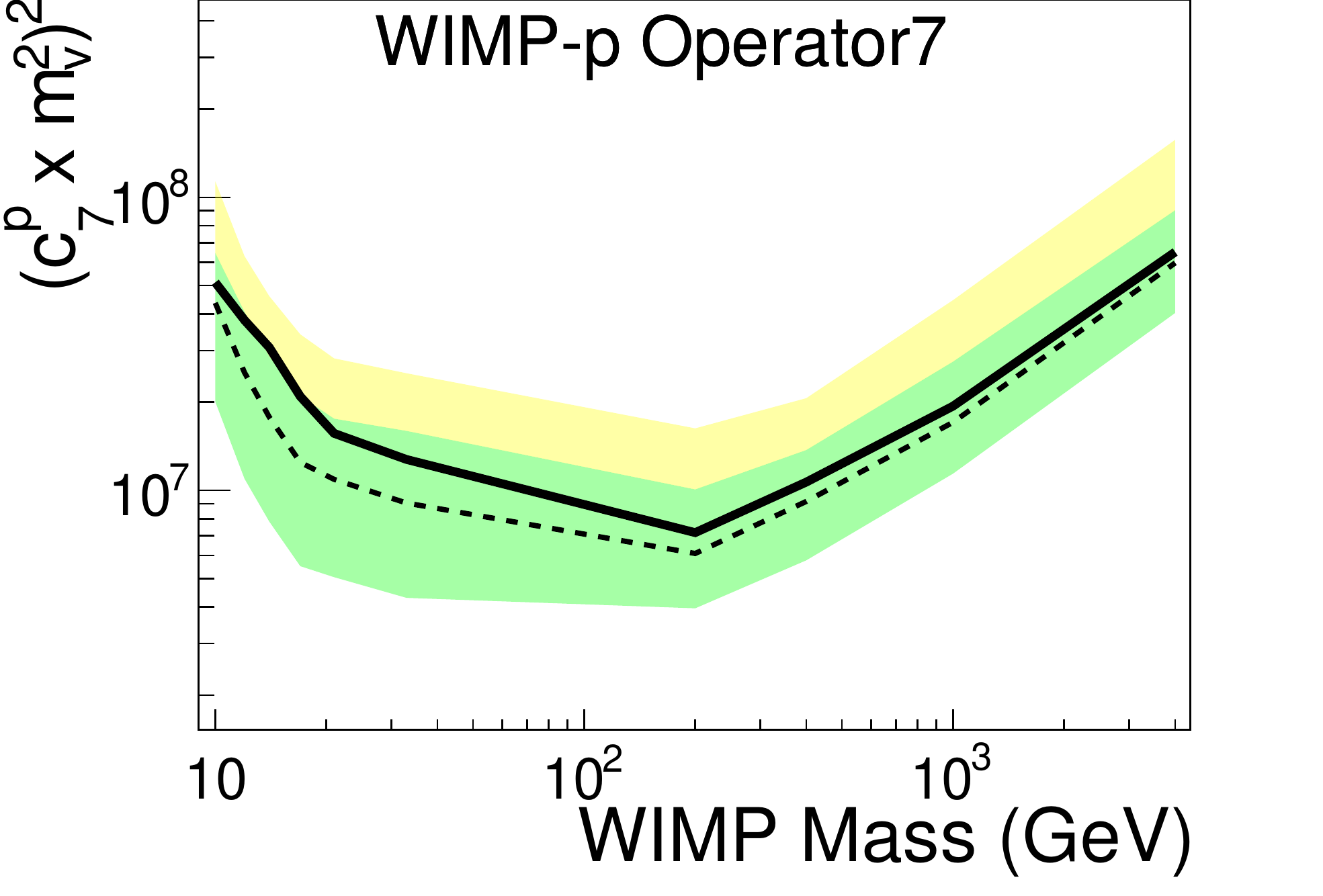}
    \includegraphics[width=0.5\columnwidth]{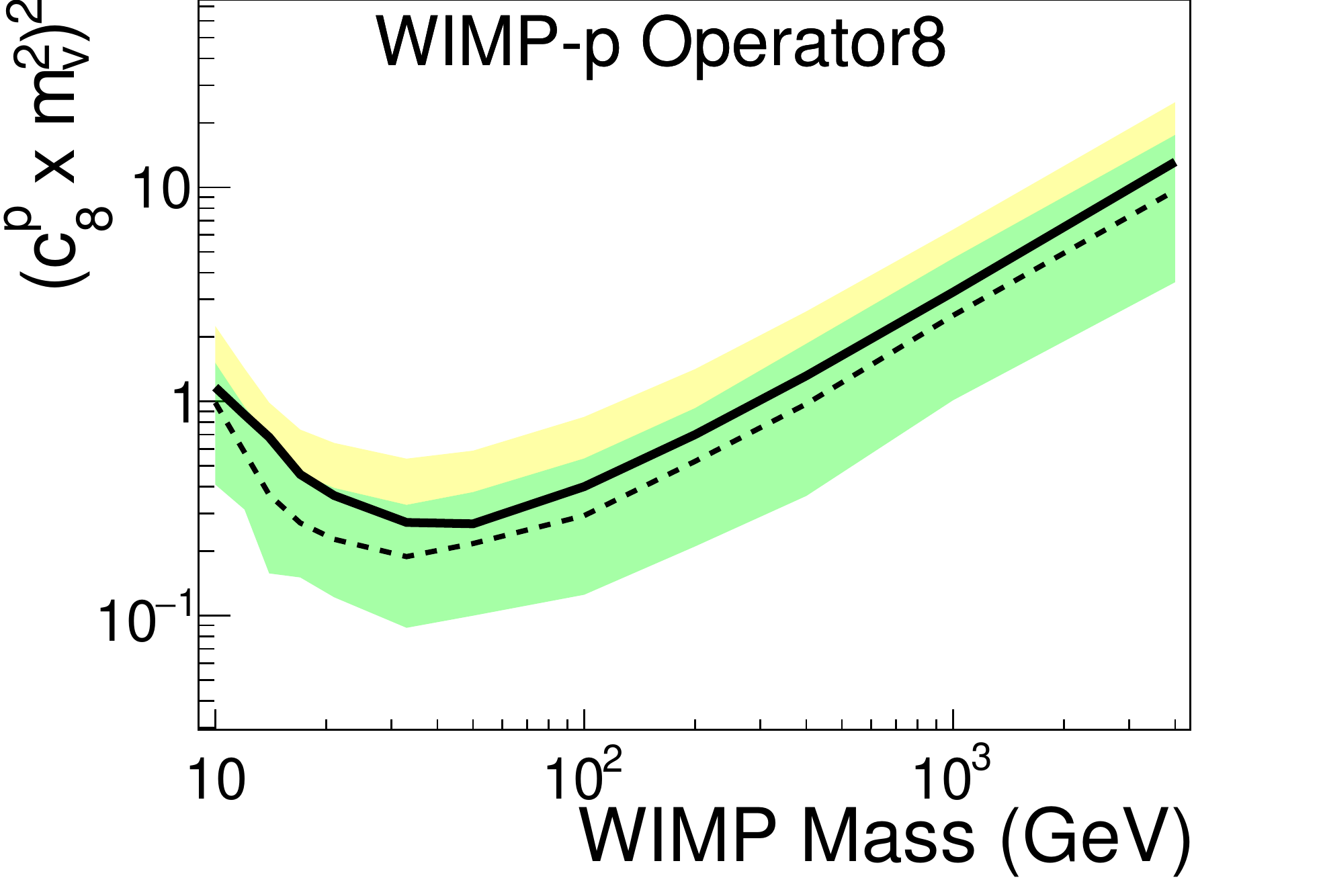}
    \includegraphics[width=0.5\columnwidth]{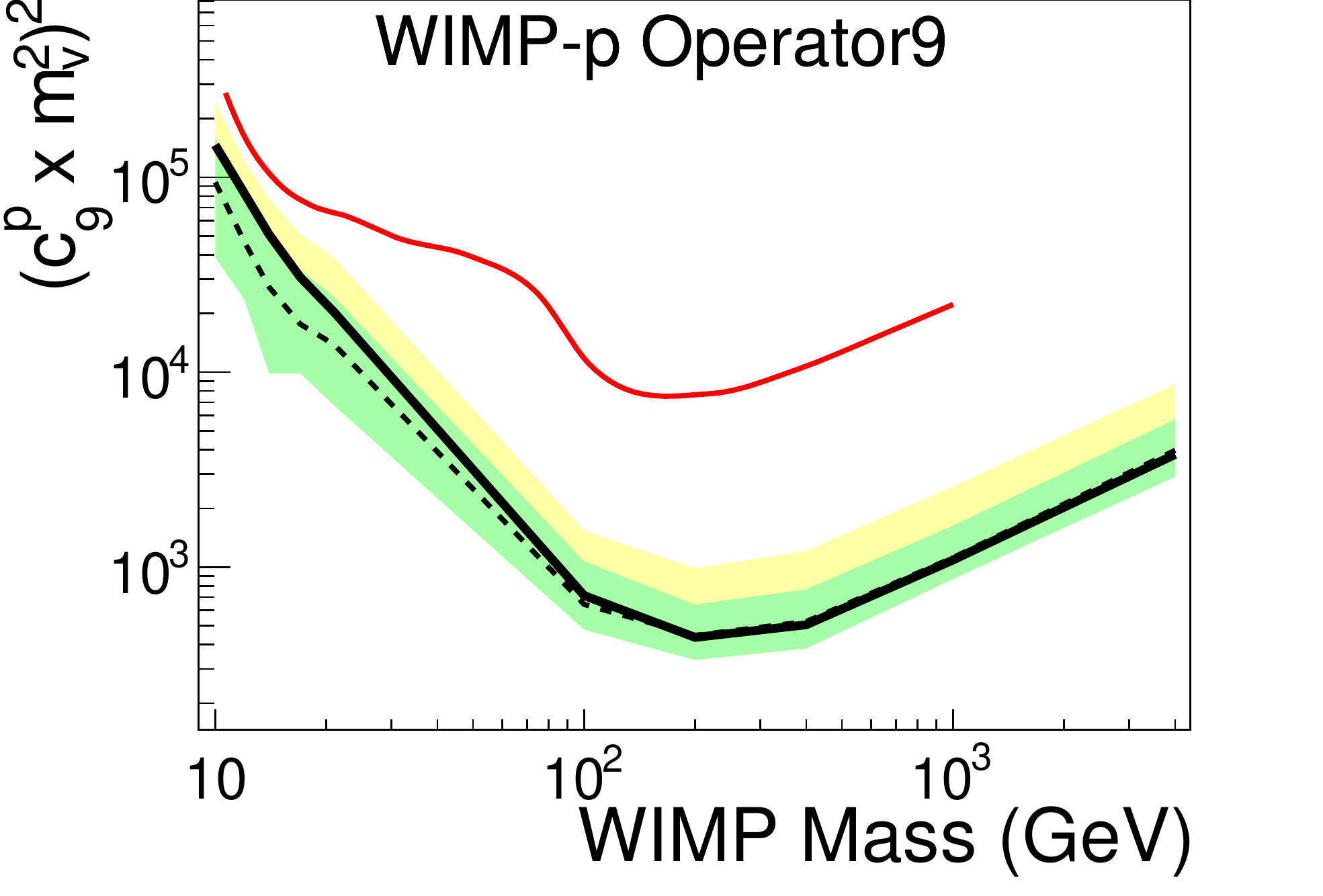}
    \includegraphics[width=0.5\columnwidth]{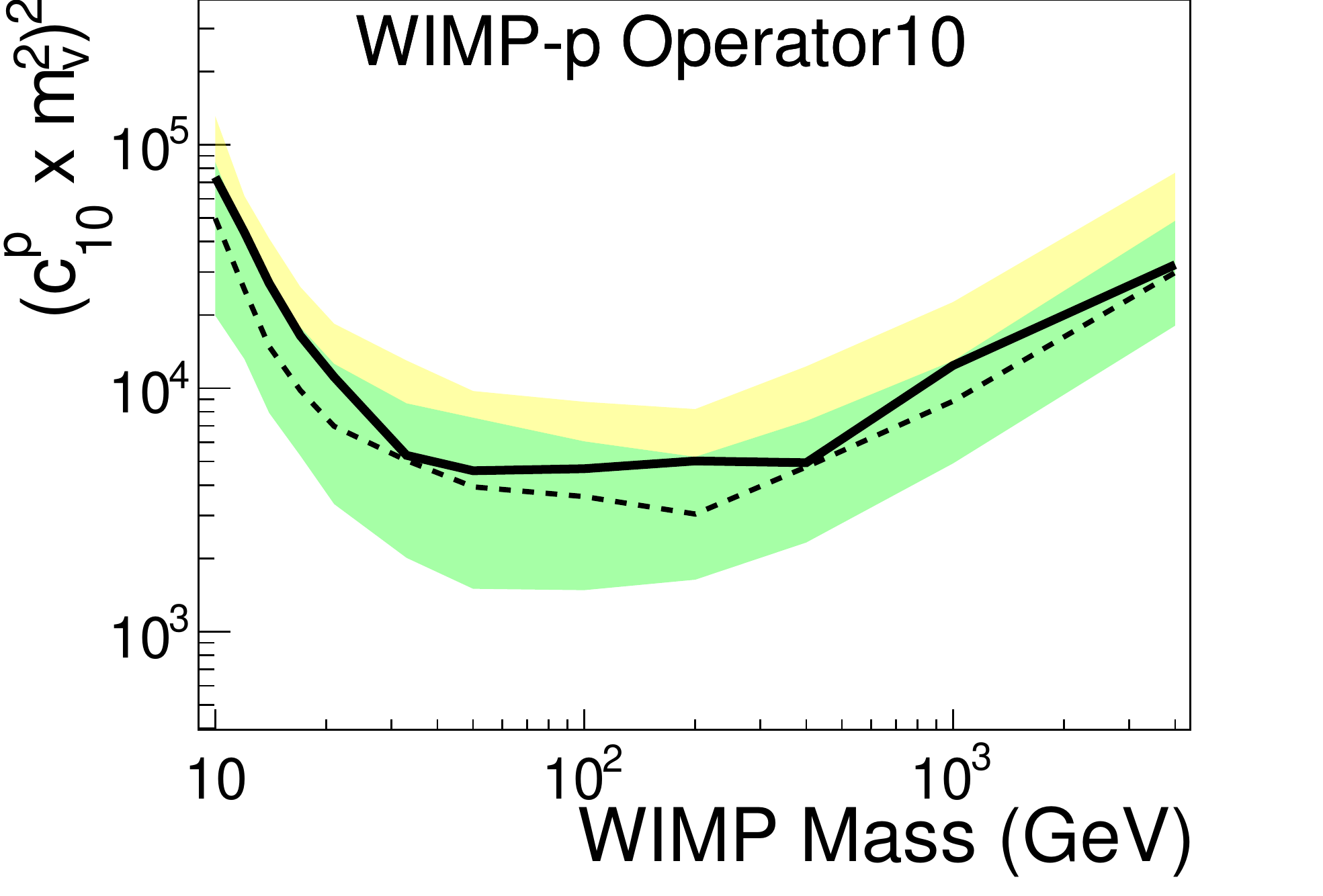}
    \includegraphics[width=0.5\columnwidth]{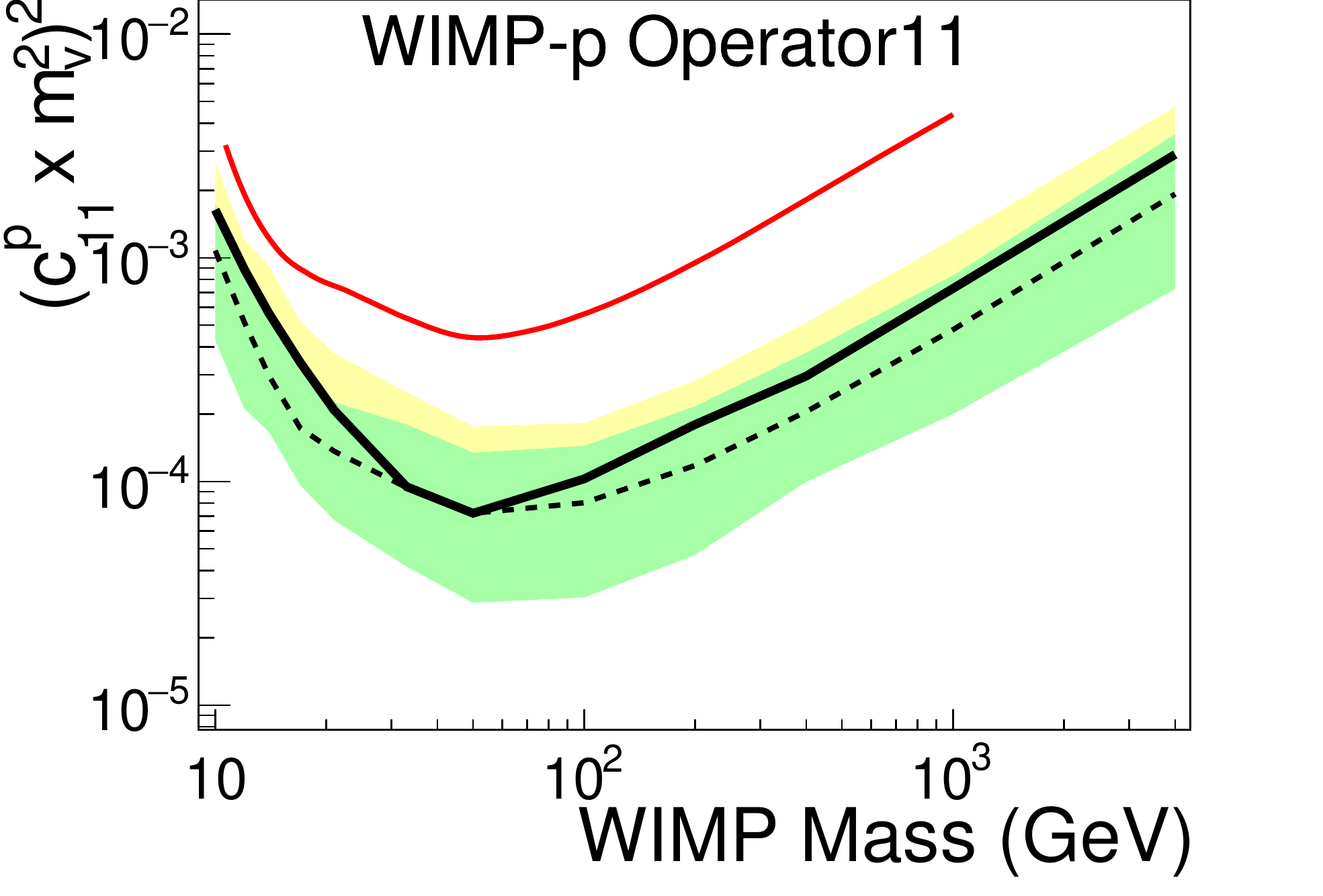}
    \includegraphics[width=0.5\columnwidth]{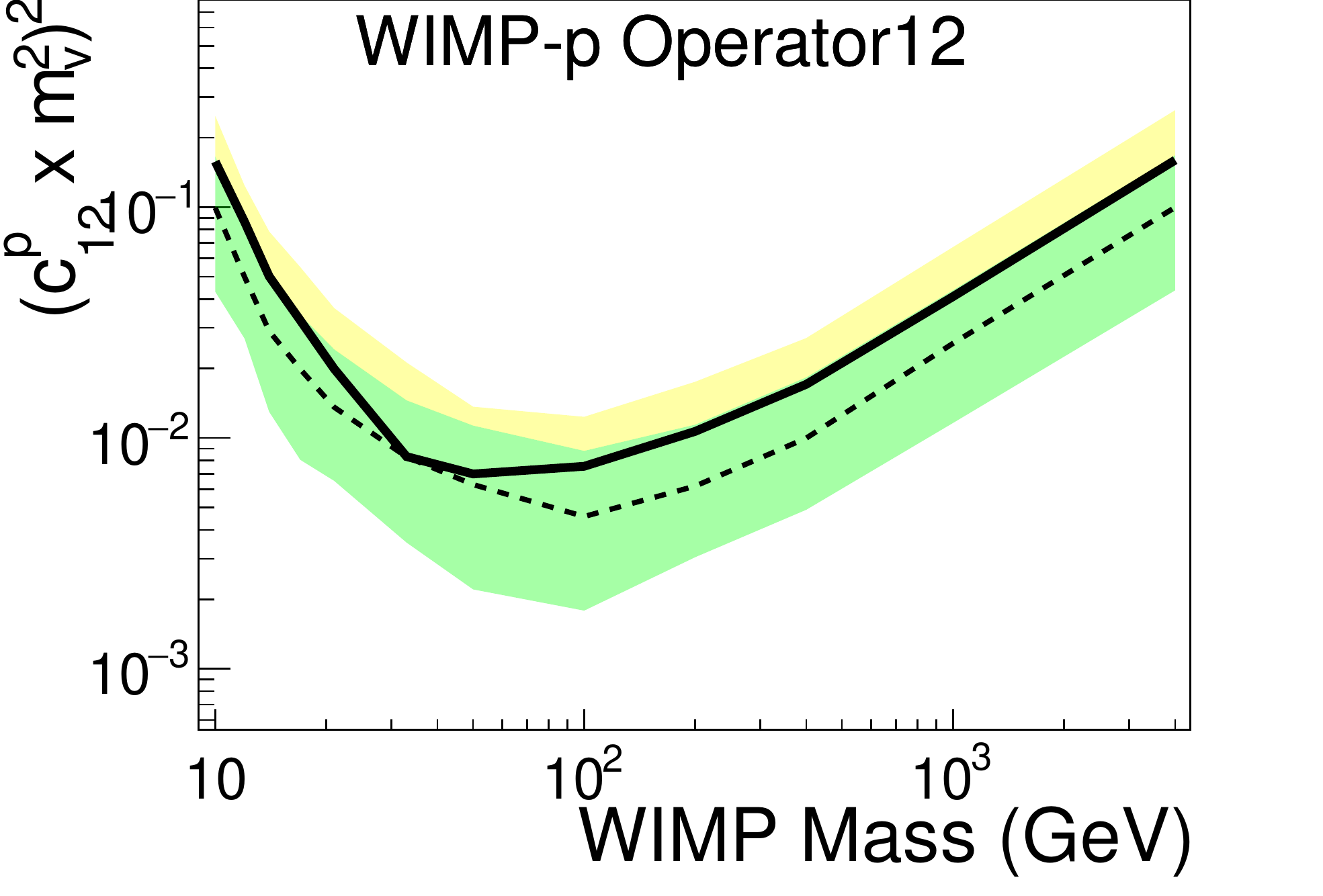}
    \includegraphics[width=0.5\columnwidth]{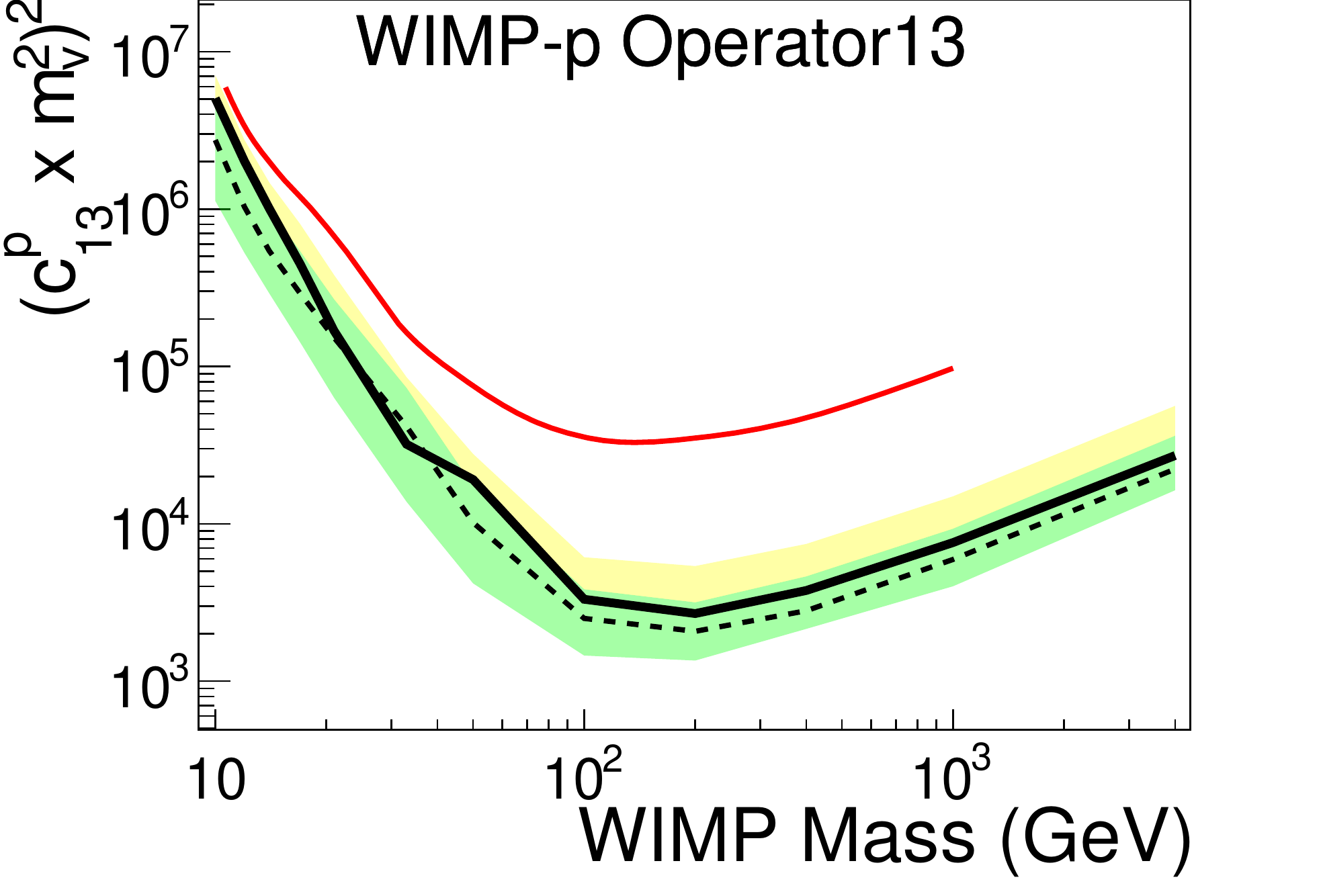}
    \includegraphics[width=0.5\columnwidth]{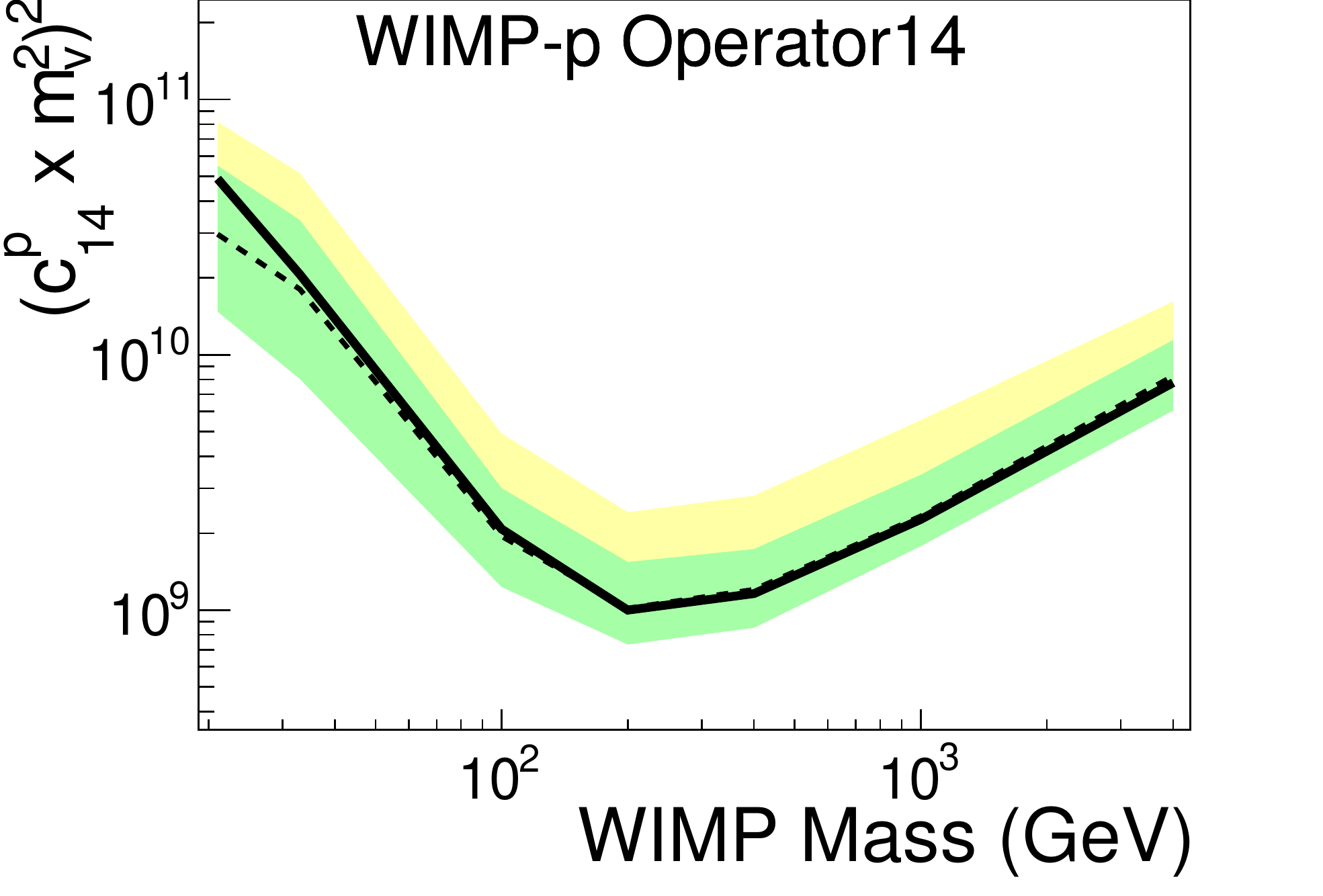}
    \includegraphics[width=0.5\columnwidth]{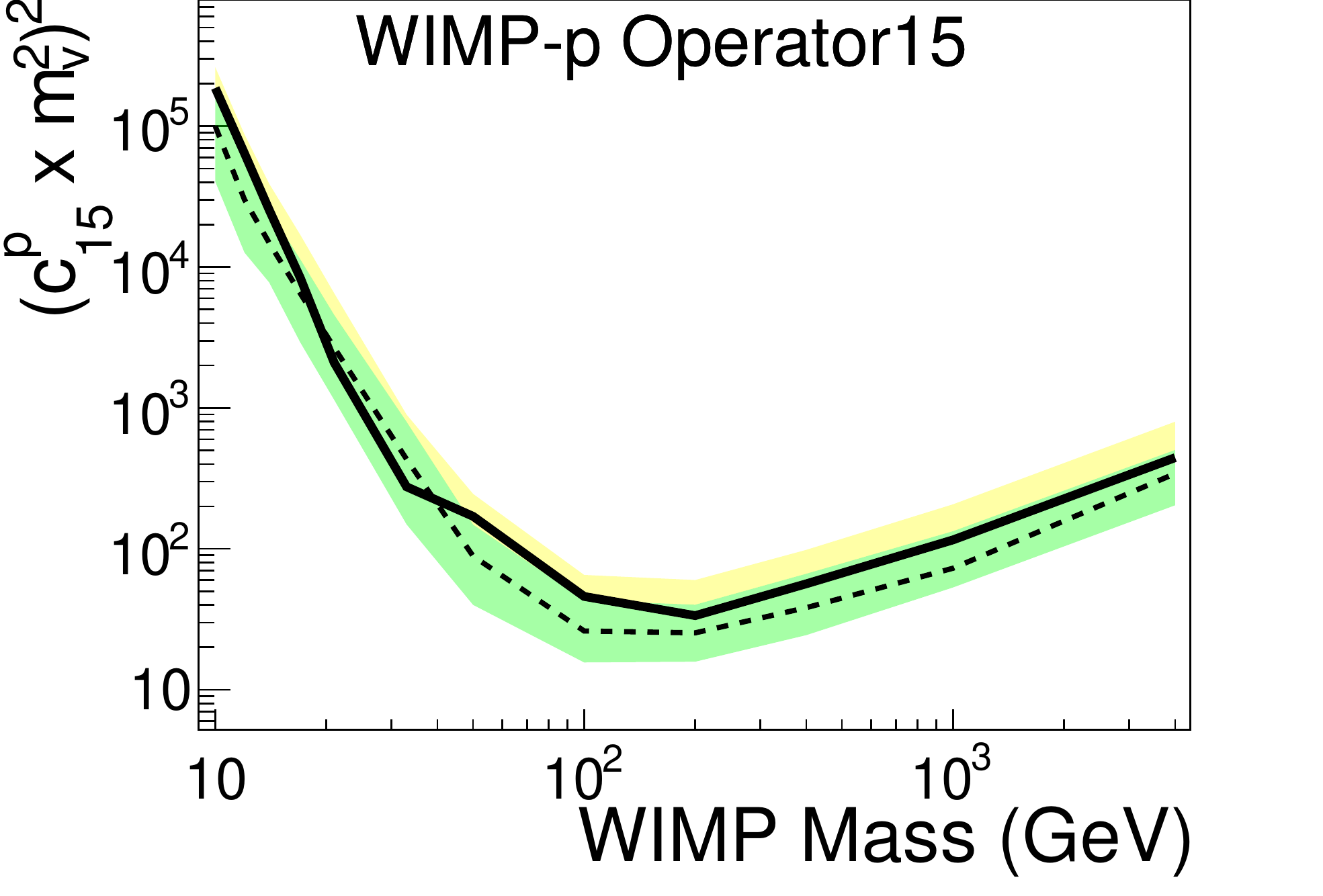}
  \caption{The LUX WS2014--16 90\% C.L. limits for WIMP-proton dimensionless couplings for each of the fourteen nonrelativistic EFT operators. Solid black lines show the limit, while dashed black indicate the expectation, with green and yellow bands indicating the $\pm1\sigma$ and $+2\sigma$ sensitivity expectations, respectively. Each plot uses mass values of 10, 12, 14, 21, 33, 50, 100, 200, 400, 1000, and 4000 GeV, with the exception of Operator 14, which begins at 21 GeV. Red lines show the upper limits from the WS2013 analysis~\cite{run3eft} .}
  \label{proton_limits}
\end{figure*}

\begin{figure*}
    \includegraphics[width=0.5\columnwidth]{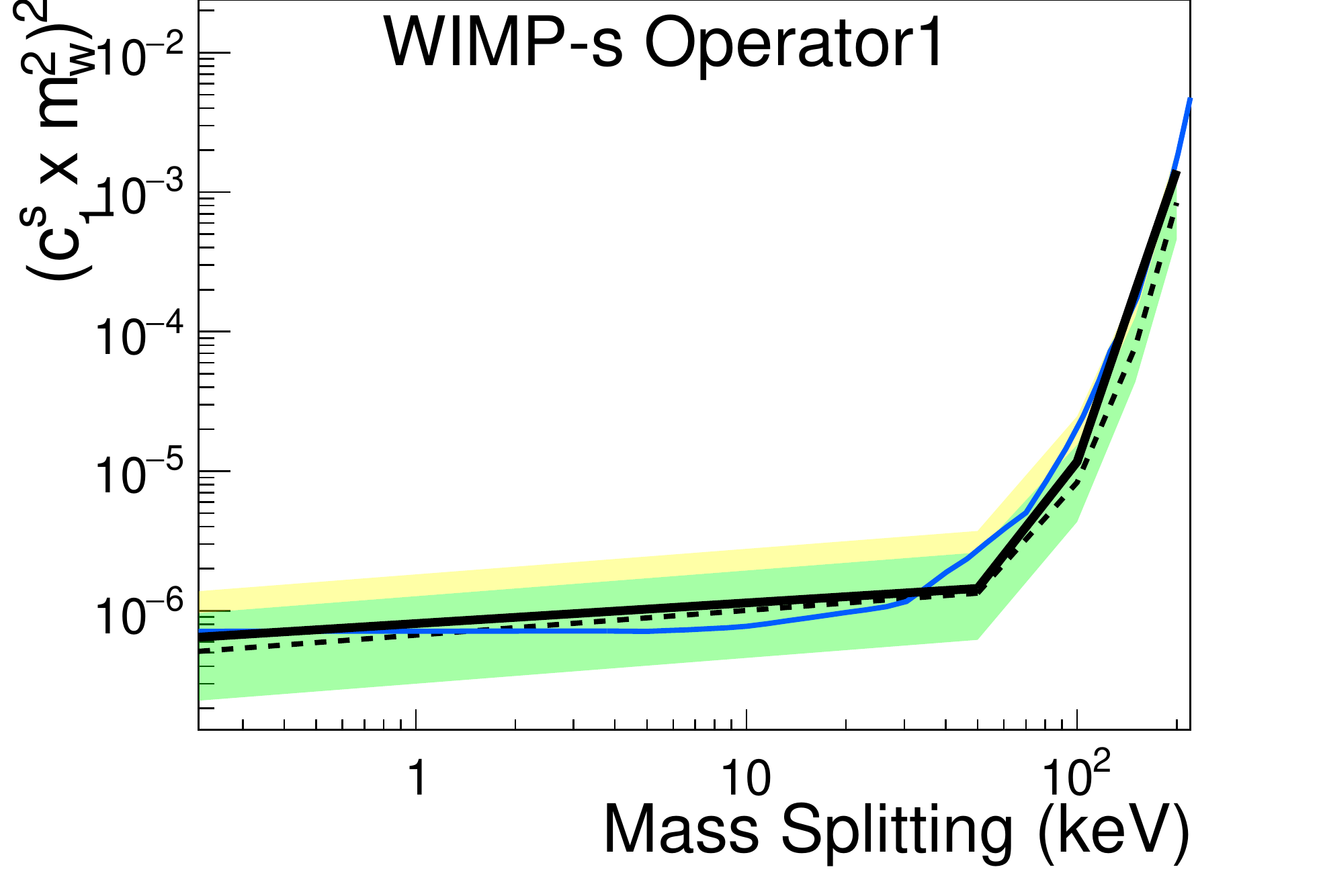}
    \includegraphics[width=0.5\columnwidth]{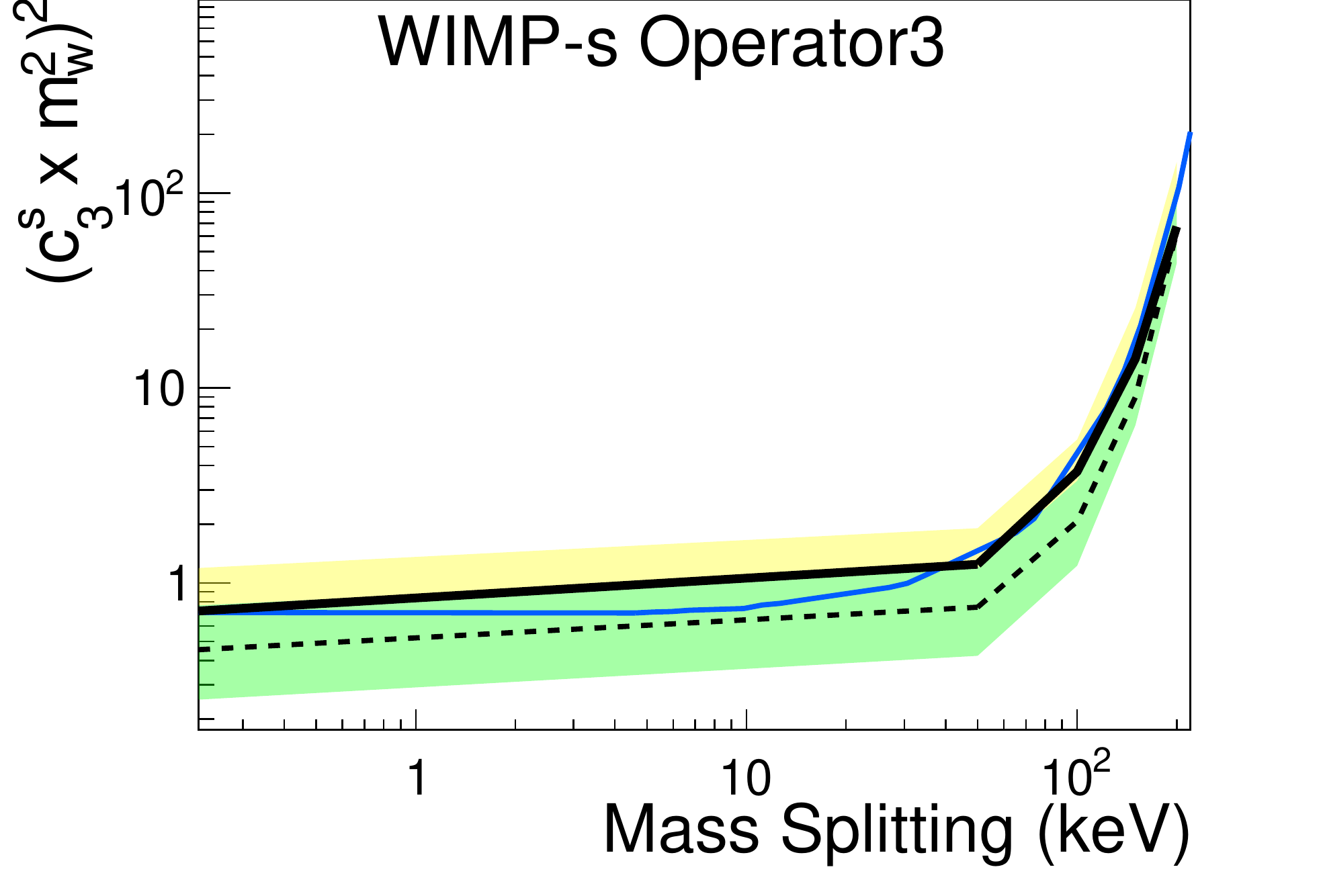}
    \includegraphics[width=0.5\columnwidth]{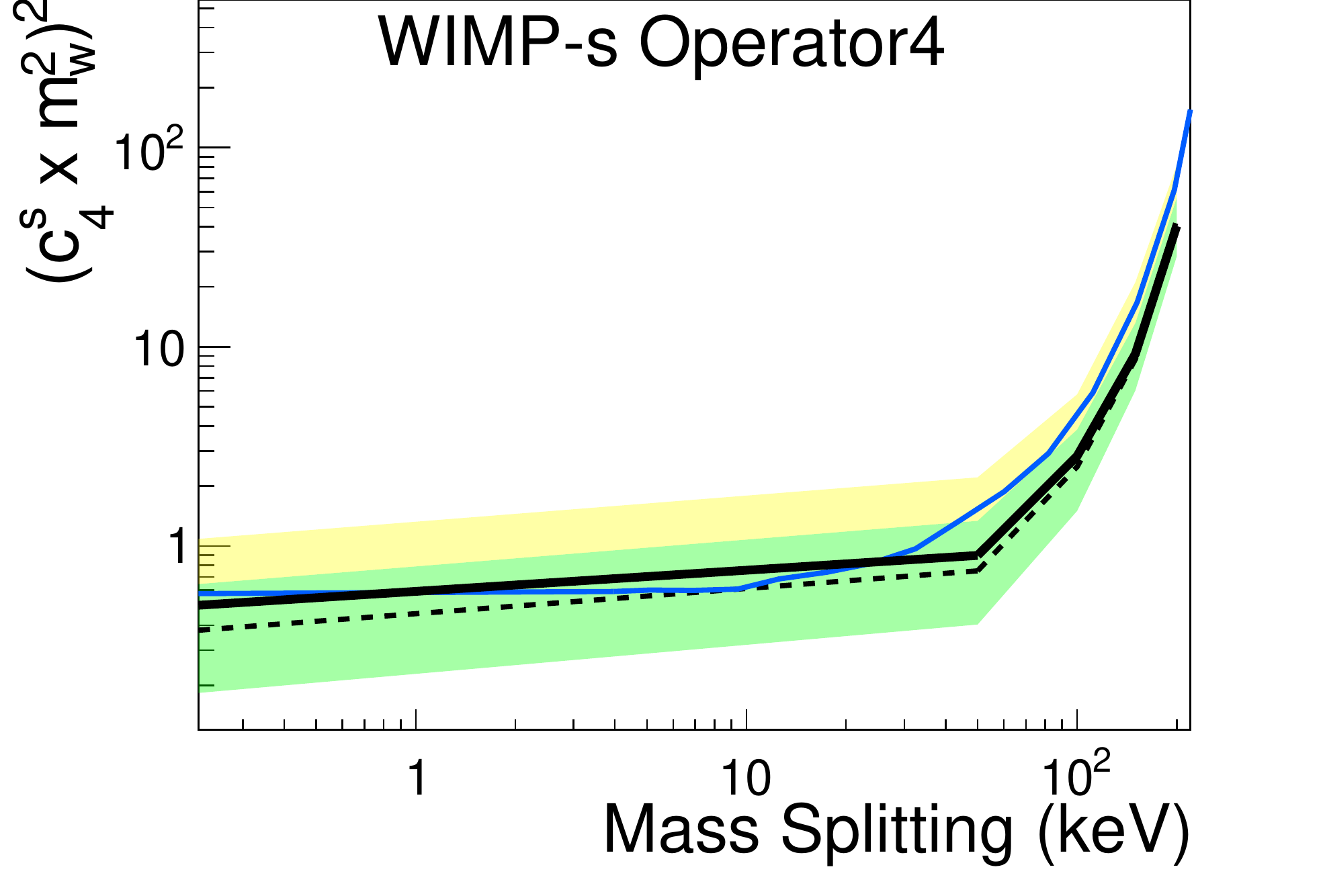}
    \includegraphics[width=0.5\columnwidth]{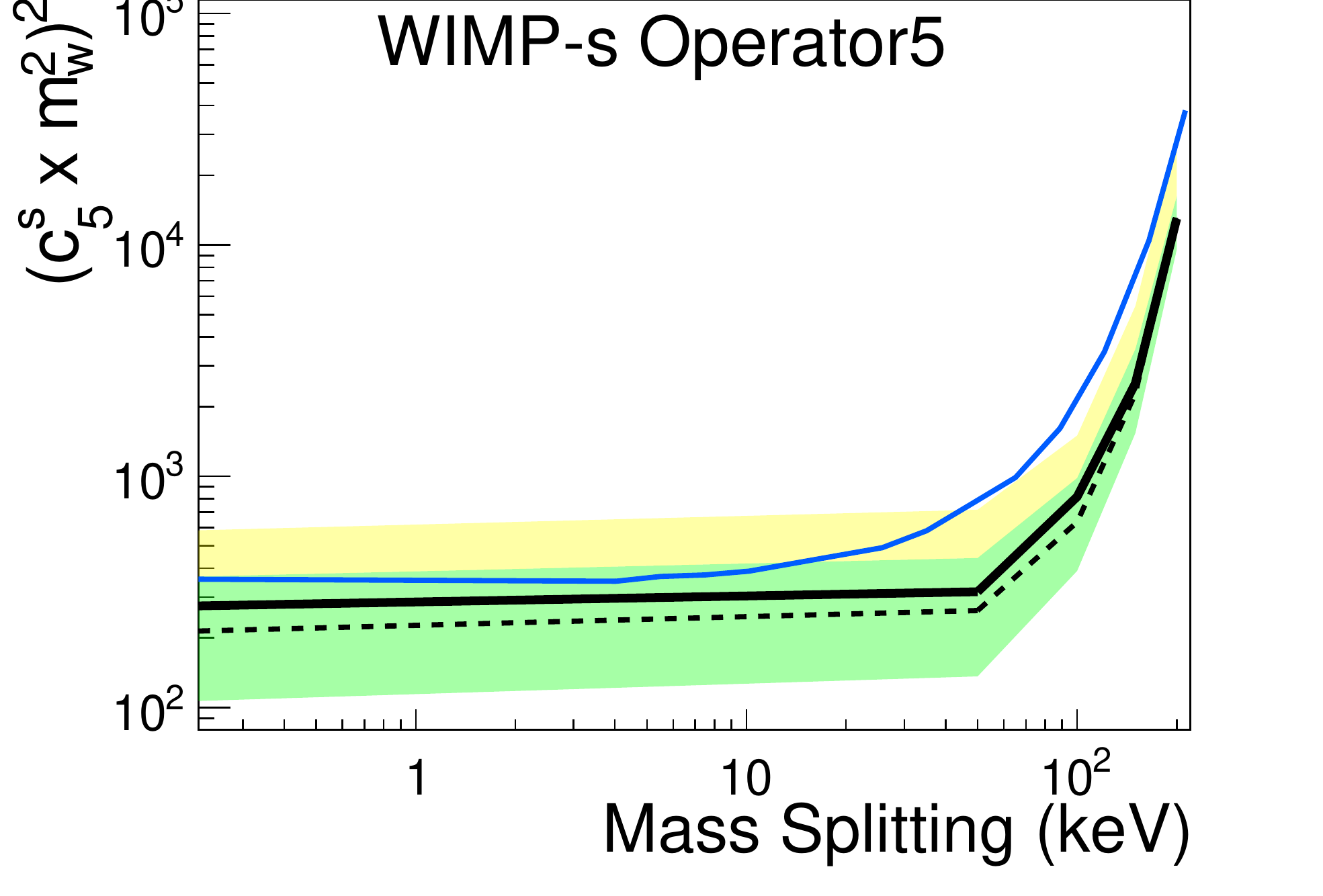}
    \includegraphics[width=0.5\columnwidth]{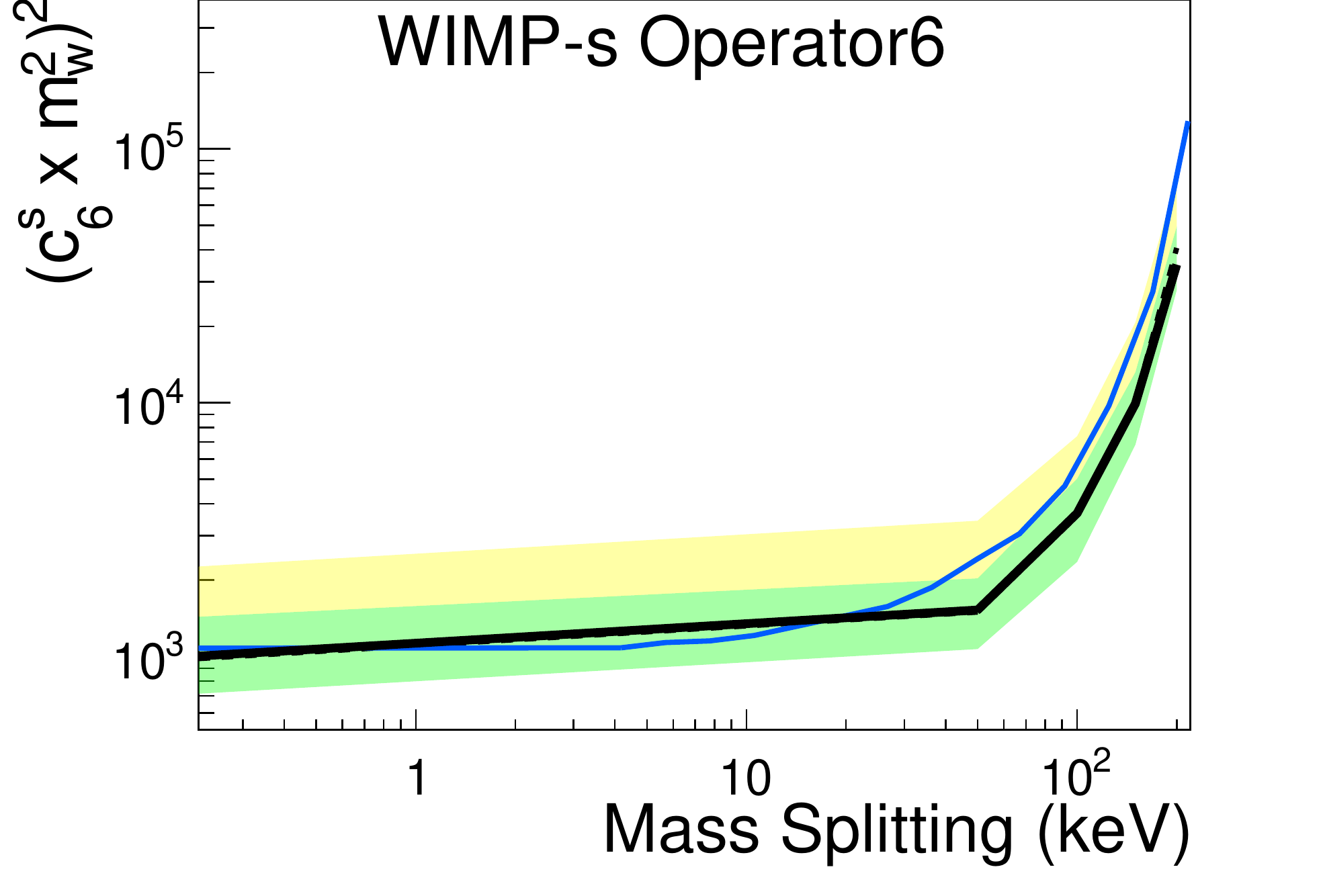}
    \includegraphics[width=0.5\columnwidth]{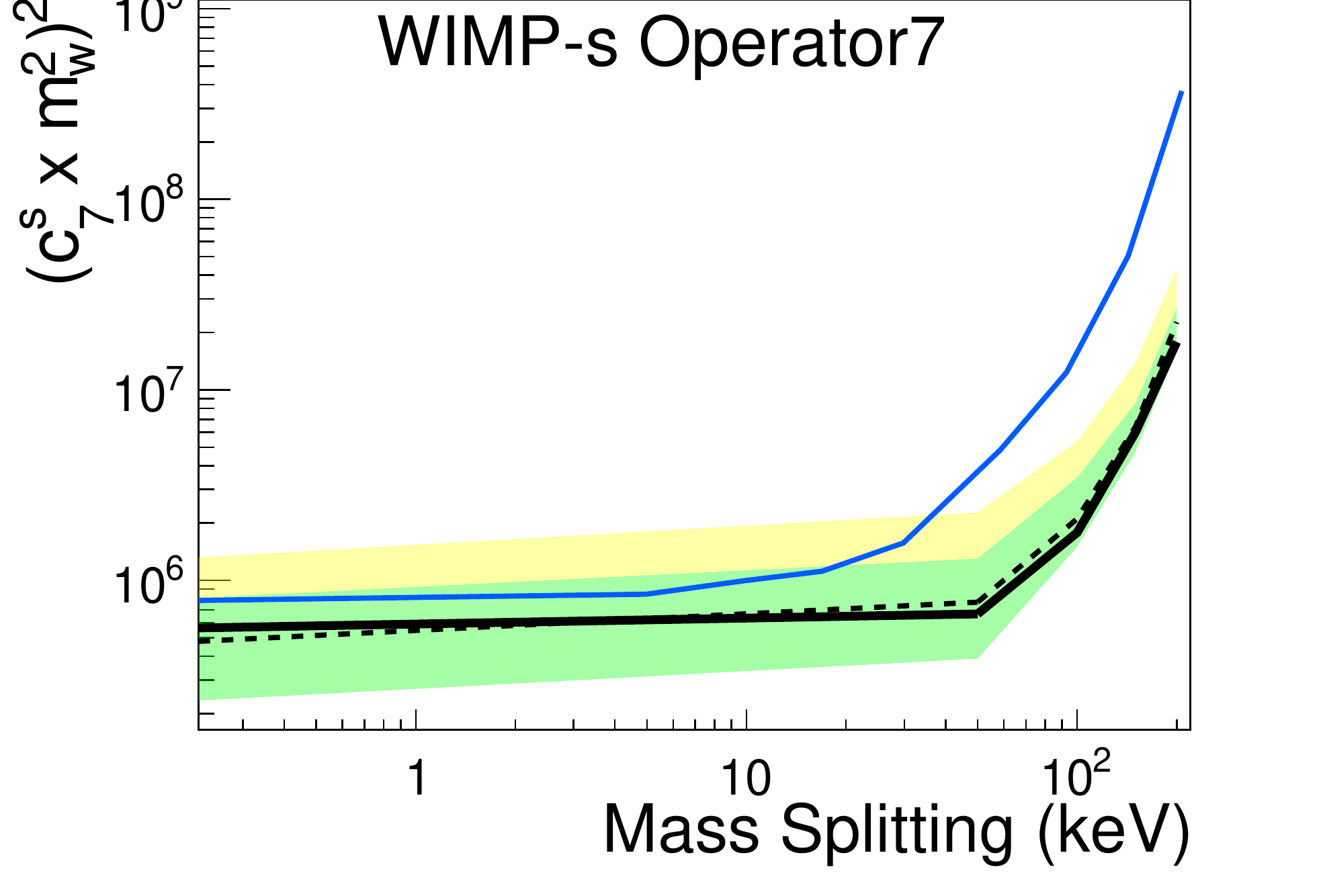}
    \includegraphics[width=0.5\columnwidth]{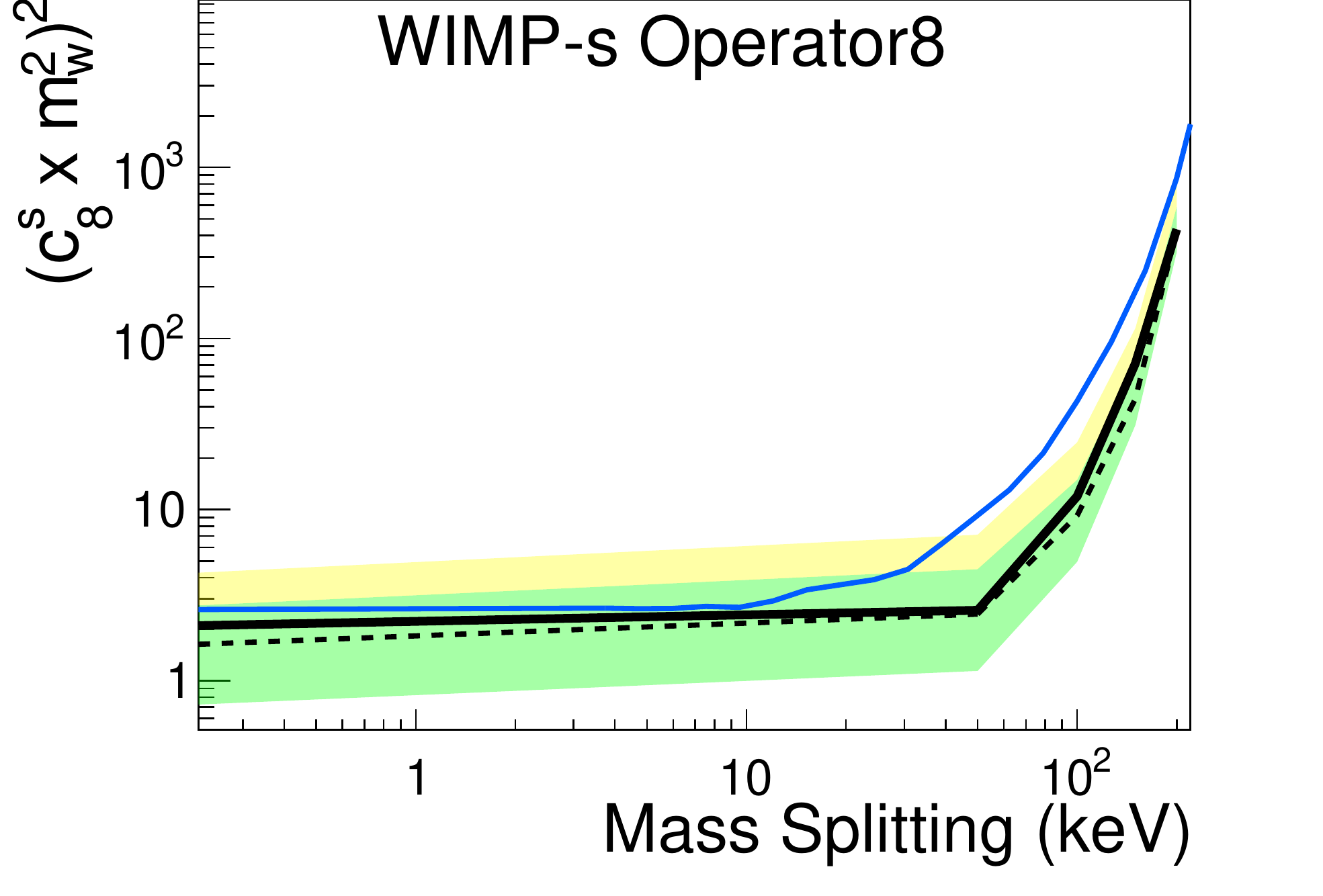}
    \includegraphics[width=0.5\columnwidth]{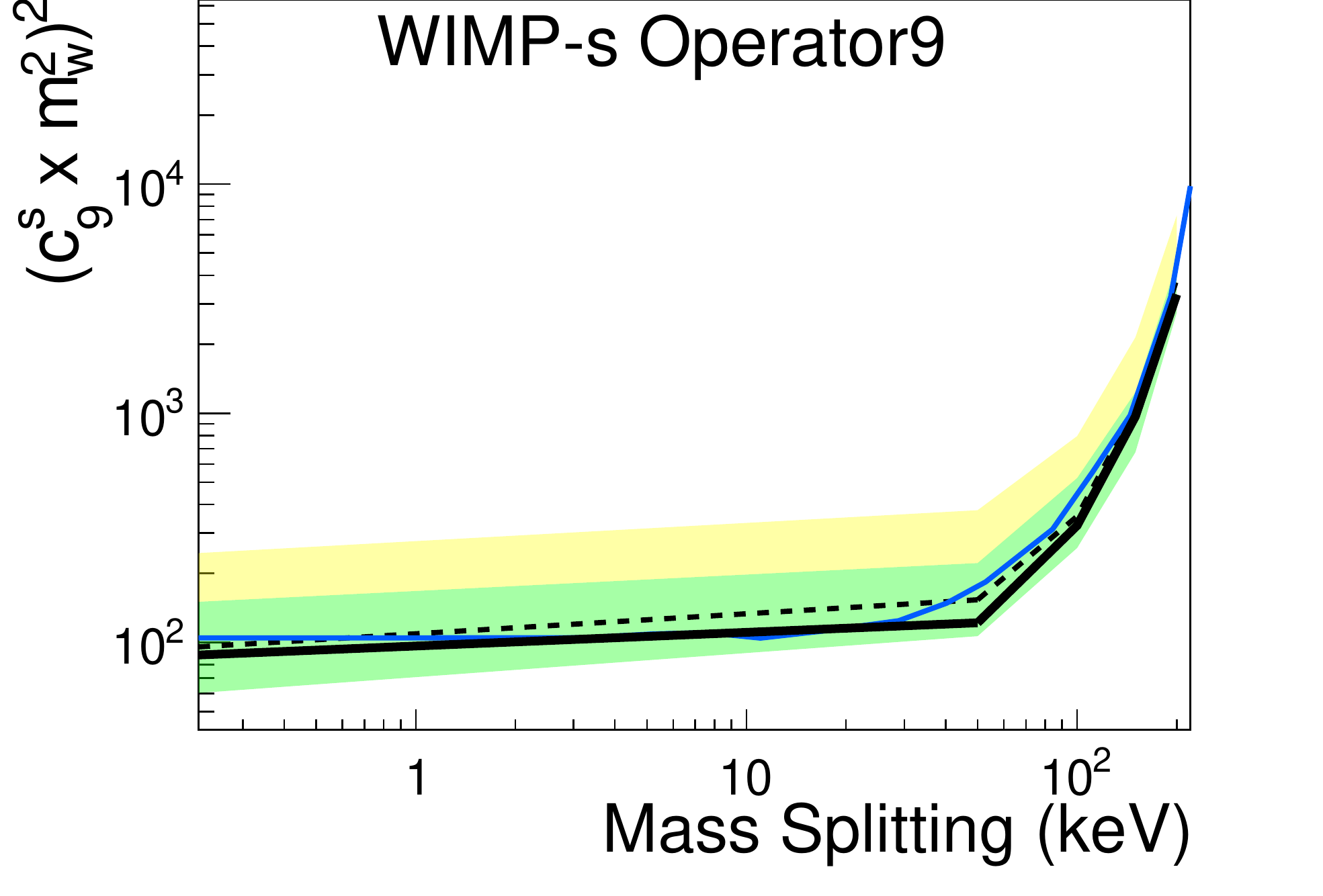}
    \includegraphics[width=0.5\columnwidth]{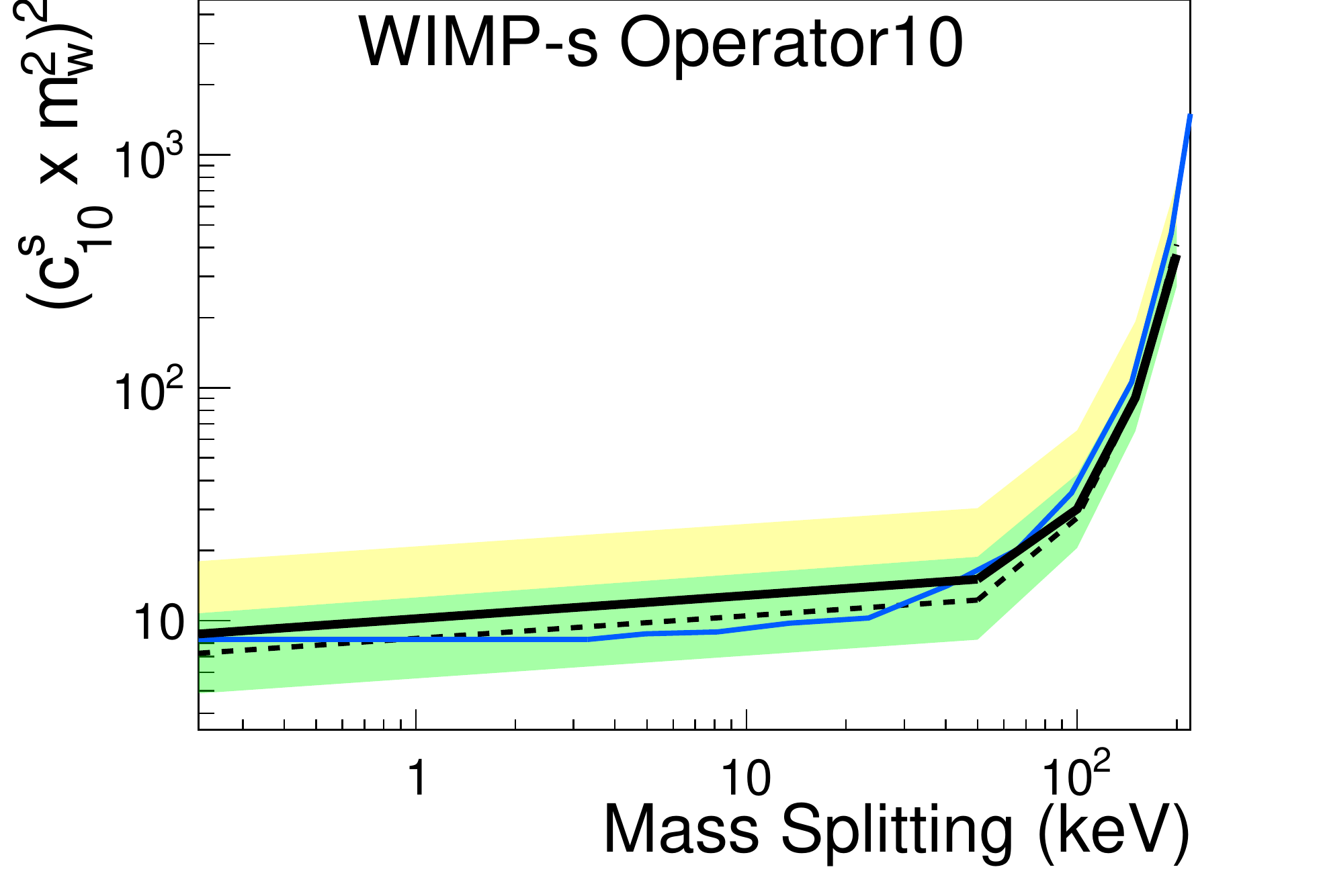}
    \includegraphics[width=0.5\columnwidth]{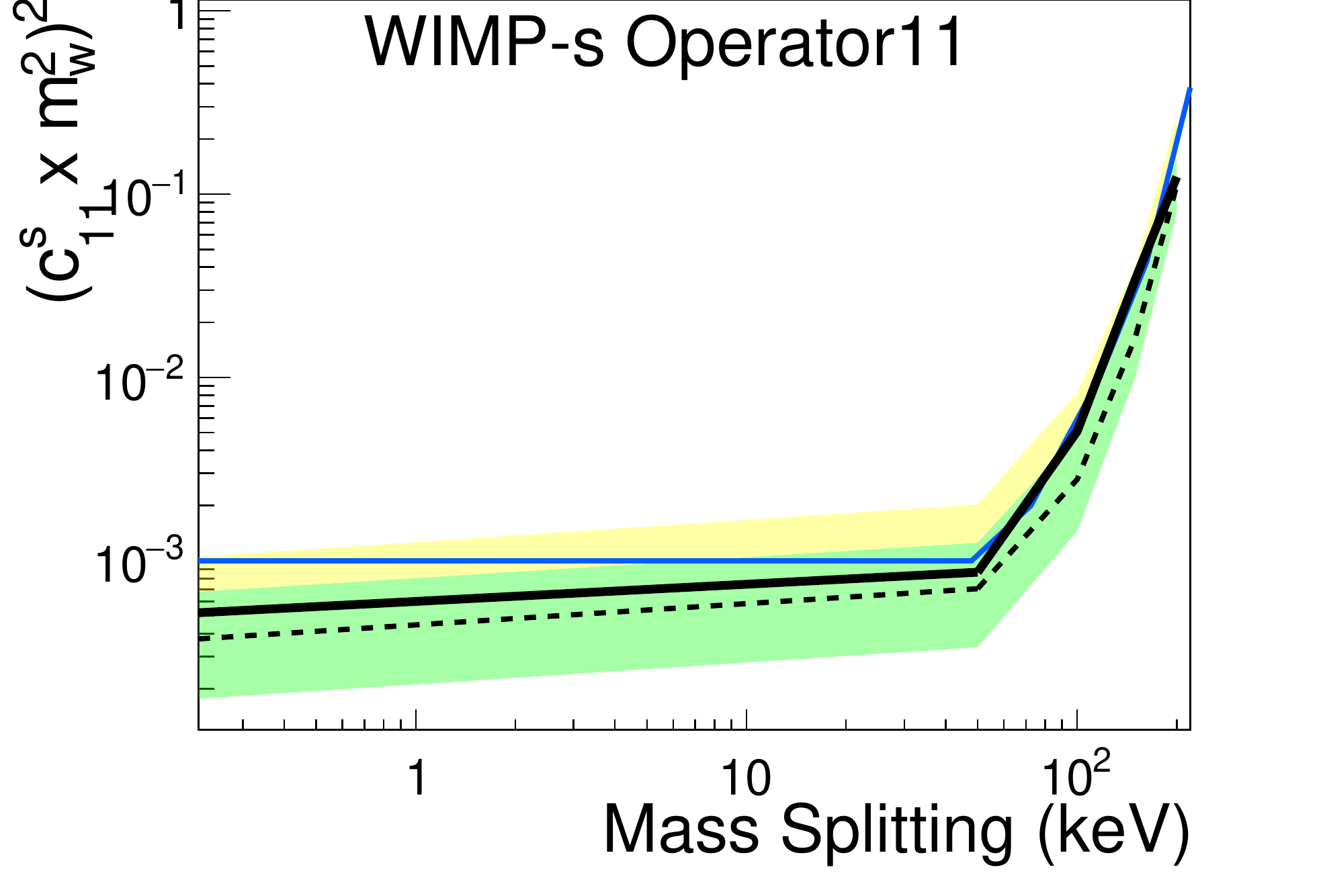}
    \includegraphics[width=0.5\columnwidth]{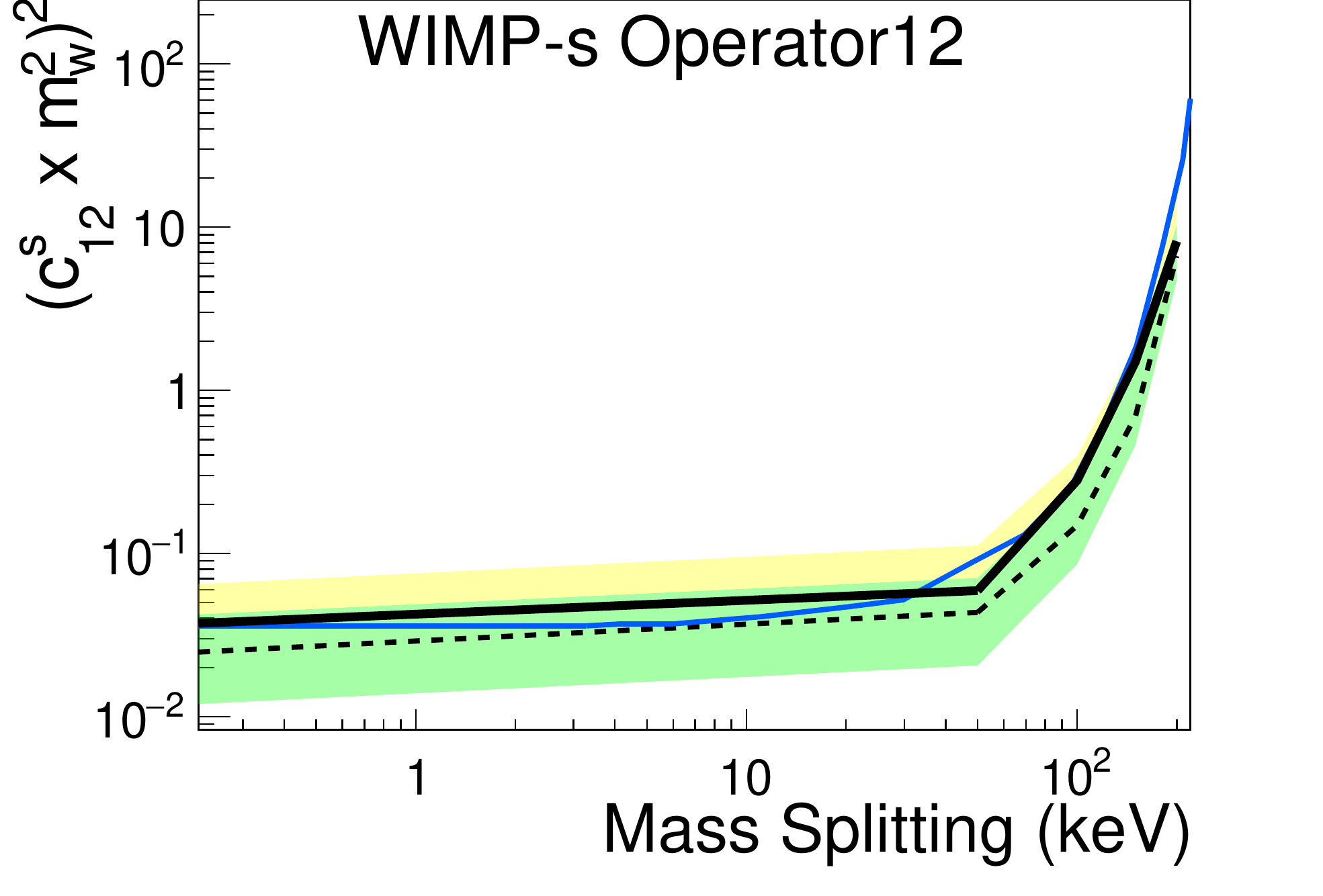}
    \includegraphics[width=0.5\columnwidth]{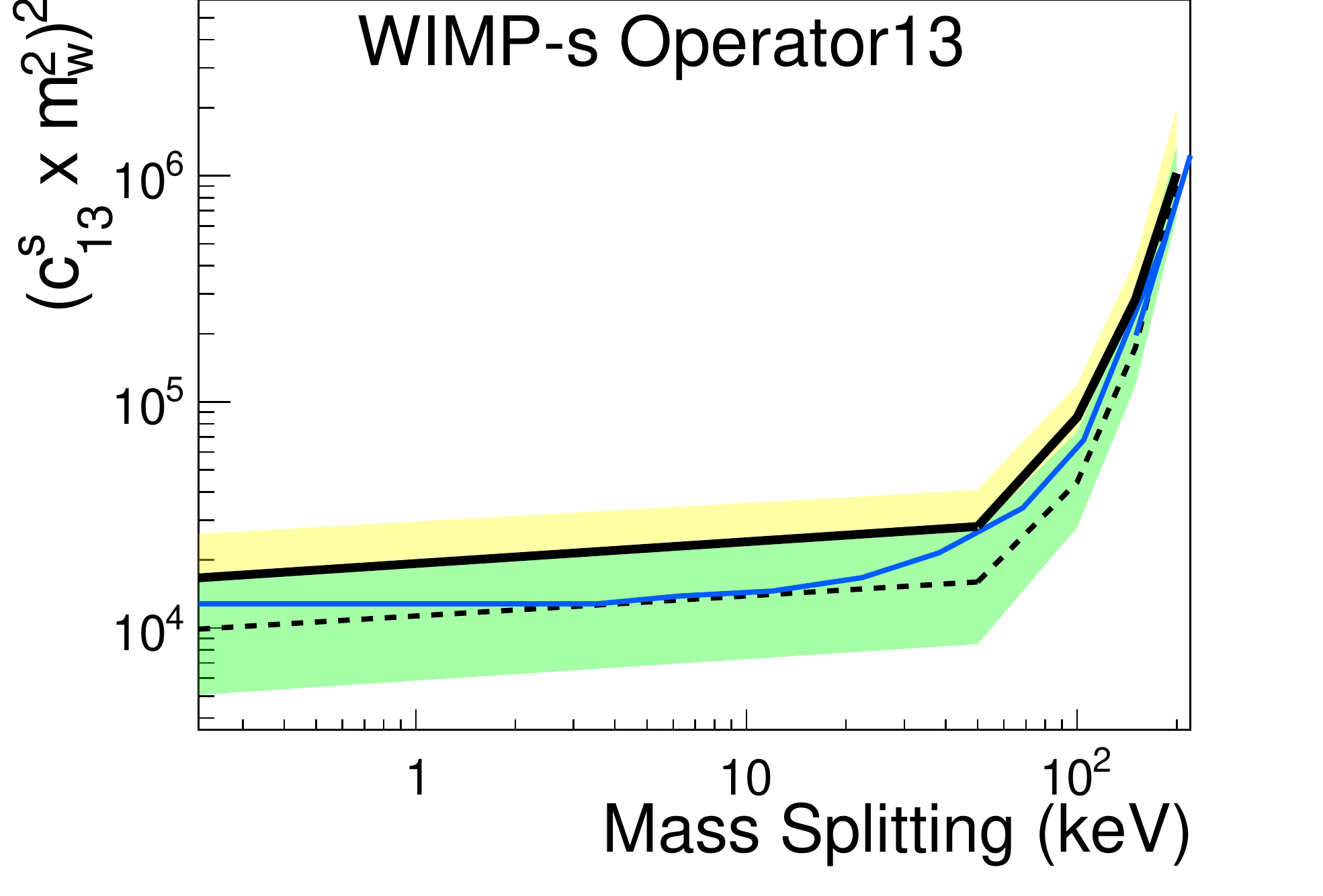}
    \includegraphics[width=0.5\columnwidth]{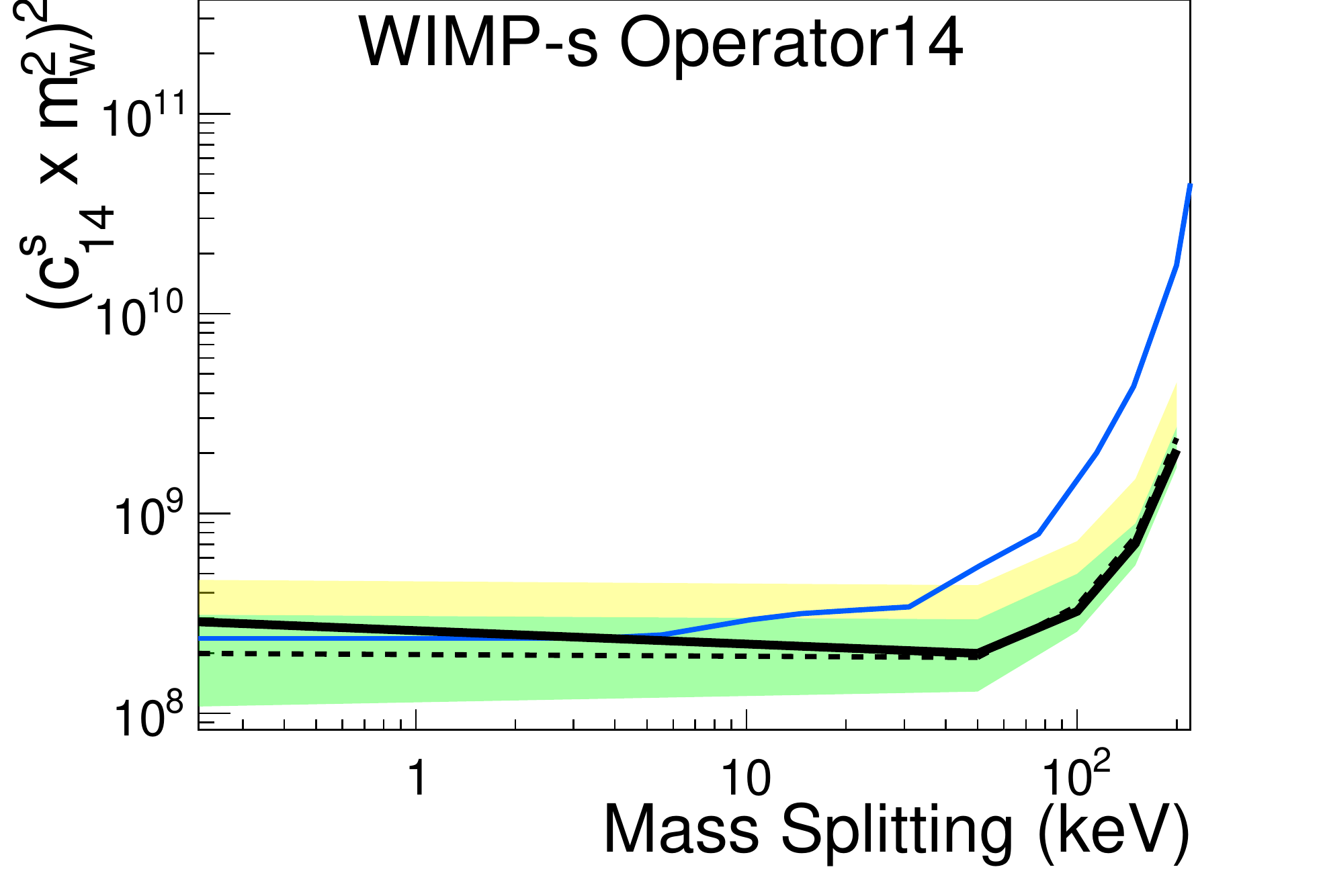}
    \includegraphics[width=0.5\columnwidth]{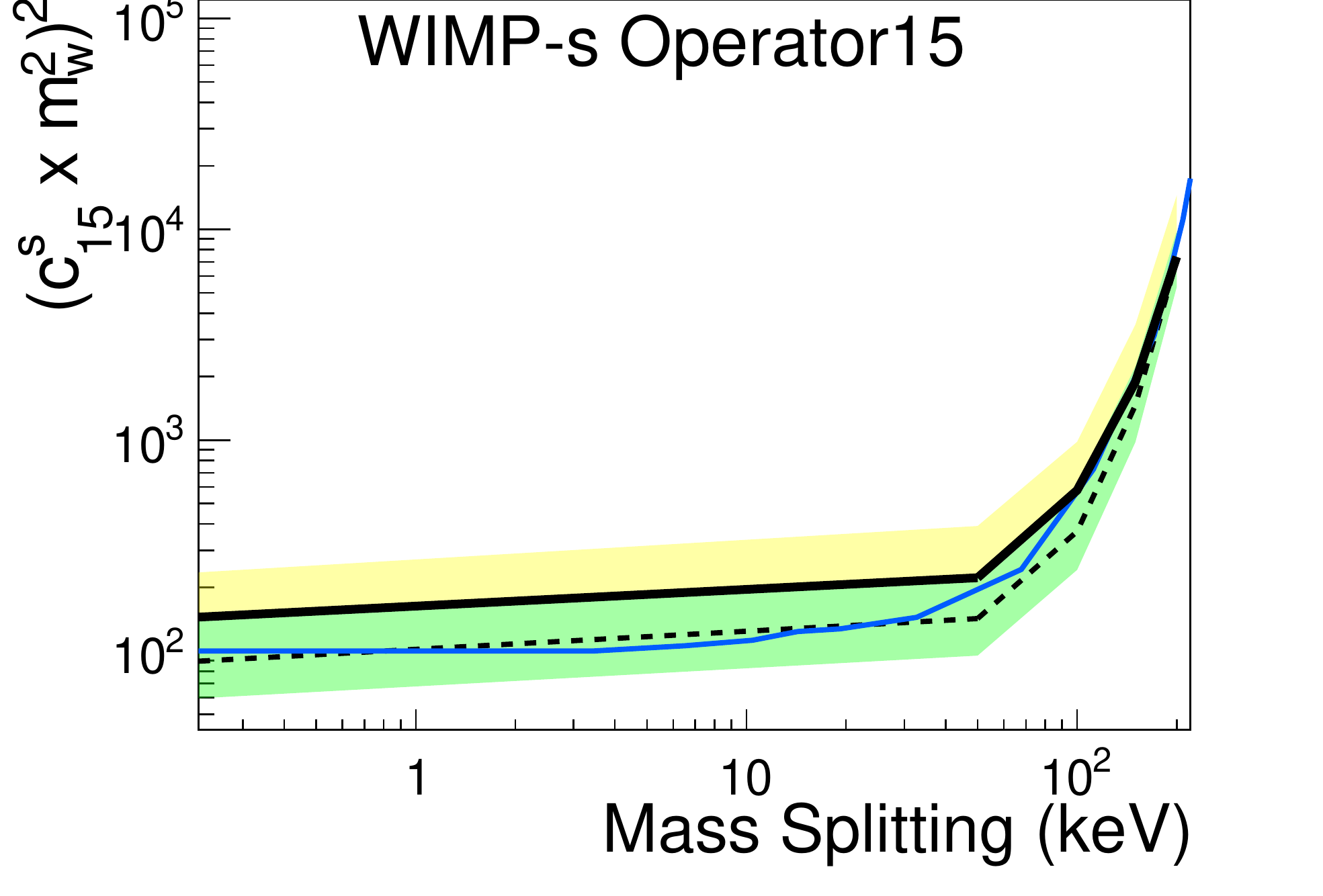}
  \caption{The LUX WS2014--16 90\% C.L. limits for isoscalar WIMP-nucleon dimensionless couplings for each of the fourteen nonrelativistic EFT operators and a fixed WIMP mass of 1 TeV. Solid black lines show the limit, while dashed black indicate the expectation, with green and yellow bands indicating the $\pm1\sigma$ and $+2\sigma$ sensitivity expectations, respectively. Each plot uses  $\delta_m$  values of 0, 50, 100, 150, and 200~keV. Blue lines show limits from XENON100~\cite{xenon100}. }
  \label{idm_limits}
\end{figure*}

Figure~\ref{dataPlots} shows the final WS2014--16 data used in this analysis, with the events used in the PLR framework highlighting the behavior of the different background models. The max ROI is the region of \{$S1_c$, $\log_{10}(S2_c)$\} space that includes at least 90\% of the expected differential rate from each signal model. Our dataset shows consistency with our background models, resulting in p-values between 0.14 and 0.50 for the 28 elastic operator/nucleon combinations at 50~GeV mass, with a median p-value of 0.28. Additionally, Kolmogorov-Smirnov tests for each of the five observables - $S1_c$, $S2_c$, $r$, $d$, and $\phi$ - compared to the background model PDFs return p-values: 0.39, 0.24, 0.60, 0.43, and 0.81, respectively. The initial constraints and final PLR fits for each nuisance parameter are shown in Table~\ref{nuisParams}, where fit values are for the background-only scenario ($\mu=0$). 

We set a 2-sided frequentist confidence interval on the value of $c_i^{(N)^2}$ using the method discussed in Sec.~\ref{sec:stats} at a 90\% confidence level ($\alpha = 0.1$). We do this for all operators, selecting values for the WIMP mass ranging from 10~GeV to 4~TeV.  Upper limits are shown in Fig.~\ref{neutron_limits} and Fig.~\ref{proton_limits} for elastic WIMP-neutron and WIMP-proton interactions, respectively.  
We explicitly note here that the $c_i^{(N)}$ have dimensionality of [mass]${}^{-2}$ as the conventions of Ref.~\cite{Anand:MathematicaEFT} use a dimensionless operator representation and normalize spinors to unity, which differs from the representation used in Ref.~\cite{Fitzpatrick:EFT}. 
Consequently, results are scaled by a factor of $m_v^2$ in order to report dimensionless values similar to the results reported in Ref.~\cite{xenon100} for convenience. Figures~\ref{neutron_limits} and~\ref{proton_limits} additionally show the available comparisons with the upper limits from the 1.4$\times 10^4$~kg$\cdot$day exposure results using LUX WS2013 data~\cite{run3eft}.

Limits for most operators remain within approximately 1$\sigma$ of our expectation, with the most significant discrepancies for $\mathcal{O}_{3}$, $\mathcal{O}_{13}$, and $\mathcal{O}_{15}$, which differ from the expected sensitivity by as much as 1.5$\sigma$ (100 GeV $\mathcal{O}_{15}$ WIMP-n). Returning to Fig.~\ref{signalSpectra}, we note that these three operators are characteristically similar: they have maximum recoil rates at nonzero energies; their differential rates are relatively flat through most of the ROI; and they exhibit a pronounced secondary peak at higher energies. Comparing Fig.~\ref{signalCompare} to the background data (Fig.~\ref{dataPlots}), it is understandable that $\mathcal{O}_{15}$ and similar operator models are the most difficult to discern from an ER background with ${}^{83\text{m}}$Kr contamination as these signal models resemble the distribution of natural ER leakage in our ROI. 

However, despite the agreement shown between our models and the background data described above, the discrepancy between some observed limits and the expected sensitivity suggests there remain slight inconsistencies between the models and the observed data. Table~2 of Ref.~\cite{GR_ERmodel} suggests that ER band widths may be slightly underestimated by our models for the two most central drift times bins, especially when compared to the ${}^{14}$C calibration data. While those reported discrepancies are small ($\sim5\%$), it is possible they are manifesting here to produce weaker observed limits than expected, most noticeably for the signal models that most resemble ER leakage. Additionally, even though models for $\gamma-X$, accidental coincidences, and ${}^{83\text{m}}$Kr have been included, uncertainties in their expected rates leads to allowed signal events from several EFT WIMP models, resulting in weaker expected limits for several combinations of operator and mass when compared to the background-only expectation. Appendix~\ref{excess_tables} contains tables detailing the deviation between our expected and observed limits for each signal model considered in this analysis. To test the effects of the underestimated ER widths, we increased the widths of the standard ER model for the two central-most drift time bins in accordance to the discrepancies with the ${}^{14}$C calibration data reported in Table~2 of Ref.~\cite{GR_ERmodel}. A single test using the 100 GeV $\mathcal{O}_{15}$ WIMP-n case was processed. The resultant excess between our expectation and observed limit for this operator/mass combination was reduced from 1.5$\sigma$ to 0.6$\sigma$, indicating that the $\mathcal{O}(1\%)$ underestimations of our ER band width are indeed largely responsible for the slightly poorer observed limits compared to our expected sensitivity.

\begin{table}
\caption{The nuisance parameters used in the PLR framework, along with their initial constraints and fit values.}
\begin{tabular}{|c|c|c|c|}
\hline
Parameter & Constraint & Fit Value \\
\hline
Standard ER  & 1510.2$\pm$187.5 & 1503.1$\pm$51.1\\
Wall-based Backgrounds & 11.3$\pm$2.8 & 10.1$\pm$2.2 \\
$\gamma-X$ & 3.4$\pm$2.5 & 5.2$\pm$2.0 \\
${}^{83\text{m}}$Kr & 5.2$\pm$1.5 & 3.5$\pm$2.0  \\
Accidental Coincidence & 0.75$^{+0.79}_{-0.75}$ & 1.08$\pm$0.63\\
\hline 
\end{tabular}
\label{nuisParams}
\end{table}

Despite the resultant limits being poorer than our expectation, we show major improvements on the previously reported LUX \{$n$,$p$\} limits from Ref.~\cite{run3eft} in the comparisons in Figs.~\ref{neutron_limits} and~\ref{proton_limits}. The improved sensitivity of these results is greater than that expected solely from the increased exposure, highlighting the benefit from reassessing data selection criteria and background models for the extended ROI and the use of five-dimensional PLR framework. Recent competitive analyses report their results using the \{isoscalar,isovector\} basis, such as CDMS, XENON100, DEAP-3600, and PandaX-II~\cite{cdms_EFT, xenon100,deap3600,PandaX_EFT}, which prohibits direct comparison. However, we note that for xenon targets, the expected event rates for WIMP-n interactions are typically larger than that for WIMP-p interactions, but the isoscalar formulation splits the differences between these. While this does not take into account the differences in signal shape, it allows for qualitative comparisons between LUX results and the reported isoscalar limits. Our observed WIMP-p limits are competitive (and sometimes more sensitive) than the isoscalar limits reported in Refs.~\cite{xenon100,deap3600}, suggesting new exclusion of EFT WIMP parameter space, regardless of the chosen basis.

We also report the WS2014--16 limits for inelastic scattering using the isoscalar basis.  Figure~\ref{idm_limits} shows the inelastic EFT WIMP-nucleon isoscalar limits as a function of $\delta_m$ for a fixed WIMP mass of 1~TeV compared to the previous limits set by XENON100~\cite{xenon100}.  At this mass, we show similar limits to XENON100 despite using a larger exposure. This is due to the regions of phase space where our background models and signal models overlap, effectively increasing upper limits on the number of WIMP scatters possible in our dataset and thus reducing the impact of the larger exposure for both observed and expected limits. Data in these overlapped regions of phase space include low-energy accidental-like events, a handful of $\gamma-X$-like events near the bottom of the fiducial volume, and ${}^{83\text{m}}$Kr events from the bottom-most drift time bin. Similarly to the elastic results, some observed limits deviate from the expected sensitivities, as described in the preceding paragraphs and suspected to be largely due to underestimated ER band widths.

Despite this, our 1~TeV inelastic limits are often competitive with XENON100, and for several combinations of $\delta_m$ and $\mathcal{O}_i$, we present improved exclusion of the possible WIMP-nucleon interactions. Additionally, some operators show world-leading exclusion limits in the elastic limit ($\delta_m=0$~keV), allowing for qualitative comparisons between our ${n,p}$ elastic limits and those reported in the isoscalar basis by other experiments. 

\section{Summary}

We have expanded and improved the LUX background models to allow for characterization of data at energies much higher than a traditional WIMP search. These backgrounds include novel characterization of multiply scattering $\gamma-X$ events disguised as single scatters, as well as the inclusion of ${}^{83\text{m}}$Kr decays in our background model. Utilization of the Noble Element Simulation Technique allowed for efficient modeling of the ER and NR LXe response, independently for each of the 16 time and drift time bins of WS2014--16 data. Additionally, NEST allows us to extrapolate the NR LXe response to higher energies than measured with \textit{in situ} calibrations, after accounting for the uncertainties in all of the light and charge yield measurements combined from beyond LUX.

We set exclusion limits for the 28 combinations of EFT operator and atomic nucleon in the \{$n$, $p$\} basis, following the precedent of previous LUX results.  While we consider this basis to be more physically intuitive as it is similar to standard spin-dependent WIMP searches, it does not allow for direct comparison with recent EFT WIMP exclusion curves in the \{isoscalar,isovector\} basis. 

We also report the results of inelastic WIMP-nucleon scattering with respect to isoscalar nucleons at 1 TeV and compare to those reported by the XENON100 Collaboration~\cite{xenon100}, excluding new parameter space for several EFT operators.

\begin{acknowledgments}

This work was partially supported by the U.S. Department of Energy (DOE) under Awards No. DE-AC02-05CH11231, No. DE-AC05-06OR23100, No. DE-AC52-07NA27344, No. DE-FG01-91ER40618, No. DE-FG02-08ER41549, No. DE-FG02-11ER41738, No. DE-FG02-91ER40674, No. DE-FG02-91ER40688, No. DE-FG02-95ER40917, No. DE-NA0000979, No. DE-SC0006605, No. DE-SC0010010, No. DE-SC0015535, and No. DE-SC0019066; the U.S. National Science Foundation under Grants No. PHY-0750671, No. PHY-0801536, No. PHY-1003660, No. PHY-1004661, No. PHY-1102470, No. PHY-1312561, No. PHY-1347449, No. PHY-1505868, and No. PHY-1636738; the Research Corporation Grant No. RA0350; the Center for Ultra-low Background Experiments in the Dakotas (CUBED); and the South Dakota School of Mines and Technology (SDSMT).

Laborat\'{o}rio de Instrumenta\c{c}\~{a}o e F\'{i}sica Experimental de Part\'{i}culas (LIP)-Coimbra acknowledges funding from Funda\c{c}\~{a}o para a Ci\^{e}ncia e a Tecnologia (FCT) through the Project-Grant No. PTDC/FIS-NUC/1525/2014. Imperial College and Brown University thank the UK Royal Society for travel funds under the International Exchange Scheme (IE120804). The UK groups acknowledge institutional support from Imperial College London, University College London, the University of Sheffield, and Edinburgh University, and from the Science \& Technology Facilities Council for PhD studentships R504737 (EL), M126369B (NM), P006795 (AN), T93036D (RT), and N50449X (UU). This work was partially enabled by the University College London (UCL) Cosmoparticle Initiative. The University of Edinburgh is a charitable body, registered in Scotland, with Registration No. SC005336.

This research was conducted using computational resources and services at the Center for Computation and Visualization, Brown University, and also the Yale Science Research Software Core.

We gratefully acknowledge the logistical and technical support and the access to laboratory infrastructure provided to us by SURF and its personnel at Lead, South Dakota. SURF was developed by the South Dakota Science and Technology Authority, with an important philanthropic donation from T. Denny Sanford. SURF is a federally sponsored research facility under Award Number DE-SC0020216.
\end{acknowledgments}

\appendix
\section{ILLUSTRATION OF DATA SEPARATED BY DATE AND DRIFT TIME BINS}\label{binSeparation}

\begin{figure*}[h!]
\begin{center}
\includegraphics[width=\textwidth,clip]{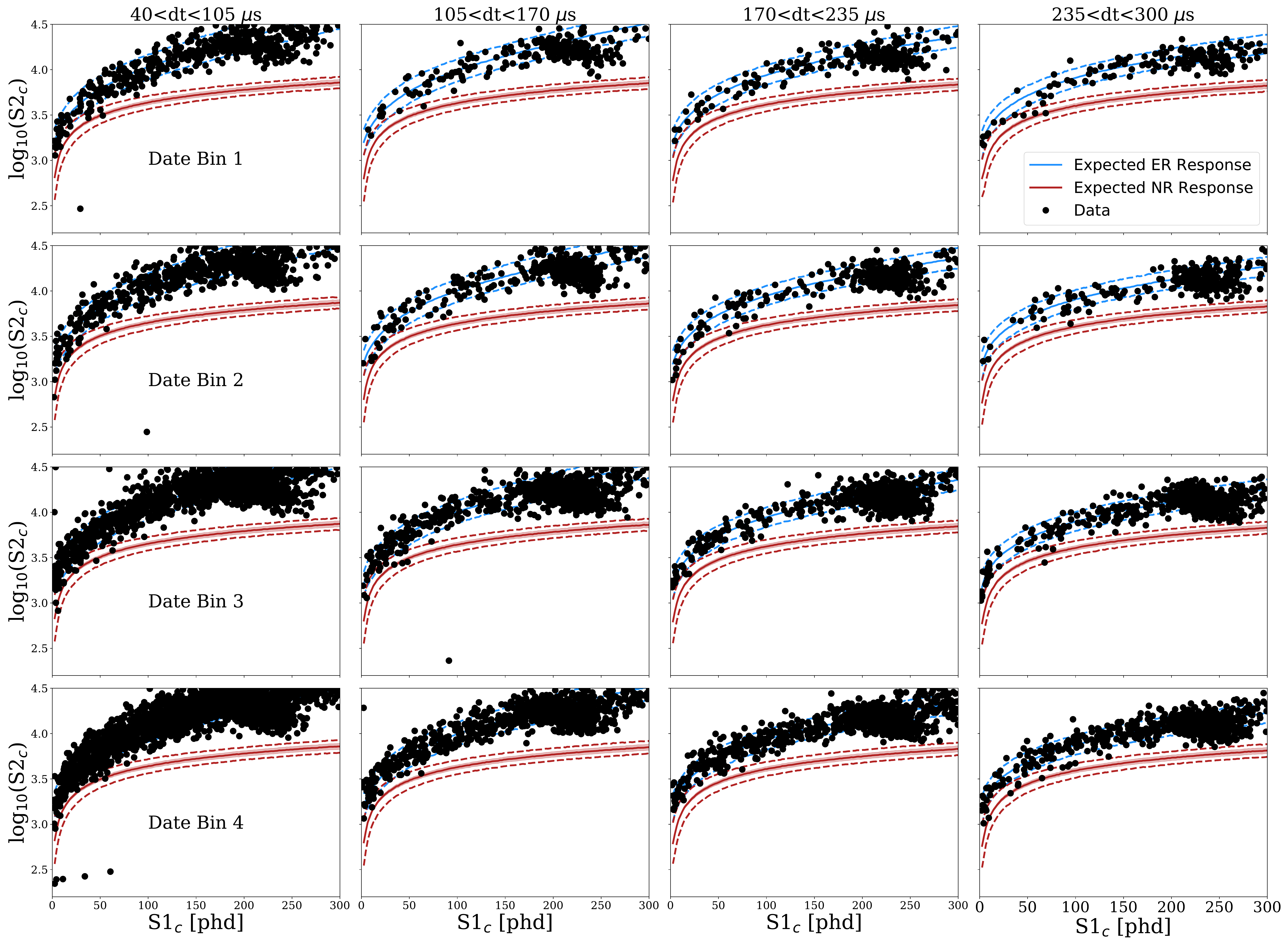}
\vspace{-10pt}
\caption{The WS2014--16 data divided into the 16 drift time bins: four temporal bins of unequal livetime (rows) subdivided further into four spatial bins (columns). Blue bands represent the mean and 90\% C.L. ER response, while red illustrates the mean and 90\% C.L. NR response. 20.8 live-days of data corresponding to significant ${}^{83\text{m}}$Kr contamination have been removed from this dataset, however, remaining ${}^{83\text{m}}$Kr events can be seen encroaching upon the signal region at higher energies. }
\vspace{-22pt}
\label{eachBin}
\end{center}
\end{figure*}

We present in this appendix the full WS2014--16 data used in this analysis, separated into the 16 drift time bins: four temporal bins, each subdivided to correspond to a 65 $\mu$s window of drift time. The livetimes for each temporal bin are: 43.9, 43.8, 85.8, and 137.7 days, respectively. Figure~\ref{eachBin} illustrates the data compared to the relevant ER and NR simulated responses; bands represent 90\% C.L. about the mean response. Despite the exclusion of exposures associated with the ${}^{83\text{m}}$Kr calibration injections, many of these events can be seen in each drift time bin. 

\section{TABLES OF EXCESSES FOR EACH ANALYZED MODEL}

\label{excess_tables}

As described in Sec.~\ref{results}, we show discrepancies between our observed exclusion limits and the expected limits due to uncertainties in the background models that lead to allowable signal events to be fitted to the data, as well as the possibility of an underestimation of the ER leakage from our background models as suggested in Table 2 of Ref.~\cite{GR_ERmodel}. We present below tables highlighting the deviations between observed and expected limits for the WIMP-n and WIMP-p elastic results (Tables~\ref{wimp-n_table} and~\ref{wimp-p_table}, respectively), as well as the WIMP-isoscalar inelastic results (Table~\ref{inelastic_table}).  

\begin{table*}[h!]
\caption{Discrepancies between observed and expected exclusion limit for each model used for the elastic WIMP-n results, expressed in terms of number of $\sigma$ from the expectation. Positive values indicate weaker sensitivity than the expected limit. Dashes are used for operator/mass combinations that were not analyzed in this report. }
\begin{tabular}{|c|c|c|c|c|c|c|c|c|c|c|c|c|}
\hline
\textbf{WIMP-n} & 10 GeV & 12 GeV & 14 GeV & 17 GeV & 21 GeV & 33 GeV & 50 GeV & 100 GeV & 200 GeV & 400 GeV & 1000 GeV & 4000 GeV \\ \hline \hline
$\mathcal{O}_1$ & 0.52 & 0.99 & 1.07 & 1.06 & 0.99 & 1.02 & 0.50 & 0.35 & 0.23 & 0.58 & 0.40 & 0.46 \\ \hline
$\mathcal{O}_3$ & 0.99 & 1.16 & 0.98 & 0.71 & 0.39 & -0.41 & 0.25 & 1.00 & 1.27 & 0.95 & 1.01 & 1.05 \\ \hline
$\mathcal{O}_4$ & 0.46 & 0.88 & 1.06 & 1.00 & 0.87 & 0.49 & 0.44 & 0.33 & 0.51 & 0.34 & 0.35 & 0.40 \\ \hline
$\mathcal{O}_5$ & 0.73 & 1.08 & 1.12 & 0.97 & 0.67 & 0.09 & 0.18 & 0.26 & 0.29 & 0.30 & 0.09 & 0.28 \\ \hline
$\mathcal{O}_6$ & 1.09 & 1.06 & 1.10 & 0.58 & 0.22 & -0.29 & -0.05 & 0.62 & 0.25 & 0.03 & -0.01 & 0.09 \\ \hline
$\mathcal{O}_7$ & 0.23 & 0.69 & 1.03 & 0.94 & 1.06 & 0.64 & 0.50 & 0.40 & 0.39 & 0.30 & 0.21 & 0.31 \\ \hline
$\mathcal{O}_8$ & 0.34 & 0.79 & 1.04 & 0.88 & 0.83 & 0.50 & 0.34 & 0.35 & 0.40 & 0.35 & 0.34 & 0.27 \\ \hline
$\mathcal{O}_9$ & 0.81 & 1.08 & 1.04 & 0.84 & 0.80 & 0.27 & 0.05 & -0.12 & -0.10 & -0.28 & -0.49 & -0.35 \\ \hline
$\mathcal{O}_{10}$ & 0.72 & 1.15 & 1.12 & 0.81 & 58 & -0.08 & 0.28 & 0.49 & 0.23 & 0.32 & 0.37 & 0.18 \\ \hline
$\mathcal{O}_{11}$ & 0.83 & 1.12 & 1.04 & 0.92 & 0.62 & 0.08 & -0.04 & 0.32 & 0.46 & 0.49 & 0.46 & 0.40 \\ \hline
$\mathcal{O}_{12}$ & - & 0.81 & 1.06 & 0.99 & 0.64 & 0.08 & 0.08 & 0.46 & 0.79 & 0.78 & 0.75 & 0.80 \\ \hline
$\mathcal{O}_{13}$ & 1.07 & 1.16 & 1.06 & 0.70 & 0.27 & -0.45 & 0.61 & 1.11 & 1.07 & 0.90 & 0.87 & 0.84 \\ \hline
$\mathcal{O}_{14}$ & - & - & - & - & 0.88 & 0.26 & 0.04 & -0.03 & -0.11 & -0.29 & -0.60 & -0.37 \\ \hline
$\mathcal{O}_{15}$ & 1.26 & 1.28 & 0.99 & 0.47 & -0.12 & -0.49 & 0.98 & 1.54 & 1.04 & 0.93 & 1.14 & 0.89 \\
\hline
\end{tabular}
\label{wimp-n_table}
\end{table*}

\begin{table*}[h!]
\caption{Discrepancies between observed and expected exclusion limit for each model used for the elastic WIMP-p results, expressed in terms of number of $\sigma$ from the expectation. Positive values indicate weaker sensitivity than the expected limit. Dashes are used for operator/mass combinations that were not analyzed in this report.}
\begin{tabular}{|c|c|c|c|c|c|c|c|c|c|c|c|c|}
\hline
\textbf{WIMP-p} & 10 GeV & 12 GeV & 14 GeV & 17 GeV & 21 GeV & 33 GeV & 50 GeV & 100 GeV & 200 GeV & 400 GeV & 1000 GeV & 4000 GeV \\ \hline \hline
$\mathcal{O}_1$ & 0.37 & 1.14 & 1.01 & 1.02 & 0.77 & 0.55 & 0.31 & 0.33 & 0.37 & 0.40 & 0.41 & 0.34 \\ \hline
$\mathcal{O}_3$ & 1.41 & 1.14 & 1.21 & 0.75 & 0.28 & -0.39 & 0.64 & 1.01 & 0.82 & 0.86 & 0.81 & 0.73 \\ \hline
$\mathcal{O}_4$ & 0.54 & 0.93 & 1.05 & 1.09 & 0.96 & 0.52 & 0.47 & 0.41 & 0.57 & 0.24 & 0.10 & 0.20 \\ \hline
$\mathcal{O}_5$ & 0.55 & 0.99 & 1.00 & 0.96 & 0.77 & 0.07 & -0.01 & 0.43 & 0.62 & 0.55 & 0.50 & 0.57 \\ \hline
$\mathcal{O}_6$ & 1.00 & 1.14 & 1.03 & 0.83 & 0.35 & -0.29 & -0.07 & 0.28 & 0.19 & 0.01 & 0.01 & 0.99 \\ \hline
$\mathcal{O}_7$ & 0.36 & 0.83 & 1.19 & 0.99 & 0.71 & 0.54 & 0.45 & 0.34 & 0.27 & 0.33 & 0.22 & 0.17 \\ \hline
$\mathcal{O}_8$ & 0.34 & 0.79 & 0.98 & 0.98 & 0.81 & 0.60 & 0.32 & 0.43 & 0.43 & 0.39 & 0.34 & 0.44 \\ \hline
$\mathcal{O}_9$ & 0.79 & 1.04 & 1.10 & 0.83 & 0.62 & 0.41 & 0.29 & 0.17 & -0.05 & -0.08 & -0.06 & -0.11 \\ \hline
$\mathcal{O}_{10}$ & 0.68 & 1.01 & 1.20 & 0.82 & 0.74 & 0.07 & 0.18 & 0.44 & 0.91 & 0.08 & 0.86 & 0.12 \\ \hline
$\mathcal{O}_{11}$ & 0.83 & 1.08 & 1.05 & 0.87 & 0.80 & 0.01 & 0.00 & 0.35 & 0.63 & 0.53 & 0.70 & 0.59 \\ \hline
$\mathcal{O}_{12}$ & 0.88 & 1.04 & 1.09 & 0.87 & 0.60 & 0.00 & 0.14 & 0.70 & 0.86 & 0.85 & 0.85 & 0.88 \\ \hline
$\mathcal{O}_{13}$ & 1.15 & 1.20 & 1.01 & 0.63 & 0.14 & -0.30 & 1.12 & 0.61 & 0.56 & 0.53 & 0.49 & 0.35 \\ \hline
$\mathcal{O}_{14}$ & - & - & - & - & 0.76 & 0.17 & 0.14 & 0.12 & -0.03 & -0.08 & -0.06 & -0.15 \\ \hline
$\mathcal{O}_{15}$ & 1.21 & 1.24 & 1.01 & 0.37 & -0.31 & -0.43 & 1.36 & 1.16 & 0.56 & 0.64 & 0.71 & 0.63 \\ \hline
\end{tabular}
\label{wimp-p_table}
\end{table*}

\begin{table*}[h!]
\caption{Discrepancies between observed and expected exclusion limit for each model used for the inelastic WIMP-isoscalar results for 1 TeV WIMPs, expressed in terms of number of $\sigma$ from the expectation. Column headers are mass-splitting values, $\delta_m$. Positive entries indicate weaker sensitivity than the expected limit.}
\begin{tabular}{|c|c|c|c|c|c|}
\hline
\textbf{WIMP-s} & 0 keV & 50 keV & 100 keV & 150 keV & 200 keV \\ 
\hline \hline
$\mathcal{O}_1$ & 0.37 & 0.08 & 0.47 & 1.31 & 1.14 \\ \hline
$\mathcal{O}_3$ & 0.78 & 1.11 & 1.15 & 0.80 & 0.14 \\ \hline
$\mathcal{O}_4$ & 0.54 & 0.25 & 0.25 & 0.12 & -0.25 \\ \hline
$\mathcal{O}_5$ & 0.42 & 0.30 & 0.52 & 0.19 & -0.27 \\ \hline
$\mathcal{O}_6$ & 0.05 & 0.05 & 0.05 & -0.10 & -0.66 \\ \hline
$\mathcal{O}_7$ & 0.39 & -0.19 & -0.24 & -0.20 & -0.67 \\ \hline
$\mathcal{O}_8$ & 0.54 & 0.06 & 0.48 & 0.95 & 0.04 \\ \hline
$\mathcal{O}_9$ & -0.05 & -0.46 & -0.19 & -0.41 & -0.79 \\ \hline
$\mathcal{O}_{10}$ & 0.43 & 0.43 & 0.17 & 0.06 & -0.42\\ \hline
$\mathcal{O}_{11}$ & 0.53 & 0.29 & 1.10 & 1.46 & 0.28 \\ \hline
$\mathcal{O}_{12}$ & 0.75 & 0.58 & 1.22 & 1.30 & 0.42 \\ \hline
$\mathcal{O}_{13}$ & 0.98 & 1.08 & 1.25 & 1.02 & 0.20 \\ \hline
$\mathcal{O}_{14}$ & 0.99 & 0.09 & -0.14 & -0.48 & -0.55 \\ \hline 
$\mathcal{O}_{15}$ & 0.90 & 0.97 & 0.88 & 0.55 & 0.10 \\ \hline
\end{tabular}
\label{inelastic_table}
\end{table*}


\bibliography{EFTBib}

\end{document}